%% file: GrapheneSolitons.tex
\documentclass[prb,reprint]{revtex4-2}

\usepackage{preamble}

\graphicspath{{./Figures/}}

\newcommand{\energy}{\varepsilon}

\newcommand{\proj}{\mathcal{P}}
\newcommand{\absm}{\abs{m}}
\newcommand{\Kronecker}{\delta}
\DeclareMathOperator{\Li}{Li}
\DeclareMathOperator{\sgn}{sgn}

\makeatletter
\newcommand{\vast}{\bBigg@{4}}
\newcommand{\Vast}{\bBigg@{5}}
\makeatother
\newcommand{\vastl}{\mathopen\vast}
\newcommand{\vastr}{\mathclose\vast}

\newcommand{\epsilonVal}{0.1}

\newcommand{\bandwidthDim}{\SI{150}{\giga\hertz}}
\newcommand{\lRefDim}{\SI{50}{\nano\meter}}
\newcommand{\nDim}{\SI{4.0e9}{\per\centi\meter\squared}} 
\newcommand{\dDim}{\SI{890}{\nano\meter}} 
\newcommand{\TDim}{\SI{60}{\kelvin}} 
\newcommand{\TPertDim}{\SI{6.0}{\kelvin}} 
\newcommand{\currentDensityDim}{\SI{2.5}{\ampere\per\meter}}
\newcommand{\uDim}{\SI{4.0e5}{\meter\per\second}} 
\newcommand{\widthDim}{\SI{2.8}{\micro\meter}}
\newcommand{\tdDim}{\SI{44}{\pico\second}}
\newcommand{\muDim}{\SI{6.1e-22}{\joule}} 
\newcommand{\voltageDim}{\SI{3.8}{\milli\volt}} 
\newcommand{\voltagePertDim}{\SI{380}{\micro\volt}} 

\newcommand{\JouleTempRateDim}{\SI{3.0}{\kelvin\per\nano\second}}
\newcommand{\tJouleDim}{\SI{2.0}{\nano\second}}
\newcommand{\solitonHeightDim}{\SI{4.0e8}{\per\centi\meter\squared}} 
\newcommand{\AADim}{\SI{2.1e-35}{\joule\meter\squared}}
\newcommand{\PDim}{\SI{5.9e-8}{\newton\per\meter}} 
\newcommand{\sigmaDim}{\SI{0.048}{\per\kilo\ohm}} 
\newcommand{\etaDim}{\SI{4.8e-20}{\kilo\gram\per\second}} 
\newcommand{\tEEDim}{\SI{0.17}{\pico\second}} 
\newcommand{\tCharDim}{\SI{6.5}{\pico\second}} 
\newcommand{\tPhDim}{\SI{280}{\pico\second}} 
\newcommand{\csMassDim}{\SI{60}{\milli\joule\per\gram\per\kelvin}} 
\newcommand{\csAreaDim}{\SI{4.5e-9}{\joule\per\centi\meter\squared\per\kelvin}} 
\newcommand{\powerRadDim}{\SI{2.9e-7}{\kilo\watt\per\centi\meter\squared}} 
\newcommand{\radTempRateDim}{\SI{65}{\kelvin\per\second}} 
\newcommand{\tRadDim}{\SI{93}{\milli\second}} 
\newcommand{\kappaDim}{1} 

\newcommand{\sigmaNondim}{0.20} 

\newcommand{\AVal}{0.88}
\newcommand{\BVal}{-0.70}
\newcommand{\CVal}{-0.060}
\newcommand{\FVal}{-1.1}
\newcommand{\GVal}{0.53}

\newcommand{\sigmaVal}{0.63} 
\newcommand{\vVal}{0.43} 
\newcommand{\uVal}{0.40} 
\newcommand{\nVal}{1.0} 
\newcommand{\cVal}{1.0} 
\newcommand{\dVal}{1.0} 
\newcommand{\AAVal}{0.22}
\newcommand{\TVal}{0.70} 
\newcommand{\PDivTVal}{1.1} 
\newcommand{\muDivnVal}{1.1} 
\newcommand{\etaVal}{1.1}
\newcommand{\alphaVal}{0.439} 

\newcommand{\nUnits}{\SI{4e10}{\per\centi\meter\squared}} 
\newcommand{\dUnits}{\SI{50}{\nano\meter}} 
\newcommand{\AAUnits}{\SI{5.3e-36}{\joule\meter\squared}} 
\newcommand{\TUnits}{\SI{150}{\kelvin}} 
\newcommand{\PUnits}{\SI{8.4e-7}{\newton\per\meter}} 
\newcommand{\muUnits}{\SI{2.1e-21}{\joule}} 
\newcommand{\sigmaUnits}{\SI{0.24}{\per\kilo\ohm}} 
\newcommand{\etaUnits}{\SI{4.2e-20}{\kilo\gram\per\second}} 

\begin{document}

\title{Effects of dissipation on  solitons in the hydrodynamic regime of graphene}

\author{Thomas Zdyrski}
\author{John McGreevy}
\affiliation{University of California San Diego, La Jolla, California,
92093, USA}

\date{13 June 2019}

\begin{abstract}
  We use hydrodynamic techniques to analyze the one-dimensional
  propagation of solitons in gated graphene on an arbitrary uniform
  background current.
  Results are derived for both the Fermi liquid and Dirac fluid
  regimes.
  We find that these solutions satisfy the Korteweg-de Vries-Burgers
  equation.
  Viscous dissipation and ohmic heating are included, causing the
  solitons to decay.
  Experiments are proposed to measure this decay and thereby quantify
  the shear viscosity in graphene.
\end{abstract}


\maketitle

\section{Introduction}
Graphene offers a promising platform to realize and explore the
hydrodynamics of electrons~\citep{lucas2018hydrodynamics}.
Graphene serves as an excellent model system for theorists due to its
simple electronic band structure; likewise, it is utilized by
experimentalists for the relative ease of manufacturing pure samples.
In certain thermodynamic regimes, the electrons in graphene become
strongly interacting; hydrodynamics is a useful tool to study strongly
interacting systems not amenable to ordinary perturbation methods.
Hydrodynamics is applicable when systems rapidly thermalize and when
both the mean-free path ($l_{\text{ee}}$) and mean-free time
($\tau_{\text{ee}}$) are short compared to the relevant length and time
scales of the
problem~\citep{landau1959fluid}.
When a system is in this regime, the main observables are conserved
quantities: these are precisely the objects tracked by hydrodynamics.

Graphene has two different hydrodynamic regimes.
When the chemical potential $\mu$ is much larger than the temperature,
$k_B T \ll \mu$, graphene behaves like an ordinary conductor and is
described by Fermi liquid theory.
First discovered by \citet{landau1956theory} in 1959,
Fermi liquid theory treats the electrons as a non-interacting Fermi
gas and then turns on interactions adiabatically; thus, Fermi liquids
exhibit \emph{weakly} interacting quasiparticles.
The excitations, no longer pure electron states, are instead described
as quasiparticles.
Though weak interactions imply long mean-free paths, graphene can
actually exhibit hydrodynamic effects in this regime.
The electrons in graphene only weakly interact with phonons (which
typically disrupt the hydrodynamic signature), so it is still possible
to have $l_{\text{ee}} \ll l_{phonon}$.
Likewise, graphene samples can be made very pure; therefore, the
impurity scattering distances can be made large compared to the
mean-free path as well ($l_{\text{ee}} \ll l_{imp}$).

In the opposite limit---\ie when $\mu \ll k_B T$---graphene enters a
strongly coupled state known as a Dirac fluid (also known as a ``quantum
critical regime'').
In the Fermi liquid regime, the presence of a Fermi surface imposes
strong kinematic constraints on the possible scattering pathways;
this prevents electrons far from the Fermi surface from interacting
strongly.
However, near charge neutrality, the Fermi surface shrinks, allowing
electrons to interact \emph{strongly}.
The bare coupling constant $\alpha_0$ gives a measure of this
interaction strength.
In the Dirac regime of graphene, $\alpha_0$ can be of order unity;
renormalization reveals the coupling to be marginally irrelevant, but
for many laboratory conditions, it can still be on the order of
\numrange{0.1}{0.5}: see \citet{lucas2018hydrodynamics} for more
details.
This strong coupling makes Dirac fluids ideal candidates for
hydrodynamic analysis.

A hydrodynamic analysis of electron motion in graphene is governed by a
number of phenomenological parameters.
A derivative expansion can be utilized to derive the hydrodynamic
equation~\citep{lucas2018hydrodynamics}.
The first-order corrections contain three such parameters: the shear
viscosity $\eta$, the bulk viscosity $\zeta$, and the ``intrinsic''
conductivity $\sigma_Q$.
These cannot be predicted from the hydrodynamic theory and must be
measured or calculated microscopically.

A number of experiments have measured the value of intrinsic
conductivity~\citep{novoselov2005two,crossno2016observation}.
Similarly, there have been a number experimental
proposals~\citep{torre2015nonlocal,tomadin2014corbino,levitov2016electron,dyakonov1993shallow}
for measuring $\eta$.
While there have been a few
measurements~\citep{bandurin2016negative,kumar2017superballistic} of
$\eta$ in the Dirac regime, many of the proposals---such as negative
nonlocal resistance measurements~\citep{levitov2016electron}---only
apply to the Fermi regime~\citep{lucas2018hydrodynamics}.
Therefore, different hydrodynamic predictions would be useful for
investigating $\eta$ in Dirac fluids.

Solitons---disturbances that propagate without changing shape, even
after interacting with each other---serve as prototypical hydrodynamics
phenomena amenable to analytic tools.
Solitons are made possible when dispersion balances
focusing-nonlinearities.
Graphene's hydrodynamic regime supports collective electron/hole sound
waves called ``first-sound'' modes~\citep{lucas2018electronic} or
``demons''~\citep{sun2016adiabatic}; these sound modes can become
solitons if dispersion balances focusing.
\Citet{akbari2013universal} analyzed solitons and periodic waves in
both the 2D and 3D completely degenerate ($T=0$) Fermi regimes.
Solitons are permitted due to the inherently nonlinear nature of the
hydrodynamic equations; to capture this behavior,
a Bernoulli pseudo-potential was used to analyze the fully nonlinear
equations.
However, while this method predicted some parameters---such as minimum
propagation speeds---it did not generate an analytic expression for the
soliton's profile.

A different approach to studying solitons was presented by
\citet{svintsov2013hydrodynamic} using standard perturbation theory.
This produced a Korteweg-de Vries (KdV) equation to describe the
solitons' propagation and generated analytic approximations to the
disturbances' shapes.
Unlike the analysis of \citet{akbari2013universal},
this linearized approach lacked a dispersive term to balance the
nonlinearities.
Instead, the graphene was placed on a gated substrate; this provided a
weak dispersive force that permitted the formation of solitons.

While the analysis of solitons by \citet{svintsov2013hydrodynamic}
provided a more concrete result, it was limited to inviscid Fermi
liquids.
The present study will extend the results to include the Dirac regime as
well.
Whereas \citet{svintsov2013hydrodynamic} used kinetic
theory, we will instead treat the system using a systematic hydrodynamic
expansion.
Additionally, this paper will extend the results of both
\citet{svintsov2013hydrodynamic} and \citet{akbari2013universal} by
including the effects of dissipation.
This allows us to propose new experiments to measure the viscosity of
the electron fluid.
The derivation presented here is applicable to either the Dirac ($\mu
\ll k_B T$) or Fermi ($k_B T \ll \mu$) regime, though it is unable to
interpolate between the two.
Nevertheless, our proposal offers an advantage over transport
measurements in that its interpretation is less theory-laden.

In \cref{sec:derivation} we will derive the governing equations.
\Cref{sec:normalization} will be devoted to the subtle aspects of
normalization.
Next, \cref{sec:stationary} will detail the perturbation expansion
for the special case of stationary solitons.
\Cref{sec:multiple_scales} extends the analysis to the more general case
of solitons on an arbitrary background flow.
We will provide a short analysis of the results in \cref{sec:analysis}.
Finally, in \cref{sec:experimental}, we will detail potential
experimental setups using these solitons to measure graphene's
viscosity.

\section{\label{sec:derivation} Governing Equations}
The electrons in graphene satisfy a pseudo-relativistic dispersion
relation~\citep{lucas2018hydrodynamics}
\begin{equation}
  \energy(\vec{p}) = \pm v_F \abs{\vec{p}} \,,
\end{equation}
with $\vec{p}$ the momentum, $v_F \approx c/300$ the Fermi velocity,
and $\energy(\vec{p})$ the energy density.
This equation is valid near a Dirac point at $\vec{p}=0$, and deviates
from linearity when $\abs{\vec{p}} a/\hbar \approx 1/2$ with $a$ the
distance between adjacent carbon atoms in the graphene.

Given the pseudo-relativistic dispersion, it is natural to write the
conserved currents in relativistic notation with $x^{\mu} =
(v_F t, \vec{x})^{\mu}$ and $\partial_{\mu} = (\partial_t/v_F,
\grad)_{\mu}$.
Ignoring impurity and phonon scattering, the equations of motion
are~\citep{lucas2018hydrodynamics}
\begin{gather}
  \partial_{\mu} J^\mu = 0 \,,\\
  \partial_{\mu} T^{\mu \nu} = \frac{1}{v_F} F^{\nu \mu} J_{\mu} \,.
\end{gather}
Here, $T^{\mu \nu}$ is the energy-momentum tensor, and $F^{\mu \nu}$ is
the
electromagnetic tensor (including self-interactions).
Additionally, $J^{\mu}$ is the charge 4-current~%
\footnote{
  Note that some of our variable definitions differ from those of
  \citet{lucas2018hydrodynamics} to better match usual conventions.
  The relevant changes (with the variables of
  \citet{lucas2018hydrodynamics} subscripted with L) are $J^{\mu} = -e
  J^{\mu}_{\text{L}}$, $F^{\mu,\nu} = -F^{\mu \nu}_{\text{L}}/e$, and
  $\sigma_Q = e^2 \sigma_{Q, \text{L}}$.
}.
Note that we will be using Gaussian units with $e = \abs{e}$ positive.
Finally, we will include a factor of $v_F$ in the time-like components
of four-vectors, like $x^{\mu} = (v_F t, \vec{x})^{\mu}$, so that the
metric $g^{\mu \nu} = \text{diag}(-1,1,1,1)^{\mu \nu}$ is dimensionless.

It is often preferable to write these equations in terms of more
conventional quantities such as the fluid 3-velocity $\vec{u}$ and the
(rest-frame) number density of charge carriers, $n = (n_{el} -
n_{hol})$, with $n_{el}$ ($n_{hol}$) the number density of electrons
(holes).
To do so, $J^{\mu}$ and $T^{\mu \nu}$ are expanded in the small
parameter $l_{\text{ee}} \delta$.
In this equation, $l_{\text{ee}}$ is the electron-electron scattering mean
free path and $\delta$ is a characteristic inverse length scale of the
observables.
Since $\delta \sim \partial$ (with the partial derivative acting on slow
observables) this is called the derivative expansion: see
\citet{lucas2018hydrodynamics} for more details.

The expansions for $T^{\mu \nu}$ and $J^{\mu}$ become unwieldy at
higher orders, but truncating at order $l_{\text{ee}} \delta$~%
\footnote{
  Note that, as mentioned previously, $\delta \sim \partial$; the factor
  of $l_{\text{ee}}$ is implicit in the definitions of the dissipative
  coefficients $\sigma_Q$, $\eta$, and
  $\zeta$~\citep{lucas2018hydrodynamics}.
}
we find~\citep{lucas2018hydrodynamics}
\begin{gather}
  J^{\mu} = - en u^{\mu} + \frac{\sigma_Q}{e} \proj^{\mu
    \nu} \left( \partial_{\nu} \mu - \frac{\mu}{T} \partial_{\nu}T+ e
    F_{\nu \rho} u^{\rho} \right) \,,\\
  \begin{aligned}
  T^{\mu \nu} &= (\energy + P) \frac{u^{\mu}}{v_F} \frac{u^{\nu}}{v_F}
    + P g^{\mu \nu} - \eta \proj^{\mu \rho}
    \proj^{\nu \alpha} \bigl( \partial_{\rho} u_{\alpha} \\
  & \quad + \partial_{\alpha} u_{\rho} - \frac{2}{d}
    g_{\rho \alpha} \partial_{\beta} u^{\beta}\bigr) - \zeta \proj^{\mu
    \nu} \partial_{\alpha} u^{\alpha} \,,
  \end{aligned}
\end{gather}
with $\energy$ the energy density $P$ pressure, $\mu$ chemical
potential, and temperature $T$ in the rest frame.
We have defined the spacelike projection operator
$\proj^{\mu \nu} \coloneqq g^{\mu \nu} + u^{\mu} u^{\nu}/v_F^2$ and used
$u^{\mu} u_{\mu} = -v_F^2$ to write the four-velocity as $u^{\mu} =
\gamma(v_F,\vec{u})$ with $\gamma = 1/\sqrt{1-(\abs{\vec{u}}/v_F)^2}$ a
Lorentz factor.
Further, we have chosen the Landau frame, where
\begin{equation}
  u_{\mu} J^{\mu} = e n v_F^2 \qq{and} u_{\mu} T^{\mu \nu} = -\energy
    u^{\mu} \,.
\end{equation}

It is sometimes more instructive to write-out four-vectors in terms of
their three-vector and time-like components.
For instance, $J^{\mu}$ is
\begin{align}
  J^0 &= -\gamma e n v_F + \frac{\sigma_Q}{e} \Bigl[ \frac{T
      \gamma^2}{v_F} \left(
      \frac{\abs{\vec{u}}^2}{v_F^2} \pdv{t} + \vec{u} \vdot \grad
      \right) \left( \frac{\mu}{T} \right) \nonumber \\
  &\qquad + \gamma e \frac{\vec{E} \vdot \vec{u}}{v_F} \Bigr] \,, \\
    \vec{J} &= -\gamma e n \vec{u} + \frac{\sigma_Q}{e}
    \Bigl[ T \left(\grad + \gamma^2 \frac{\vec{u}}{v_F^2} D\right)
    \left( \frac{\mu}{T} \right) \nonumber \\
  &\qquad + \gamma e \left( \vec{E} + \frac{\vec{u}}{v_F} \cross \vec{B}
    \right) \Bigr] \,.
\end{align}
where $D \coloneqq \partial_t + \vec{u} \vdot \grad$ is a material
derivative.

To facilitate comparison with the existing literature, it is
useful to re-write the spacelike components as $v_F \partial_{\nu} T^{i
\nu} - u^i \partial_{\nu} T^{0 \nu} = v_F F^{\mu i} J_{\mu} - u^i
F^{\mu 0} J_{\mu}$.
Thus, our system becomes
\begin{gather}
  \partial_{\mu} J^{\mu} = 0 \,,
  \label{eq:charge_cons}\\
  \partial_{\nu} T^{0 \nu} = F^{\mu 0} J_{\mu} \,,
  \label{eq:energy_cons}\\
  v_F \partial_{\nu} T^{i \nu} - u^i \partial_{\nu} T^{0 \nu} =
    v_F F^{\mu i} J_{\mu} - u^i F^{\mu 0} J_{\mu}
  \label{eq:mom_cons} \,.
\end{gather}

\subsection{Ideal Fluid}
It is illuminating to temporarily consider the dissipationless case
$\sigma_Q = \eta = \zeta = 0$.
We are then able to write \crefrange{eq:charge_cons}{eq:mom_cons} in
three-vector notation as
\begin{gather}
  \pdv{t}(\gamma n) + \div(\gamma n \vec{u}) = 0 \,,\\
  \begin{aligned}
  &\pdv{t}(\gamma^2 (\energy+P)) + \div(\gamma^2
    (\energy + P) \vec{u}) - \pdv{P}{t}=\\
  & \qquad -\gamma n e \vec{E} \vdot \vec{u} \,,
  \end{aligned}\\
  \begin{aligned}
  &\gamma^2 \frac{(\energy + P)}{v_F^2} \left(\pdv{\vec{u}}{t} +
    \vec{u}\vdot\grad\vec{u}\right) + \left(\frac{\vec{u}}{v_F^2}
    \pdv{P}{t} + \grad{P}\right) =\\
  &\qquad- n e \gamma \left(\vec{E} + \frac{\vec{u}}{v_F} \cross \vec{B}
    - \frac{\vec{u}}{v_F} \vec{E} \vdot \frac{\vec{u}}{v_F} \right) \,.
  \end{aligned}
\end{gather}
Then, it is clear that \crefrange{eq:charge_cons}{eq:mom_cons}
represent charge, energy, and $3$-momentum conservation, respectively.

\subsection{\label{sec:phonons} Phonons and Heat Flow}
We have neglected the interactions (emission, absorption, and
scattering) with phonons in our governing equations,
\crefrange{eq:charge_cons}{eq:mom_cons}; we will now attempt to justify
that choice.
First, we consider the momentum equation \cref{eq:mom_cons}.

The hydrodynamic regime is relevant when the
electron-electron interaction time $t_{\text{ee}}$ is the smallest
timescale: $t_{\text{ee}} \ll t_{\text{char}} \ll t_d$ with
$t_{\text{char}}$ the soliton's propagation timescale and
$t_d$ its dissipation timescale.
Following the standard
prescription~\citep{gurzhi1963minimum,crossno2016observation,bandurin2018fluidity,lucas2018hydrodynamics},
we will neglect phonon-induced momentum relaxation in the momentum
conservation equation, \cref{eq:mom_cons}, if the phonon-induced
momentum-relaxation time $t^{(p)}_{\text{e-ph}}$ is much longer than the
other timescales of interest, $t_{\text{ee}} \ll t_{d} \ll t_d \ll
t^{(p)}_{\text{e-ph}}$.

To support the claim that such a regime exists, we now present
sample numerical values that satisfy such a timescale hierarchy.
Nevertheless, we stress that this is simply an example; the derivation
in the remainder of the paper will be valid over a wide range of
experimental parameters; see \cref{sec:nondim} for further details.

The electron-electron scattering time in the Dirac regime
is~\citep{lucas2018hydrodynamics}
\begin{equation}
  t_{\text{ee}} \sim \SI{0.1}{\pico\second} \times \pqty{
    \frac{\SI{100}{\kelvin}}{T}} \,.
\end{equation}
At $T=\TDim$, this gives $t_{\text{ee}} = \tEEDim$.
Using the sample values chosen in \cref{sec:experimental}, we find (\cf
\cref{sec:source_signal}) a characteristic propagation time of
$t_{\text{char}} = \tCharDim$.
In that same section, we calculate a decay time of $t_d \approx \tdDim$.
Finally, the electron-phonon momentum-relaxation time for acoustic
phonons (with speed $v_s = \SI{2e4}{\meter\per\second}$) is given
by~\citep{stauber2007electronic}
\begin{equation}
  t^{(p)}_{\text{e-ph}} \sim
    \frac{\SI{10}{\pico\second}}{\pqty{T/\SI{100}{\kelvin}}
    \sqrt{n/(\SI{e12}{\per\centi\meter\squared})}} \,.
\end{equation}
This yields $t^{(p)}_{\text{e-ph}} = \tPhDim$.
Therefore, we see that we have $t_{\text{ee}} \ll t_{\text{char}} \ll
t_d \ll t_{\text{e-ph}}$.
Thus, with the experimental values chosen here, phonon-induced momentum
relaxation can be neglected from \cref{eq:mom_cons}.

Importantly, as shown in recent
experiments~\citep{crossno2016observation}, there does appear to
exist an experimentally realizable regime where the requisite
hydrodynamic condition $t_{\text{ee}} \ll t_{char} \ll
t^{(p)}_{\text{e-ph}}$ holds.
Indeed, these experiments motivate us to suggest that such an
approximation might be valid.
Nevertheless, it would be useful to have a more refined estimate of the
rate at which momentum and energy are lost to phonons.

\subsubsection{\label{sec:iso_vs_adi} Isothermal \vs Adiabatic}
Now we consider the effect of phonons on the energy conservation
equation \cref{eq:energy_cons}.
The energy conservation equation implicitly assumes our system is
adiabatic: that is, the absence of energy sources/sinks presumes that
heat neither enters nor leaves the system.
In general, we could include terms (such as coupling to phonons)
representing heat gain/loss.
Instead, we could consider the opposite limit involving rapid heat
transfer with the environment resulting in isothermal conditions.
Under this assumption, the energy conservation equation is no longer
needed; rather, the thermodynamic relations of \cref{sec:thermo} could
be used to relate our dynamic variables $P$ and $n$, since $T$ would no
longer be dynamical.
Therefore, (as in the case of Newton's calculation of sound-speed in
air) it is important to determine whether adiabatic or isothermal
conditions are more applicable.

The most likely thermalization pathway would involve energy loss to
phonons: the soliton's location in the middle of the sample minimizes
heat advection through the edge contacts; similarly, radiative cooling
is far too slow to thermalize the system on relevant timescales~%
\footnote{
  The Stefan-Boltzmann law would give a power loss rate of $P_r = \sigma
  \epsilon \bqty{(T_0+ T_1)^4-T_0^4} \approx 4 \sigma \epsilon T_0^3
  T_1$, with $\sigma =
  \SI{5.67e-8}{\watt\per\meter\squared\per\kelvin\tothe{4}}$ and
  $\epsilon \le 1$ graphene's emissivity.
  Using $\epsilon \approx \SI{1}{\percent}$~\citep{freitag2010thermal},
  $T_0 = \TDim$, and $T_1 = \epsilonVal T_0 = \TPertDim$, we find a
  power loss density of $P_r = \powerRadDim$.

  As we will calculate in \cref{sec:joule}, graphene has a specific heat
  of $c_s = \csAreaDim$.
  Therefore, the soliton's temperature will change at a rate of $P_r/c_s
  = \radTempRateDim$.
  Hence, it would take approximately $T_1 c_s / P_r = \tRadDim$
  for the system to thermalize with the environment via radiation.
}.
Indeed, if the graphene is placed on a substrate, phonons are
responsible for the majority of the heat transfer to the
environment~\citep{ong2011effect,chen2017coupled}.

For the isothermal condition to be applicable, the electrons
must quickly lose energy to the environment: that is, the
energy-relaxation time $t^{(\energy)}_{\text{e-ph}}$ must satisfy
$t^{(\energy)}_{\text{e-ph}} \le t_{\text{ee}} \ll t_{\text{char}} \ll
t^{(p)}_{\text{e-ph}}$.
However, single-phonon interactions are unlikely to extract heat quickly
enough.
Each phonon with wavenumber $k$ carries a momentum $\hbar k$ while the
electron fluid has momentum density $u (\energy+P)/v_F^2 \sim u
\energy/v_F^2$.
Likewise, phonons have energy $\hbar k v_s$ with sound speed $v_s$,
while the electrons have energy density $\energy$.
Recall that we require the electron-electron momentum exchange rate
$\dot{p}_{\text{ee}}$ to be much greater than the electron-phonon momentum
relaxation rate $\dot{p}_{\text{e-ph}}$ in order for hydrodynamics to be
valid: $\dot{p}_{\text{ee}} \gg \dot{p}_{\text{e-ph}}$.
However, multiplying by $v_s$ and re-writing in terms of the energy
exchange rates yields $\dot{\energy}_{\text{ee}} u v_s / v_F^2 \gg
\dot{\energy}_{\text{e-ph}}$.
Given that $v_s \approx \SI{2e4}{\meter\per\second} \ll v_F$ for
acoustic phonons~\citep{virtanen2014energy} and $u \approx \uDim \sim
v_F$ for our system, we see that $\dot{\energy}_{\text{ee}} \gg
\dot{\energy}_{\text{e-ph}}$.
Hence, if phonon-induced momentum relaxation can be neglected, so can
phonon-induced energy relaxation.

For isothermal conditions to be applicable, other thermalization
pathways must be available.
For instance, multiphonon supercollisions~\citep{virtanen2014energy} can
increase the energy flux relative to the momentum flux.
However, under the assumption of weak phonon coupling, we can ignore the
influence of multiphonon processes.
Therefore, in the absence of other energy-relaxation mechanisms, it
appears that adiabatic conditions are more appropriate for our system,
with $t_{\text{ee}} \ll t_{\text{char}} \ll t^{(p)}_{\text{e-ph}} \ll
t^{(\energy)}_{\text{e-ph}}$.

In the body of this paper, we will use isothermal conditions: these are
more common in the
literature~\citep{akbari2013universal,svintsov2013hydrodynamic} and are
somewhat simpler.
Nevertheless, adiabatic conditions appear to be more practical and are
used for the derivation in \cref{sec:adiabatic}.

\subsection{\label{sec:thermo} Thermodynamics}
Currently, our system, \cref{eq:charge_cons,eq:mom_cons}, is
underdetermined.
This can be remedied by including a thermodynamic equation of state to
relate $\energy$ and $P$.

In graphene, the photon-like dispersion relation for the electrons
gives the pressure as $P = \energy d$, with $d$ the dimension of the
system ($d=2$ for graphene)~\citep{lucas2018hydrodynamics}.
Graphene has a natural energy scale at which the band structure's
curvature becomes relevant.
However, for temperatures much lower than this scale, $\Lambda \sim
\SI{e4}{\kelvin}$, there are only two energy scales in the problem: $k_B
T$ and $\mu$.
Therefore, from dimensional analysis, the pressure must be expressed
as~\citep{lucas2018hydrodynamics}
\begin{equation}
  P(\mu, T) = \frac{(k_B T)^{d+1}}{(\hbar v_F)^d} F\pqty{\frac{\mu}{k_B
  T}} \,,
\end{equation}
for a function $F$ subject to constraints imposed by the positivity of
the entropy density $s = \pdv*{P}{T} \ge 0$.
Additionally, since our system is charge conjugation symmetric with $\mu
\to -\mu$, $F$ must be an even function.

{
\renewcommand{\theequation}{Dirac: \arabic{equation}}
In the Dirac regime ($\mu \ll k_B T$), $P$ can be expanded as
\begin{align}
  P(\mu, T) &= \frac{(k_B T)^{d+1}}{(\hbar v_F)^d} \Bigl[\mathcal{C}_0^D
    + \mathcal{C}_1^D \pqty{\frac{\mu}{k_B T}}^2 \nonumber \\
  &\qquad + \mathcal{C}_2^D \pqty{\frac{\mu}{k_B T}}^4 + \ldots \Bigr] \,.
\end{align}
Similarly, the carrier density can be expressed as
\begin{align}
  n(\mu, T) &= \pdv{P}{\mu} = \frac{(k_B T)^d}{(\hbar v_F)^d}
    \frac{\mu}{k_B T} \Bigl[2 \mathcal{C}_1^D  \nonumber \\
  &\qquad + 4 \mathcal{C}_2^d \pqty{\frac{\mu}{k_B T}}^2 + \ldots \Bigr]
      \,.
\end{align}
}

{
\renewcommand{\theequation}{Fermi: \arabic{equation}}
Instead, in the Fermi regime ($\mu \gg k_B T$), we can write $P$ as
\begin{align}
  P(\mu, T) &= \frac{\abs{\mu}^{d+1}}{(\hbar v_F)^d} \Bigl[
    \mathcal{C}_0^F + \mathcal{C}_1^F \pqty{\frac{k_B T}{\mu}}^2
    \nonumber \\
  &\qquad + \mathcal{C}_2^F \pqty{\frac{k_B T}{\mu}}^4 + \ldots \Bigr] \,.
\end{align}
Likewise, the carrier density is given by
\begin{align}
  n(\mu, T) &= (d+1) \frac{\abs{\mu}^d \sgn{\mu}}{(\hbar v_F)^d} \Bigl[
    \mathcal{C}_0^F + \frac{d-1}{d+1} \mathcal{C}_1^F \pqty{\frac{k_B
    T}{\mu}}^2 \nonumber \\
  &\qquad + \frac{d-3}{d-1} \mathcal{C}_2^F \pqty{\frac{k_B T}{\mu}}^4 +
    \ldots \Bigr] \,.
\end{align}
}

Throughout the remainder of this paper, we will generically write
$\mathcal{C}_0$, $\mathcal{C}_1$, \etc; the current regime of interest
will determine whether to use $\mathcal{C}^D$ or $\mathcal{C}^F$.
Explicit expressions for these coefficients are given in
\cref{sec:thermo_coeffs}.
It is important to reiterate that, for our isothermal system, $T$ is not
a dynamical quantity dependent on space or time, but is merely a
parameter.

\subsection{\label{sec:electrodynamics} Electrostatics}
While our electron fluid moves in $d$-dimensions ($d=2$ for graphene),
we will assume the electromagnetic field propagates in $d+1$ dimensions
(\ie 3-space for graphene, as usual).
We are only concerned with the electric potential $\phi$ since the
magnetic terms are smaller by a factor of $v_F/c \approx 1/300$.
The self-interaction of the charge distribution
$n(x,t)$ generates an electric potential in the Lorenz gauge as
\begin{equation}
  -\frac{1}{c^2} \pdv[2]{\phi}{t} + \laplacian{\phi} = -4 \pi J^0 = -4
  \pi [-e n(x,t) \gamma ] \,.
\end{equation}
Note that we are using the $d+1$-dimensional Laplacian.
Neglecting the $1/c^2$ time derivative gives Poisson's equation.
For instance, with $d=2$; this gives
\begin{equation}
  \phi(\vec{x},t) = -e \int \frac{n(y,t) \gamma}{\abs{x-y}}
    \dd[3]{y} \,.
\end{equation}
Making the quasi-static approximation that $\partial_t/\partial_x \ll
c$---so we can neglect electrodynamic effects like $\partial_t
\vec{A}$---we find
\begin{equation}
  \vec{E} = e \int \frac{(\vec{x} - \vec{y}) n(y,t)
    \gamma}{\abs{x-y}^{3}} \dd[3]{y} \,.
\end{equation}
This equation is highly non-local in $n$, and using it in the
energy-momentum tensor equation would produce a complicated
integro-differential equation.
While we can deal with this (via a Fourier transform) for the linear
approximation, going to higher orders would necessarily involve
convolutions.

The main problem with this setup is that the Coulomb force is
long-ranged; we can simplify this by using conducting gates.
Since the electric field lines must be normal to conductors, placing
conductors directly above and below the graphene will force $\vec{E}$ to
be nearly normal to the
graphene~\citep{svintsov2013hydrodynamic,govorov1999solitons}.
Therefore, the $x$-component $E_x$ will necessarily be small and can be
handled perturbatively.

We impose gates a distance $d_1$ above and $d_2$ below the sample and
fill the intervening space with a dielectric of relative permittivity
$\kappa$.
This gives a potential (in $d=2$) of the
form~\citep{svintsov2013hydrodynamic}
\begin{equation}
  \phi = \frac{-\alpha \hbar v_F d_1 d_2}{e \kappa(d_1 + d_2)} \left( 1
    + \frac{d_1 d_2}{3} \pdv[2]{x} \right) (\gamma n) + \order{d_i
    \partial_x}^4 \,.
\end{equation}
Naturally, the electric field is given by the negative gradient of
$\phi$.
Here, we have assumed that $d_i \partial_x \ll 1$.
Furthermore, we have replaced $4 \pi e^2/\hbar v_F$ with $\alpha(T)$, the
renormalized coupling constant; this accounts for the effect of
screening and is given by~\citep{lucas2018hydrodynamics}
\begin{equation}
  \alpha(T) = \frac{4}{(4/\alpha_0) + \ln(\SI{e4}{\kelvin}/T)} \,,
  \label{eq:alpha_renorm}
\end{equation}
with $\alpha_0 \approx 1$ depending on the graphene's substrate.
For the Dirac regime at $T = \TDim$ considered throughout
this paper, this gives $\alpha \approx \alphaVal$.

For convenience, we will define the collection of coefficients
\begin{equation}
  A \coloneqq \frac{\alpha \hbar v_F d_1 d_2}{\kappa (d_1 + d_2)} \,,
  \label{eq:A_def}
\end{equation}
so that the potential is given as
\begin{equation}
  \phi = -\frac{A}{e} \left( 1 + \frac{d_1 d_2}{3} \pdv[2]{x} \right)
    (\gamma n) + \order{d_i \partial_x}^4 \,.
  \label{eq:electric_potential}
\end{equation}
While \cref{eq:A_def} only applies for $d=2$, we will use $\phi$ given
by \cref{eq:electric_potential} for arbitrary dimension, with
an appropriately chosen $A$.

The first term on the right-hand side of \cref{eq:electric_potential}
represents the electric potential from a uniform charge density.
The second term is a weakly non-local correction that causes a weak
dispersion.

\section{\label{sec:normalization} Dimensions, Units, and Regime of
Interest}
It will be helpful in the following sections to be rather precise
in specifying a nondimensionalization scheme.
For convenience, we will choose units where $k_B = \hbar = v_F = e = 1$.
We still have one dimension unspecified; in order to fully specify our
unit system, we will choose an arbitrary reference length $l_{\text{ref}} =
\lRefDim$; this is chosen so that $T$ is nondimensionalized to roughly
unity (see below)~
\footnote{
  After choosing $\hbar=v_F=k_B=e=1$, all quantities will be expressed
  in various powers of length.
  If the parameters have been chosen correctly, there will exist a
  characteristic length $\Xi$ shared by all quantities.
  It is most convienent to choose $l_{\text{ref}}=\Xi$,
  though it is not strictly necessary---choosing $l_{\text{ref}}$
  otherwise will multiply all terms in each equation by the same factor
  of $l_{\text{ref}}/\Xi$.
}.

In later sections, we will be performing a perturbation expansion to
solve the nonlinear system of equations.
There, we will use expansions of the form $f = f_0 + \epsilon f_1 + f_2
\epsilon^2 + \ldots$ with $\epsilon \ll 1$ a small parameter
representing the size of perturbations.

Choosing the order of the problem's variables is very important.
When collecting terms in perturbation theory, we assume that all
variables and constants are order $\order{1}$; the relative magnitude of
terms is given solely by powers of $\epsilon$.
Let us emphasize that, unlike the choice of parameters to normalize
above, this choice of nondimensionalization \emph{is} physically
relevant and determines our regime of interest.

Nondimensionalization sets the relative size of different terms and
corresponds to a specification of our location in parameter space.
Indeed, this choice dictates which terms and processes are relevant and
which are negligible.
Equivalently, this process can be viewed through the lens of
dimensional analysis.
Our system has seventeen variables (5 dynamic $n$, $u$, $\energy$,
$P$, and $\mu$; 11 static: $x$, $t$, $k_B T$, $d_i$,
$\kappa$, $\sigma_Q/e^2$, $\eta$, $\zeta$, $\hbar$, $v_F$, and
$l_{\text{ref}}$;
and the previously defined perturbation scale $\epsilon$).
In total, there are 3 independent physical units (mass, length, and time).
Therefore, the Buckingham Pi theorem implies there are 14 dimensionless
parameters.

However, these 14 dimensionless parameters are not all independent.
Our 3 thermodynamic equations ($\energy = Pd$, as well as the
definitions of $P$ and $n$) reduce this number to 11.
Furthermore, we have not yet specialized to solitons:
in \cref{sec:nondim}, we will use dominant balance to impose 4
additional restrictions arising from our conservation equations,
\crefrange{eq:charge_cons}{eq:mom_cons}.
This leaves a total of 7 independent nondimensional parameters:
$\epsilon$, $m$, $p$, $q$, $\order{\sigma_Q \hbar}$,
$\order{\eta l_{\text{ref}}^d/\hbar}$, and $\order{\zeta l_{\text{ref}}^d/\hbar}$, as
defined in \cref{sec:nondim}~%
\footnote{
  As discussed in \cref{sec:diss_order}, we could introduce three
  additional microscopic equations and eliminate $\eta$, $\zeta$, and
  $\sigma_Q/e^2$ as independent quantities.
  However, we will refrain from doing so.
}.

Naturally, investigations of the Fermi and Dirac regimes entail
different nondimensionalizations.
Additionally, even without a set regime, there are different
nondimensionalization choices highlighting different areas of parameter
space.
\Cref{sec:nondim} outlines a general nondimensionalization using
dominant balance that encompasses various parameter spaces in both the
Dirac and Fermi regimes.
For concreteness, we will examine one particular nondimensionalization
in the Dirac regime in this section.
Nevertheless, the equations and solutions generated in the remainder of
the paper are largely similar for both the Dirac and Fermi regimes; we
will explicitly highlight the few terms that do differ between the two
regimes.
The nondimensionalization utilized in the Fermi regime is laid out in
\cref{sec:param_choice}.

\subsection{Dirac Nondimensionalization}
We will denote nondimensional variables with a caret.
Restricting to the Dirac regime and using a bit of foresight, we will
choose to nondimensionalize the dynamical and thermodynamic variables as
follows:
\begin{equation}
  \begin{aligned}
    n &= \epsilon^{(d+2)/4} \hat{n} l_{\text{ref}}^{-d} \,, &  u &= \hat{u} v_F
      \,, \\
    \energy &= \epsilon^{(d+1)/4} \hat{\energy} \hbar v_F l_{\text{ref}}^{-d-1}
      \,,
    & P &= \epsilon^{(d+1)/4} \hat{P} \hbar v_F l_{\text{ref}}^{-d-1} \,, \\
    \mu &= \epsilon^{3/4} \hat{\mu} \hbar v_F l_{\text{ref}}^{-1} \,, \qq{and}
    & T &= \epsilon^{1/4} \hat{T} \hbar v_F l_{\text{ref}}^{-1} k_B^{-1} \,.
  \end{aligned}
\end{equation}
Here, we made use of the fact that we are in the Dirac regime
($\mu/T \ll 1$) and the thermodynamic equations of \cref{sec:thermo} by
ensuring
\begin{equation}
  \order{n l_{\text{ref}}^d} = \order{\frac{\mu l_{\text{ref}}^{d+1}}{\hbar v_F}
    \pqty{\frac{T l_{\text{ref}}}{\hbar v_F}}^{d-1} }
\end{equation}
and
\begin{equation}
  \order{\frac{P l_{\text{ref}}^{d+1}}{\hbar v_F}} =
  \order{\frac{k_B T l_{\text{ref}}}{\hbar v_F}}^{d+1}.
\end{equation}
Note that we took $\mu$ to be small but finite; as we will see later,
taking $\mu$ to be identically zero causes disturbances to be ``frozen''
in place.

The gating distance will be normalized as $d_i = l_{\text{ref}} \hat{d}_i
\epsilon^{-(d+3)/4}$.
The electrostatic coefficient $A$ [defined for $d=2$ in \cref{eq:A_def}]
is normalized as $A = \hat{A} \epsilon^{-(d+3)/4} \hbar l_{\text{ref}}^{d-1} v_F$.

The dissipative ``intrinsic'' conductivity $\sigma_Q \hbar/e^2$
represents another non-dimensional parameter in our problem.
In the hydrodynamic regime for $d=2$, we have~\citep{fritz2008quantum}
\begin{equation}
  \frac{\sigma_Q}{e^2} \approx \frac{0.760}{2 \pi \hbar \alpha(T)^2} \,,
\end{equation}
with $\alpha(T)$ given by \cref{eq:alpha_renorm}.
We see that for $T \approx \TDim$, we have $\sigma_Q = \sigmaNondim
e^2/\hbar$.
Therefore, $\sigma_Q \hbar/e^2$ is now a second small parameter (in
addition to $\epsilon$).
To make progress with our perturbation expansion we need to fix the
magnitude of $\sigma_Q \hbar/e^2$ relative to $\epsilon$.
Since we will later choose $\epsilon \sim \epsilonVal$, we see that
$\hat{\sigma}_Q = \sigmaNondim \approx \sqrt{\epsilonVal}$.
Thus, we will nondimensionalize $\sigma_Q$ as $\sigma_Q = \hat{\sigma}_Q
\epsilon^{1/2} e^2 l_{\text{ref}}^{2-d}/\hbar$.

According to \citet{lucas2018hydrodynamics}, near the charge neutrality
point with $d=2$, the shear viscosity is given by
\begin{equation}
  \eta \approx 0.45 \frac{(k_B T)^2}{\hbar v_F^2 \alpha(T)^2} \,.
\end{equation}
For $T \approx \TDim$, we have $\eta l_{\text{ref}}^2 / \hbar =
\etaVal$.
Therefore, we will choose $\eta = \epsilon^0 \hat{\eta} \hbar
l_{\text{ref}}^{-d}$.
Though the bulk viscosity $\zeta$ is expected to be much smaller than
$\eta$ (due to approximate scale invariance), our setup is only
sensitive to $\zeta + 2\eta(1-1/d)$; therefore, we will simply choose
$\zeta = \epsilon^0 \hat{\zeta} \hbar l_{\text{ref}}^{-d}$ as well.
We can safely take $\hat{\zeta} \to 0$ without affecting the derivation.

In performing a derivative expansion, it is assumed that the relevant
variables ($n$, $\energy$, \etc) vary on length scales $\xi \gg
l_{\text{ee}}$.
If we normalize the length scales by $\xi$ as $x = \hat{x}
\xi$, then the derivatives are normalized according to \cref{sec:nondim}
as
\begin{equation}
  \pdv{x} = \frac{1}{\xi} \pdv{\hat{x}} = \frac{l_{\text{ref}}}{\xi}
  \frac{1}{l_{\text{ref}}} \pdv{\hat{x}} = \epsilon^{(d+5)/4} \frac{1}{l_{\text{ref}}}
  \pdv{\hat{x}} \,.
\end{equation}
For the remainder of this paper, carets denoting normalized variables
will be dropped for convenience.

Note that, in addition to our perturbation expansion in terms of
$\epsilon$, we have already made use of two other expansions: one
for $\phi$ expanding in $(\partial_x d_i)^2$ and one for $P(\mu,T)$
expanding in $(\mu/T)^2$.
Using these normalizations, we see that both $(\partial_x d_i)^2$
and $(\mu/T)^2$ are of order $\epsilon$, so all perturbation expansions
in the problem have the same accuracy.

\section{\label{sec:stationary} Perturbation Expansion}
To analyze \crefrange{eq:charge_cons}{eq:mom_cons}, it will be
useful to expand the dependent variables in a perturbation series:
\begin{align}
  \vec{u} &= \vec{u}_0 + \epsilon \vec{u}_1 + \epsilon^2 \vec{u}_2 +
    \ldots \,, \\
  P &= P_0 + \epsilon P_1 + \epsilon^2 P_2 + \ldots \,, \\
  n &= n_0 + \epsilon n_1 + \epsilon^2 n_2 + \ldots \,.
\end{align}

\subsection{Perturbative Thermodynamics}
We will be using the thermodynamic relationships of \cref{sec:thermo} to
write $\mu$ and $T$ in terms of $n$ and $P$; however, since $T$ is
non-dynamical, it will only have a constant $T_0$ component, but not a
$T_1(x,t)$ contribution.
It is useful to define $m$ as the order of $(\mu_0/T_0)^2$; that is,
$\epsilon^{m} \coloneqq \order{\mu/(k_B T)}^2$.
For the nondimensionalization specified in \cref{sec:normalization}, $m
= 1$.

{
\renewcommand{\theequation}{Dirac: \arabic{equation}}
Expanding the thermodynamic variables and collecting powers of
$\epsilon$ yields the following relations for the Dirac regime:
\begin{gather}
  P_0 = T_0^{d+1} \mathcal{C}_0 \,, \\
  n_0 = 2 T_0^{d-1} \mu_0 \mathcal{C}_1 \,, \\
  P_1 = P_0 \bqty{
    \frac{\mathcal{C}_1}{\mathcal{C}_0} \pqty{\frac{\mu_0}{T_0}}^2
    \Kronecker_{m,1}} \,, \\
  n_1 = n_0 \bqty{\frac{\mu_1}{\mu_0} + 2
    \frac{\mathcal{C}_2}{\mathcal{C}_1} \pqty{\frac{\mu_0}{T_0}}^2
    \Kronecker{m,1}} \,, \displaybreak[0] \\
  \begin{aligned}
    &P_2 = P_0 \Biggl[\frac{T_2}{T_0} (d+1) \frac{(d+1) d}{2} \\
    &\qquad + 2 \frac{\mathcal{C}_1}{\mathcal{C}_0}
      \frac{\mu_1}{\mu_0} \pqty{\frac{\mu_0}{T_0}}^2 \Kronecker{m,1} \\
    &\qquad +
      \frac{\mathcal{C}_2}{\mathcal{C}_0} \pqty{\frac{\mu_0}{T_0}}^4
      \Kronecker_{m,1} + \frac{\mathcal{C}_1}{\mathcal{C}_0}
      \pqty{\frac{\mu_0}{T_0}}^2 \Kronecker_{m,2} \Biggr] \,,
  \end{aligned} \displaybreak[0] \\
  \begin{aligned}
    &n_2 = n_0 \Biggl[\frac{\mu_2}{\mu_0} + \frac{T_2}{T_0} (d-1) \\
   &\qquad + 6 \frac{\mathcal{C}_2}{\mathcal{C}_1} \frac{\mu_1}{\mu_0}
    \pqty{\frac{\mu_0}{T_0}}^2 \Kronecker_{m,1} \\
   &\qquad + 3
    \frac{\mathcal{C}_3}{\mathcal{C}_1} \pqty{\frac{\mu_0}{T_0}}^4
    \Kronecker_{m,1} + 2 \frac{\mathcal{C}_2}{\mathcal{C}_1}
    \pqty{\frac{\mu_0}{T_0}}^2 \Kronecker_{m,2} \Biggr] \,,
  \end{aligned}
\end{gather}
with $\Kronecker_{a,b}$ the Kronecker delta function.
}

{
\renewcommand{\theequation}{Fermi: \arabic{equation}}
Similarly, for the Fermi regime, we find
\begin{gather}
  P_0 = \abs{\mu_0}^{d+1} \mathcal{C}_0 \,, \\
  n_0 = \abs{\mu_0}^d \sgn(\mu_0) \mathcal{C}_0 (d+1) \,, \\
  P_1 = P_0 \bqty{\frac{\mu_1}{\mu_0} (d+1) +
    \frac{\mathcal{C}_1}{\mathcal{C}_0} \pqty{\frac{T_0}{\mu_0}}^2
    \Kronecker_{m,-1}} \,, \\
  n_1 = n_0 \bqty{\frac{\mu_1}{\mu_0} d +
    \frac{\mathcal{C}_1}{\mathcal{C}_0} \frac{d-1}{d+1}
    \pqty{\frac{T_0}{\mu_0}}^2 \Kronecker_{m,-1}} \,, \displaybreak[0] \\
  \begin{aligned}
    &P_2 = P_0 \Biggl[\frac{\mu_2}{\mu_0} (d+1) +
      \frac{\mu_1^2}{\mu_0^2} \frac{(d+1) d}{2} \\
    &\qquad + \frac{\mathcal{C}_1}{\mathcal{C}_0} (d-1)
      \frac{\mu_1}{\mu_0}
      \pqty{\frac{T_0}{\mu_0}}^2 \Kronecker_{m,-1} \\
    &\qquad + \frac{\mathcal{C}_2}{\mathcal{C}_0} \pqty{\frac{T_0}{\mu_0}}^4
      \Kronecker_{m,-1} + \frac{\mathcal{C}_1}{\mathcal{C}_0}
      \pqty{\frac{T_0}{\mu_0}}^2 \Kronecker_{m,-2} \Biggr] \,,
  \end{aligned} \displaybreak[0] \\
  \begin{aligned}
    &n_2 = n_0 \Biggl[\frac{\mu_2}{\mu_0} d + \frac{\mu_1^2}{\mu_0^2}
      \frac{d (d-1)}{2} \\
    &\qquad + \frac{\mathcal{C}_1}{\mathcal{C}_0}
      \frac{(d-1)(d-2)}{d+1} \frac{\mu_1}{\mu_0}
      \pqty{\frac{T_0}{\mu_0}}^2 \Kronecker_{m,-1} \\
    &\qquad +
      \frac{\mathcal{C}_2}{\mathcal{C}_0} \frac{d-3}{d+1}
      \pqty{\frac{T_0}{\mu_0}}^4 \Kronecker_{m,-1} +
      \frac{\mathcal{C}_1}{\mathcal{C}_0} \frac{d-1}{d+1}
      \pqty{\frac{T_0}{\mu_0}}^2 \Kronecker_{m,-2} \Biggr] \,.
  \end{aligned}
\end{gather}
}

Using these equations, we can now write $\mu$ and $P$ in terms of $n$ at
each order.
In particular, we find
\begin{equation}
  \frac{P_1}{P_0} = \frac{n_1}{n_0} K_0 +
    \frac{\mathcal{C}_1}{\mathcal{C}_0} \pqty{\frac{\mu_0}{T_0}}^{2 m}
    \pqty{\Kronecker_{m,1} + \frac{1}{d} \Kronecker_{m,-1}} \,.
  \label{eq:K0_def}
\end{equation}
Here, we have defined $K_0$ as
\begin{equation}
  K_0 =
  \begin{cases}
    0 & \text{$m > 0$ (Dirac regime)} \\
    (d+1)/d & \text{$m < 0$ (Fermi regime)}
  \end{cases}
  \,.
\end{equation}
As a side note, it is straightforward to show with thermodynamic
identities that $K_0$ is the leading order term in the ratio of bulk
modulus $B$ to pressure $P$; that is, $K_0 = B_0 / P_0$.

\subsection{Conservation Equations}
First, let us investigate a scenario with a constant, uniform background
flow $u_0 \neq 0$ chosen such that the perturbations are stationary in
the laboratory frame.
This will both simplify the mathematics and be experimentally
interesting.
To accomplish this, we will only permit variations on long timescales
(this will be important when including dissipation).
Mathematically, we accomplish this by normalizing the time variable as
$t = \epsilon \hat{t}_1 \xi/v_F$ such that $\partial_{t_1} = \epsilon
\order{\partial_{\hat{x}}}$.

Expanding the governing equation, we find
\begin{widetext}
Leading Order:
\begin{subequations}
\begin{align}
  \pdv{x}(\gamma^2 n_0 u_1 + u_0 n_1) &= 0 \,, \label{eq:pert_charge_1}\\
  \pdv{x}(\gamma P_1 + u_0 \gamma^3 (\energy_0 + P_0) u_1 + \gamma A n_0 n_1
    + \gamma^3 A n_0^2 u_0 u_1) &= 0 \,. \label{eq:pert_mom_1}
\end{align}
\end{subequations}
First-Order Correction:
\begin{subequations}
\begin{align}
  &\begin{aligned}
  &\pdv{x}(\gamma^2 n_0 u_2 + u_0 n_2) = \\
  &\qquad - \gamma^2 \pqty{u_0 \gamma^2 (2+u_0^2) n_0 u_1 + n_1 + n_0 u_0
    u_1} \pdv{u_1}{x} - \gamma^2 u_1 \pdv{n_1}{x} \\
  &\qquad + \gamma u_0 A \sigma_Q
    \pdv{n_1}{t_0}{x} + \gamma A \sigma_Q \pdv[2]{n_1}{x} + \gamma^3
    u_0 A \sigma_Q n_0 \pqty{\pdv[2]{u_1}{x} + u_0 \pdv{u_1}{t_0}{x}} \\
  &\qquad + \Theta(-m) \gamma \sigma_Q \pdv[2]{x} \pqty{\mu_1 -
    \frac{T_0}{\mu_0} T_1} \,,
  \end{aligned} \label{eq:pert_charge_2} \\
  &\begin{aligned}
  &\pdv{x}(\gamma P_2 + u_0 \gamma^3 (\energy_0 + P_0) u_2 + \gamma A n_0 n_2
    + \gamma^3 A n_0^2 u_0 u_2) = \\
  &\qquad -\gamma^3 \pqty{ u_0 (\energy_1 + P_1) + (1+u_0^2) u_1 \gamma^2
    (\energy_0 + P_0)} \pdv{u_1}{x} - A n_1 \gamma \pdv{n_1}{x} + A n_0 u_0
    u_1 \gamma^3 \pdv{n_1}{x} - \gamma A n_0 \frac{d_1 d_2}{3}
    \pdv[3]{n_1}{x} \\
  &\qquad - A n_0^2 (1+u_0^2) u_1 \gamma^5 \pdv{u_1}{x} - A n_0^2 u_0
    \gamma^3 \frac{d_1 d_2}{3} \pdv[3]{u_1}{x} -A n_0 u_0 u_1
    \gamma^3 \pdv{n_1}{x} - 2 A n_0 u_0 n_1 \gamma^3 \pdv{u_1}{x} \\
  &\qquad +\gamma^4 \bqty{\zeta + 2 \eta \pqty{1 - \frac{1}{d}}} \pqty{u_0^2
    \pdv[2]{u_1}{t_0} + 2 u_0 \pdv{u_1}{t_0}{x} + \pdv[2]{u_1}{x}} \,.
  \end{aligned} \label{eq:pert_mom_2}
\end{align}
\end{subequations}
\end{widetext}
Here, we have defined $\gamma = 1/\sqrt{1-u_0^2}$ (with $v_F=1$) and
used the electrostatic coupling $A$ according to
\cref{eq:electric_potential}.
Additionally, we have used the Heaviside function
\begin{equation}
  \Theta(-m) =
  \begin{cases}
    0 & \text{$m > 0$ (Dirac regime)} \\
    1 & \text{$m < 0$ (Fermi regime)}
  \end{cases}
  \,.
\end{equation}

\subsection{Leading Order Equations}
Using the thermodynamic relation $\energy = P d$, the leading order
equations can be manipulated as
\begin{equation*}
	\gamma^2 d \pqty{A n_0 + \frac{P_0 K_0}{n_0}}
    \mbox{$\bigl[$\cref{eq:pert_charge_1}$\bigr]$} - \gamma d u_0
    \mbox{$\bigl[$\cref{eq:pert_mom_1}$\bigr]$}
\end{equation*}
yielding
\begin{equation}
	0 = \gamma^2 d \bqty{A n_0^2 + \gamma^2 P_0 \pqty{K_0-u_0^2
    (d+1)}} u_1 \,.
\end{equation}
We want nontrivial perturbations $u_1 \neq 0$, so we require the terms
in square brackets to vanish.
We see that this gives an equation for $u_0$ required to make the
leading order solutions time-independent:
\begin{equation}
  u_0 = \pm \sqrt{\frac{[K_0/(d+1)]+[A n_0^2/P_0
    (d+1)]}{1+[A n_0^2/P_0 (d+1)]}} \,.
\end{equation}
It is easy to check that $u_0^2 < 1$ for $d \neq 1$; this is
required, otherwise $\gamma = 1/\sqrt{1-u_0^2}$ would be imaginary.

Additionally, if we restrict to solutions bounded in $x$, we can require
each term inside $\partial_x$ from \cref{eq:pert_charge_1,eq:pert_mom_1}
to be zero, giving
\begin{equation}
  u_1 = - \frac{u_0}{\gamma^2 n_0} n_1 + U_1 \,.
\end{equation}
Here, we have included a constant, uniform current $U_1(x,t_0,t_1) =
U_1$; this will allow us---at the next order---to cancel the
disturbance's propagation speed (similar to our use of $u_0$ at this
order).

\subsection{First-Order Corrections}
Now, we can do the same for the first-order corrections.
Manipulating them as before,
\begin{equation*}
  \gamma^2 d \pqty{A n_0 + \frac{P_0 K_0}{n_0}}
    \mbox{$\bigl[$\cref{eq:pert_charge_2}$\bigr]$} - \gamma d u_0
    \mbox{$\bigl[$\cref{eq:pert_mom_2}$\bigr]$} \,,
\end{equation*}
gives
\begin{equation}
  \gamma^2 d \bqty{A n_0^2 + \gamma^2 P_0 \pqty{K_0-u_0^2
    (d+1)}} u_2 = \text{RHS} \,.
\end{equation}
Here, the right-hand side (RHS) depends only on $n_1$, $u_1$,
$\energy_1$ and $P_1$.
However, inserting our solution for $u_0$ causes the left-hand side to
vanish, giving us our desired compatibility condition on $n_1$.
Thus, we have the compatibility equation
\begin{equation}
  \mathcal{A} \pdv{n_1}{t_1} + \mathcal{F} \pdv{n_1}{x} + \mathcal{B}
    n_1 \pdv{n_1}{x} + \mathcal{C} \pdv[3]{n_1}{x} = \mathcal{G}
    \pdv[2]{n_1}{x} \,,
    \label{eq:KdVB}
\end{equation}
with
\begin{subequations}
\begin{gather}
  \mathcal{A} = 2 \gamma^2 \frac{P_0 d}{n_0} u_0^2 (d+1-K_0) \,, \\
  \begin{aligned}
    \mathcal{B} &= -\gamma^2 \frac{P_0}{n_0^2 d}
      u_0 \Bigl(d^2 u_0^2 [4(d+1)-K_0(d+3)] \\
    &\qquad + (d+1)\Theta(-m) - K_0 d^2 \Bigr) \,,
  \end{aligned} \\
  \mathcal{C} = -A d \frac{d_1 d_2}{3} n_0 u_0 \,, \\
  \begin{aligned}
  \mathcal{F} &= \gamma^2 \frac{P_0 d}{n_0}
    u_0 \Biggl(2 U_1 \gamma^2 (d+1-K_0) u_0 \\
  &\qquad + \frac{\mathcal{C}_1}{\mathcal{C}_0}
    \pqty{\frac{\mu_0}{T_0}}^{2m}
    \Biggl[u_0^2 (d+1) \pqty{\frac{1}{d} \Kronecker_{m,-1} +
    \Kronecker_{m,1}} \\
  &\qquad - \pqty{\frac{d-1}{d^2} \Kronecker_{m,-1}
    +2 \Kronecker_{m,1}} \Biggr] \Biggr) \,,
  \end{aligned}\\
  \begin{aligned}
    \mathcal{G} &= \frac{\gamma^3}{n_0} \Biggl(
      \sigma_Q \gamma^2 \pqty{\frac{P_0}{n_0}}^2 u_0^2 (d+1) (d+1-K_0) \\
    &\qquad \times \bqty{d u_0^2 + \underbrace{\Theta(-m)
      - K_0 \frac{d}{d+1}}_{=0}} \\
    &\qquad + d u_0^2 \bqty{\zeta+2 \eta \pqty{1-\frac{1}{d}}} \Biggr) \,,
  \end{aligned}
\end{gather}
\label{eq:stationary_coeffs}
\end{subequations}
This is known as the KdV-Burgers (KdVB) equation.
Note the underbraced term in $\mathcal{G}$ vanishes in both the Dirac
and Fermi regimes.

\subsection{\label{sec:strained} Ideal Fluid}
Before tackling the full KdVB equation, it is beneficial to consider the
simpler inviscid problem with $\sigma_Q = \eta = \zeta = 0$.
In this case, we find $\mathcal{G} = 0$ and the KdV-Burgers equation
reduces to the KdV equation.
The KdV equation has soliton solutions of the form
\begin{equation}
  \begin{aligned}
    &n_1(x,t_1) = c_1 \sgn(\mathcal{BC}) \sech^2 \Biggl( \sqrt{\frac{c_1
      \abs{\mathcal{B}}}{12 \abs{\mathcal{C}}}} \\
    &\qquad \times \biggl[x - \pqty{\frac{c_1 \abs{\mathcal{B}}}{3
      \abs{\mathcal{A}}} \sgn(\mathcal{AC})  +
      \frac{\mathcal{F}}{\mathcal{A}}} t_1 \biggr] \Biggr) \,,
  \end{aligned}
  \label{eq:kdv_solution}
\end{equation}
for arbitrary, order-$1$ constant $c_1>0$.

Substituting the coefficients, we find
\begin{equation}
  n = n_0 + \epsilon c_1 \sgn(\mathcal{BC}) \sech[2](\frac{x+vt}{W}) \,,
  \label{eq:ideal_solution}
\end{equation}
with
\begin{equation}
  v = -\epsilon \pqty{\frac{c_1 \abs{\mathcal{B}}}{3
    \abs{\mathcal{A}}} \sgn(\mathcal{AC})  +
    \frac{\mathcal{F}}{\mathcal{A}}} \,,
  \label{eq:ideal_v}
\end{equation}
and
\begin{equation}
  W = \sqrt{\frac{12 \abs{\mathcal{C}}}{c_1 \abs{\mathcal{B}}}} \,.
\end{equation}

Let us seek a soliton which is stationary in the laboratory frame; we have
already accomplished $\partial_{t_0} n_1 = 0$ by a choice of $u_0$; we
can similarly set $\partial_{t_1} n_1 = 0$ by an appropriate choice of
$U_1$.
If we choose $U_1$ so that $\mathcal{F} = - c_1 \mathcal{B}/3
\sgn \mathcal{BC}$, then the soliton is stationary:
\begin{equation}
  n = n_0 + \epsilon c_1 \sgn(\mathcal{BC}) \sech[2](\frac{x}{W}) \,.
\end{equation}

\subsection{\label{sec:dissipation} Dissipation}
Now, we return to the full KdVB equation \cref{eq:KdVB}.
It does not appear that the KdV-Burgers equation with $\mathcal{G} \neq
0$ has an analytic, solitonic solution.
However, if $\mathcal{G} \ll (\mathcal{A}, \mathcal{B}, \mathcal{C})$,
then an approximate solution is given by \cref{eq:ideal_solution} but
with time-dependent $c_1$, as described in~\citet{mei2005theory}.
For clarity, we can factor out this smallness as $\mathcal{G} = \delta
\tilde{\mathcal{G}}$ so that $\delta \ll 1$ and $\tilde{\mathcal{G}}$ is the same
order as $\mathcal{A}$.
Then, another short multiple scales expansion for $n_1$ can be done in
$\delta = \order{\mathcal{G}/\mathcal{A}}$.
To be consistent with our original perturbation series, we require that
$\epsilon \ll \delta \ll 1$.

As usual, we expand $n_1$ as $n_1 = n_1^{(0)} + \delta n_1^{(1)}$ and
$\partial_{t_1} = \partial_{\tau_0} + \delta \partial_{\tau_1}$.
Then, to leading order, the equation
\begin{align}
  \mathcal{L}_0 n_1^{(0)} &\coloneqq \mathcal{A} \partial_{\tau_0}
    n_1^{(0)} + \mathcal{F} \partial_x
    n_1^{(0)} + \frac{\mathcal{B}}{2}
    \partial_{x} \pqty{ n_1^{(0)} }^2 \\
  &\qquad + \mathcal{C} \partial_{x}^3 n_1^{(0)} = 0 \,,
  \label{eq:decay_first_order}
\end{align}
where we have again defined the linear operator $\mathcal{L}_1$ acting
on $n_1^{(1)}$.
This is the ordinary KdV equation; therefore, $n_1^{(0)}$ has the
solution given by \cref{eq:kdv_solution} with order-$1$ free parameter
$c_1>0$.

At next order in $\delta$, we must allow the constant $c_1$ to become
time-dependent on a slow time-scale $c_1 = c_1(\tau_1)$.
Then, our equation is
\begin{align}
  \mathcal{L}_1 n_1^{(1)} &\coloneqq \mathcal{A} \partial_{\tau_0} n_1^{(1)} +
    \mathcal{F} \partial_{x} n_1^{(1)}+\mathcal{B}
    \partial_{x} \pqty{n_1^{(0)} n_1^{(1)} } \nonumber \\
  &\qquad + \mathcal{C} \partial_{x}^3 n_1^{(1)} \nonumber \\
  & = -\mathcal{A} \partial_{\tau_1} n_1^{(0)} + \tilde{\mathcal{G}}
    \partial_{x}^2 n_1^{(0)} \,,
  \label{eq:decay_second_order}
\end{align}
where we have again defined the linear operator $\mathcal{L}_1$ acting
on $n_1^{(1)}$.

For certain inhomogeneous terms in \cref{eq:decay_second_order}, it is
possible to generate secular (\ie unbounded) growth; since this is
clearly no longer a localized solution, we wish to avoid this.
Here, we will utilize a multiple scales approach, though it will differ
slightly from the method used in \cref{sec:multiple_scales}
since the homogeneous operator $\mathcal{L}_0$ is nonlinear.
Following the example of \citet{mei2005theory}, we note that
$\mathcal{L}_0$ and $-\mathcal{L}_1$ are adjoints:
\begin{equation}
  \int \dd{x} \pqty{ n_1^{(1)} \mathcal{L}_0 n_1^{(0)} + n_1^{(0)}
  \mathcal{L}_1 n_1^{(1)} } = 0 \,.
\end{equation}
Then, substituting the right-hand sides of
\cref{eq:decay_first_order,eq:decay_second_order}, we get the
compatibility condition
\begin{equation}
  \int n_1^{(0)} \pqty{\mathcal{A}\partial_{\tau_1}n_1^{(0)} -
    \tilde{\mathcal{G}}\partial_{x}^2 n_1^{(0)}} \dd{x} = 0 \,.
\end{equation}
Inserting the soliton solution for $n_1^{(0)}$, we get an equation for
$c_1(\tau_1)$:
\begin{equation}
  \dot{c}_1 = -\frac{c_1^2
    \abs{\mathcal{B}}\tilde{\mathcal{G}}}{\abs{\mathcal{C}}\mathcal{A}}
    \frac{4}{45} \,.
\end{equation}
Then, solving this equation and converting back to time $t_1$ gives
\begin{equation}
  c_1(t_1) = \frac{c_1(0)}{1 + \frac{t_1}{t_d}}
  \qq{with}
  t_d = \frac{45 \mathcal{A} \abs{\mathcal{C}}}{4 c_1(0) \mathcal{G}
    \abs{\mathcal{B}}} \,,
\end{equation}
with $c_1(0)$ the initial value of the parameter $c_1(t_1)$.
Recall that this is derived under the assumption that
$\epsilon \ll \order{\mathcal{G}/\mathcal{A}} \ll 1$.

\begin{figure*}
  \centering
  \includegraphics{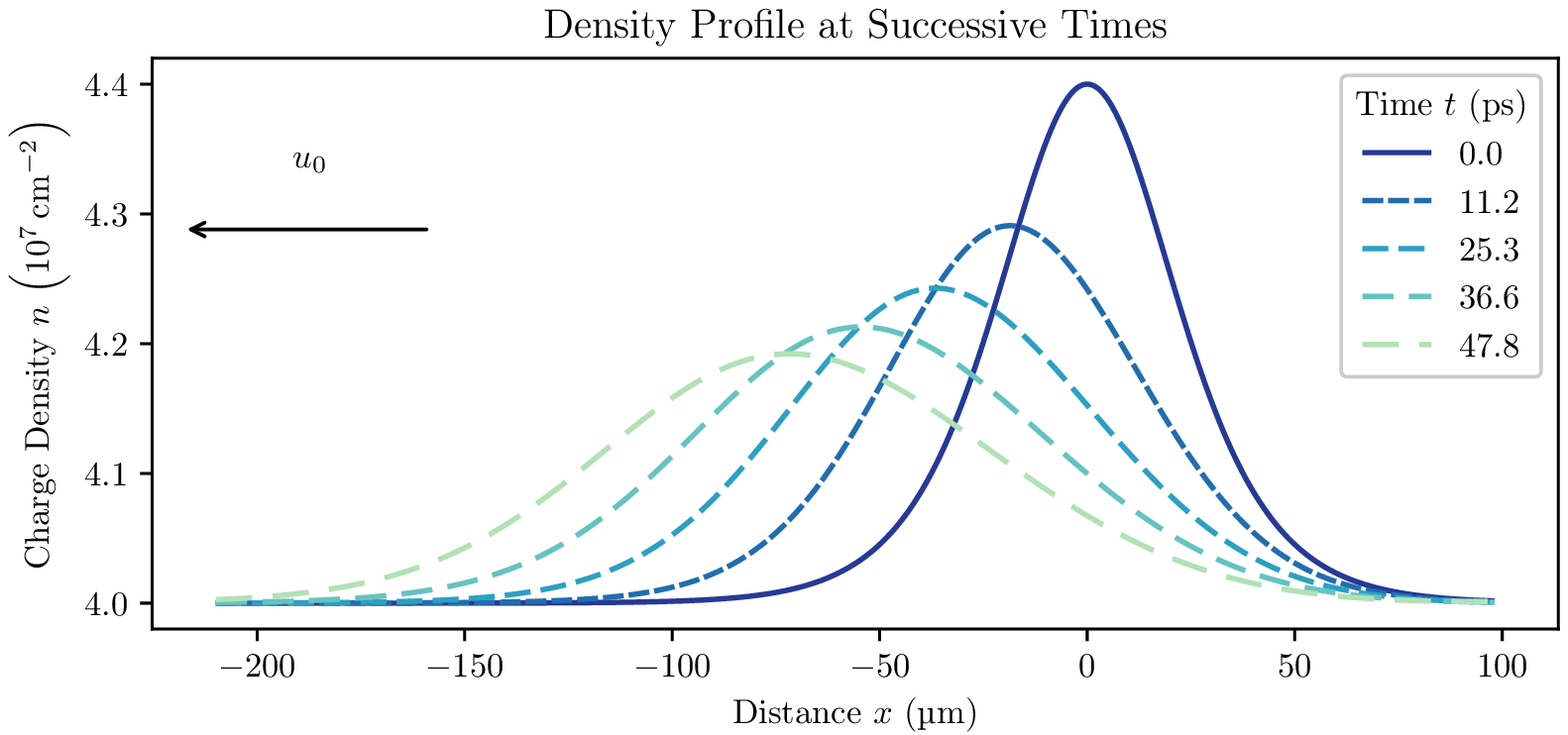}
  \caption{(Color online) Solitonic solution to KdV-Burgers.
    Values used were $\mathcal{A}=\AVal$, $\mathcal{B}=\BVal$,
    $\mathcal{C}=\CVal$, $\mathcal{F}=\FVal$, and $\mathcal{G}=\GVal$
    with the height normalized to \solitonHeightDim{}.
    This choice of parameters gives a soliton propagating in the $+x$
    direction and a counter-current $u_0$ in the $-x$ direction
    (indicated by the arrow).
    }
  \label{fig:soliton}
\end{figure*}

\begin{figure}
  \centering
  \includegraphics{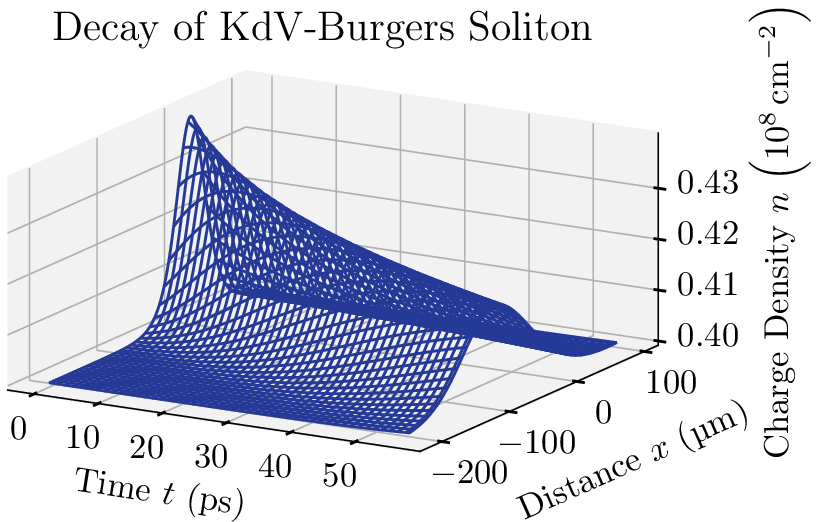}
  \caption{(Color online) Solitonic solution to KdV-Burgers showing
    decay as a function of time.
    Values used were $\mathcal{A}=\AVal$, $\mathcal{B}=\BVal$,
    $\mathcal{C}=\CVal$, $\mathcal{F}=\FVal$, and $\mathcal{G}=\GVal$
    with the height normalized to \solitonHeightDim{}.
    }
  \label{fig:soliton3d}
\end{figure}

Additionally, we can solve the KdV-Burgers equation numerically for
arbitrary $\mathcal{G}$; this shows similar behavior to the analytic
approximation~(\cf \cref{fig:soliton,fig:soliton3d}).
That is, the soliton slowly decays as it progresses.

\section{\label{sec:multiple_scales} Multiple Scales Expansion}
Now, we wish to study the previous solitonic solution in more
generality.
Here, we will allow for an arbitrary uniform, time-independent
background current $u_0$.

As we have seen previously, the nonlinearities affect the propagation
velocity $v$ (\cf \cref{eq:ideal_v}).
This is an example of a singular perturbation and requires the use of
singular perturbation theory.
Singular methods such as Poincar\'{e}-Lindstedt are only applicable to
steady or periodic solutions.
Since we are interested in decaying solutions, we need to make use of
the method of multiple scales.
Note that this approach is similar to that employed by
\citet{akbari2012higher} in the study of partially degenerate
electron-ion plasmas.

First, unlike the previous section, we will nondimensionalize the
timescale so that $\partial_t = \partial_x$.
Now, if we introduce a series of timescales $t_0 = t$, $t_1 = \epsilon
t$, $t_2 = \epsilon^2 t$, \ldots each presumed independent, the chain
rule gives
\begin{equation}
  \pdv{t} = \pdv{t_0} + \epsilon \pdv{t_1} + \epsilon^2 \pdv{t_2} +
  \ldots \,.
\end{equation}
Further, we now assume that each variable is a function of all time
scales: $n = n(x,t_0,t_1,t_2,\ldots)$.

If we again restrict to 1D motion and collect terms by powers of
$\epsilon$ we get the following equations:
\begin{widetext}
\noindent
\begin{minipage}{\linewidth}
Leading Order:
\begin{subequations}
\begin{align}
  \pdv{n_1}{t_0} + \gamma^2 n_0 u_0 \pdv{u_1}{t_0} + u_0
    \pdv{n_1}{x} + n_0 \gamma^2 \pdv{u_1}{x} &= 0 \,,
    \label{eq:MMS_charge_1} \\
  \gamma^3 (\energy_0 + P_0) \pdv{u_1}{t_0} + \gamma u_0
    \pdv{P_1}{t_0} + u_0 \gamma^3 (\energy_0 + P_0) \pdv{u_1}{x} +
    \gamma \pdv{P_1}{x} + A n_0 \gamma \pdv{n_1}{x} + A n_0^2 u_0
    \gamma^3 \pdv{u_1}{x} &= 0 \,, \label{eq:MMS_mom_1}
\end{align}
\end{subequations}
\label{eq:MMS_1}
First-Order Correction:
\begin{subequations}
\begin{align}
  \pdv{n_2}{t_0} + \gamma^2 n_0 u_0 \pdv{u_2}{t_0} + u_0
    \pdv{n_2}{x} + n_0 \gamma^2 \pdv{u_2}{x} &= \text{RHS} \,,
    \label{eq:MMS_charge_2} \\
  \gamma^3 (\energy_0 + P_0) \pdv{u_2}{t_0} + \gamma u_0
    \pdv{P_2}{t_0} + u_0 \gamma^3 (\energy_0 + P_0) \pdv{u_2}{x} +
    \gamma \pdv{P_2}{x} + A n_0 \pdv{n_2}{x} &= \text{RHS} \,.
    \label{eq:MMS_mom_2}
\end{align}
\end{subequations}
\end{minipage}
\end{widetext}
Again, we have used the electrostatic coupling $A$ according to
\cref{eq:electric_potential}.
See \cref{sec:full_equations} for the terms on the right-hand side.

Notice that, as is often the case for multiple scales analyses, the
linear operator acting on $n_1$, $u_1$, \etc{} in
\cref{eq:MMS_charge_1,eq:MMS_mom_1} is identical to the linear operator
acting on $n_2$, $u_2$, \etc{} in \cref{eq:MMS_charge_2,eq:MMS_mom_2}.
Furthermore, since this operator is linear, we do not need to employ the
operator formalism of \cref{sec:dissipation}, but can instead use a
linear algebraic approach similar to \cref{sec:stationary} (with the
addition of another timescale, $t_1$).

\subsection{Leading Order Equations}
Using $\energy = P d$ and combining equations like
\begin{align*}
  & \gamma^2 d \Biggl[A n_0 \pdv{x} + \frac{P_0 K_0}{n_0} \pqty{u_0
    \pdv{t_0} + \pdv{x}} \Biggr]
    \mbox{$\bigl[$\cref{eq:MMS_charge_1}$\bigr]$} \\
  &\qquad - \gamma d \pqty{\pdv{t_0} + u_0 \pdv{x}}
    \mbox{$\bigl[$\cref{eq:MMS_mom_1}$\bigr]$}
\end{align*}
gives
\begin{equation}
  \begin{aligned}
    0 &= \gamma^2 d \Biggl( -\gamma^2 P_0 (d+1 - u_0^2 K_0)
      \pdv[2]{u_1}{t_0} \\
    &\quad -2 \gamma^2 P_0 u_0 (d+1-K_0) \pdv{u_1}{t_0}{x} \\
    &\quad + \Bqty{ A n_0^2 + \gamma^2 P_0 \bqty{-u_0^2 (d+1) + K_0} }
    \pdv[2]{u_1}{x} \Biggr) \,.
  \end{aligned}
\end{equation}
This wave equation has solutions $f(x+v_0 t_0) + g(x-v_0 t_0)$ with
$v_0$ given by
\begin{align}
  & v_0^{(\pm)} = \frac{-u_0 (d+1-K_0)}{d+1-u_0^2 K_0} \pm
    \frac{1}{\gamma \pqty{d+1-u_0^2 K_0}} \\
  &\qquad \times \sqrt{\frac{K_0 (d+1)}{\gamma^2} + \frac{A n_0^2}{P_0}
    \pqty{d+1-u_0^2 K_0}} \,.
\end{align}
We will take the $(+)$ sign so that $v_0 = v_0^{(+)}$; the other can be
recovered by taking $u_0 \to -u_0$ and $v_0 \to -v_0$.
Further, we restrict to unidirectional solutions $u_1(x,t_0,t_1) = f(x
\pm v_0 t_0, t_1)$ for a definite choice of $\pm$; here, we choose $(+)$
as well---the other propagation direction can be recovered by taking
$v_0 \to -v_0$.

For stationary perturbations ($v_0 = 0$), we can solve for $u_0$ to
recover the result from \cref{sec:stationary}:
\begin{equation}
  u_0 = \pm \sqrt{\frac{[K_0/(d+1)]+[A n_0^2/P_0
  (d+1)]}{1+[A n_0^2/P_0 (d+1)]}} \,.
\end{equation}
For reference, the velocity of propagation in the absence of a
background flow ($u_0 = 0$) is
\begin{equation}
  v_0 = \pm \sqrt{\frac{1}{d+1}} \sqrt{K_0 + \frac{A n_0^2}{P_0}} \,.
\end{equation}

In general, $n_1$, $u_1$, and $P_1$ have traveling wave solutions;
neglecting solutions of the form $f(x-u_0 t_0,t_1)$ that are simply
advected by the background current, we find solutions given by
\begin{subequations}
\begin{gather}
  n_1(x,t_0,t_1) = n_1(x + v_0 t_0, t_1) + F_1(t_1) \,, \\
  \begin{aligned}
    u_1(x,t_0,t_1) &= -\frac{(u_0 + v_0)}{n_0 \gamma^2 (1+u_0 v_0)} n_1(x
      + v_0 t_0, t_1) \\
    &\qquad + F_2(t_1) \label{eq:u_from_n} \,,
  \end{aligned} \,.
\end{gather}
\end{subequations}
Here, we have arbitrary functions $F_1(t_1)$ and $F_2(t_2)$; by imposing
boundary conditions $n_1 = 0$ at $x = \pm \infty$, we set $F_1 = 0$.
We will allow $U_1(t_2) \coloneqq F_2(t_2)$ to remain arbitrary; this
uniform background current can be superimposed on the soliton solution
as in \cref{sec:stationary} if desired~%
\footnote{
  Note that it is possible to generate a stationary soliton by
  appropriate choice of $F_1$ instead, though the resulting
  coefficients will be different.
}.

Now, we can also see why it was important to take $\mu_0 \ll T_0$ small
but finite.
Had $\mu = 0$ identically, then the thermodynamic relations would
require $n_0 = 0$.
Then, the leading order charge conservation equation
\cref{eq:MMS_charge_1} would give $\partial_{t_0} n_1 + u_0 \partial_x
n_1 = 0$; \ie charge density perturbations are simply advected along by
the background flow.
That is, the density perturbations lack any dynamic propagation and are
``frozen-in.''
Since the other dependent variables are proportional to $n_1$, we see
$P_1$ and $u_1$ are similarly affected.
Hence, if we want a dynamic disturbance, we require $\mu_0 \neq 0$;
intuitively, this is understandable as there are no net charge carriers
at the Dirac point.

\subsection{First-Order Corrections}
Now considering the first-order corrections, preventing
secular growth of the higher-order terms (\ie $n_2$, $u_2$, \etc)
requires imposing a compatibility condition on the lower-order terms
(\ie $n_1$, $u_1$, \etc).
We can manipulate the system as
\begin{align*}
  & \gamma^2 d \Biggl[A n_0 \pdv{x} + \frac{P_0 K_0}{n_0} \pqty{u_0
    \pdv{t_0} + \pdv{x}} \Biggr]
    \mbox{$\bigl[$\cref{eq:MMS_charge_2}$\bigr]$} \\
  &\qquad - \gamma d \pqty{\pdv{t_0} + u_0 \pdv{x}}
    \mbox{$\bigl[$\cref{eq:MMS_mom_2}$\bigr]$}
\end{align*}
which gives
\begin{equation}
  \begin{aligned}
    &\gamma^2 d \Biggl( -\gamma^2 P_0 (d+1 - u_0^2 K_0)
      \pdv[2]{u_2}{t_0} \\
    &\quad -2 \gamma^2 P_0 u_0 (d+1-K_0) \pdv{u_2}{t_0}{x} \\
    &\quad + \Bqty{ A n_0^2 + \gamma^2 P_0 \bqty{-u_0^2 (d+1) + K_0} }
    \pdv[2]{u_2}{x} \Biggr) \\
    & = \text{LOT} \,,
  \end{aligned}
  \label{eq:HOT_inhom}
\end{equation}
where LOT represents lower-order terms (\ie $n_1$, $u_1$, \etc).

It is instructive here to change variables to
$\chi^{(\pm)}_0=x+v_0^{(\pm)} t_0$.
Then, the equation becomes
\begin{equation}
  \begin{aligned}
    &\gamma^4 P_0 d \pqty{d+1-u_0^2 K_0} \pqty{v_0^{(+)}-v_0^{(-)}}^2 \\
    &\qquad \times \pdv{\chi^{(-)}_0} \pdv{\chi^{(+)}_0} u_2 \\
    & = \text{LOT}
  \end{aligned}
\end{equation}
This is where we encounter an apparent problem.
Upon inserting our solutions for the lower-order terms, we find the
right-hand side depends on products and derivatives of
$f\pqty{\chi^{(+)}_0}$.
This implies that the LOT are solely functions of $\chi^{(+)}_0$.

However, we see that functions of the form $f(\chi^{(+)})$ are also
solutions to the homogeneous equation in \cref{eq:HOT_inhom} due to the
presence of the $\partial_{\chi_0^{(-)}}$ operator.

So, products and derivatives of $f(\chi^{(+)}_0)$ appear as inhomogeneous
forcing terms that give rise to secular terms.
For instance, terms proportional to $f^{(4)}\pqty{\chi^{(+)}_0}$ give
rise to solutions of the form $\chi^{(-)}_0 f^{(3)}\pqty{\chi^{(+)}_0}$.
This grows unbounded in $\chi^{(-)}_0$---and hence, in time $t$.
This will eventually cause $\abs{u_2} > \abs{u_1}$, invalidating the
perturbation expansion.
Thus, unless the LOT vanish identically, they will give rise to
$\chi^{(\pm)}_0$-secular terms in $u_2$---\ie solutions growing
unbounded in $t_0$ or $x$.

Hence, we require the right-hand side to vanish and we are left with the
desired compatibility equation:
\begin{equation}
  0 = \pdv{\chi_0^{(+)}} (\text{KdVB}[n_1])\,.
  \label{eq:MMS_compat}
\end{equation}
Here, (KdVB$[n_1]$) represents the Korteweg-de Vries-Burgers equation,
discussed earlier, acting on $n_1$:
\begin{equation}
  \begin{aligned}
    &\mathcal{A}' \pdv{n_1}{t_1} + \mathcal{F}' \pdv{n_1}{\chi^{(+)}_0}
      + \mathcal{B}' n_1 \pdv{n_1}{\chi^{(+)}_0} \\
    &\qquad + \mathcal{C}' \pdv[3]{n_1}{{\chi^{(+)}_0}} -
      \mathcal{G}' \pdv[2]{n_1}{{\chi^{(+)}_0}} n_1  = 0 \,;
  \end{aligned}
  \label{eq:kdv_burgers}
\end{equation}
see \cref{sec:kdv_burgers} for the functional form of the coefficients.

The solution to the KdV-Burgers equation was already derived in
\cref{sec:dissipation} and is simply reiterated here for convenience:
\begin{align}
  &n_1\pqty{\chi_0^{(+)},t_1} = c_1\pqty{t_1}
    \sgn(\mathcal{B'C'}) \sech^2
    \Biggl( \sqrt{\frac{c_1 \abs{\mathcal{B}'}}{12 \abs{\mathcal{C}'}}}
      \nonumber \\
  &\qquad \times \biggl[\chi_0^{(+)} - \pqty{\frac{c_1
    \abs{\mathcal{B}'}}{3 \abs{\mathcal{A}'}} \sgn(\mathcal{A'C'})  +
    \frac{\mathcal{F}'}{\mathcal{A}'}} t_1 \biggr] \Biggr) \,,
    \label{eq:soliton}
\end{align}
where
\begin{equation}
  c_1(t_1) = \frac{c_1(0)}{1 + t_1/t_d}
  \label{eq:MMS_c1}
\end{equation}
with
\begin{equation}
  t_d = \frac{45 \mathcal{A}' \abs{\mathcal{C}'}}{4 c_1(0)
    \abs{\mathcal{B}'} \mathcal{G}'} \,,
  \label{eq:MMS_tau_d}
\end{equation}
with $c_1(0)$ the initial amplitude of the soliton.

\section{\label{sec:analysis} Analysis}
Nondimensionalizing helped ensure that all quantities were order
$\order{1}$ and any information about their magnitude was solely
contained in $\epsilon$ prefactors.
However, having ordinary, dimensional expressions is more useful for
comparing with experiments or existing literature.
Therefore, the KdV-Burgers coefficients are written in terms of
ordinary, dimensional variables in
\cref{sec:kdv_burgers,sec:kdv_burgers_adi}~%
\footnote{
  A few terms were simplified using Kronecker deltas in
  \cref{sec:kdv_burgers,sec:kdv_burgers_adi}.
  For instance, substituting the dimensional expressions into
  $\mathcal{G}'$ generates an $\epsilon^{-q}$ term multiplying
  $\sigma_Q$ and an $\epsilon^{-p}$ term multiplying $\eta$ and $\zeta$.
  However, these can be neglected: as mentioned at the end of
  \cref{sec:nondim}, $\sigma_Q$ carries an implicit $\Kronecker_{q,0}$
  while $\eta$ and $\zeta$ have implicit $\Kronecker_{q,0}$ and
  $\Kronecker_{q,0} \order{\zeta}/\order{\eta}$, respectively.
  Similarly, the thermodynamic contribution of $\mathcal{F}'$ has a
  factor of $\epsilon^{-m^2}$; however, given the presence of the
  Kronecker deltas, this is equivalent to $\epsilon^{-1}$.
}.
Note that the coefficients are still dimensionless and order unity~%
\footnote{
  Actually, as written, the coefficients in
  \cref{sec:kdv_burgers,sec:kdv_burgers_adi} have all had a common
  factor of $\epsilon^{p/2-q/2} \sqrt{\order{\sigma_Q}/\order{\eta}}$
  removed for brevity.
}.

The observables that characterize the system, to this order, are
the amplitude, width, speed, and decay period of the soliton.
The amplitude is simply given by
\begin{equation}
  \norm{n_1} = l_{\text{ref}}^{-d} \epsilon^{(d+2)/4} \epsilon c_1(t)
    \sgn(\mathcal{B'C'})
  \coloneqq n_{\text{max}} \,.
\end{equation}
We can use $n_{\text{max}}$ to eliminate $c_1$ in the following
expressions~%
\footnote{
  Hence, $c_1$ is the normalized, order-unity analog of
  $n_{\text{max}}$.
}.
Furthermore, we will factor out the explicit factors of $\epsilon$ and
$l_{\text{ref}}$ from the KdV-Burgers coefficients; we will denote the
original, order unity, coefficients with a caret.
Then, we can write the speed as
\begin{align}
  v &\coloneqq v_0 - \epsilon v_F \pqty{ \frac{c_1
    \hat{\mathcal{B}}'}{3 \hat{\mathcal{A}}'} \sgn(\mathcal{B' C'}) +
    \frac{\hat{\mathcal{F}}'}{\hat{\mathcal{A}}'}} \nonumber \\
  &= v_0 - \frac{n_{\text{max}} \mathcal{B}'}{3 \mathcal{A}'}
    - v_F \frac{\mathcal{F}'}{\mathcal{A}'} \,.
\end{align}
Similarly, the width is given by
\begin{equation}
  W \coloneqq \xi \sqrt{\frac{12 \abs{\hat{\mathcal{C}}'}}{c_1
    \abs{\hat{\mathcal{B}}'}}}
  =  \sqrt{\frac{12 \mathcal{C}'}{n_{\text{max}} \mathcal{B}'}} \,.
\end{equation}
Finally, the soliton decays with
\begin{equation}
  n_{\text{max}}(t) \coloneqq l_{\text{ref}}^{-d} \epsilon^{(d+6)/4} c_1(t)
    \sgn(\mathcal{B'C'})
  = \frac{n_{\text{max}}(0)}{1 + t/t_d} \,,
\end{equation}
and decay period
\begin{equation}
  t_d \coloneqq \frac{1}{\epsilon} \frac{45 \hat{\mathcal{A}}'
    \abs{\hat{\mathcal{C}}'} \xi}{4 c_1(0) \hat{\mathcal{G}}'
    \hat{\abs{\mathcal{B}}'} v_F}
  = \frac{45 \mathcal{A' C'}}{4 n_{\text{max}}(0) v_F \mathcal{G' B'}}
    \,.
\end{equation}
Here, $n_{\text{max}}(0)$ is the initial value of $n_{\text{max}}$.
The factor of $\epsilon$ in the first equality came from converting
our $\hat{t}_1/\hat{t}_d$ to $t \epsilon \xi / \hat{t}_d v_F \coloneqq
t/t_d$.

We see that, upon re-dimensionalizing, $c_1$ and $\epsilon$ never appear
alone.
Therefore, simply defining $n_{\text{max}}$ as their combination causes
all $\epsilon$ and $c_1$ to drop out, showing that this is a
one-parameter family of solutions.
Note that these results hold in general for all nondimensionalizations
specified in \cref{sec:nondim}.
Similarly, notice that the factors of $l_{\text{ref}}$ have all canceled:
the observables are all independent of $l_{\text{ref}}$, as they must be
since $l_{\text{ref}}$ is arbitrary.

As mentioned in \cref{sec:normalization}, not all of the system's
parameters are independent.
It is helpful to re-iterate here which can be set freely.
Taking into account the thermodynamic relations, one experimentally
useful set of independent parameters would be $T_0$, $n_0$,
$n_{\text{max}}(0)$, $u = u_0+\epsilon U_1$, $d_1$, $d_2$, and $\kappa$.

\subsection{Relation to Previous Results}
As mentioned in the introduction, \citet{svintsov2013hydrodynamic}
performed a similar perturbative analysis of solitons, though that
analysis was restricted to the inviscid, Fermi liquid regime.
It is straightforward to compare the inviscid results presented
in \cref{sec:strained} to those of \citet{svintsov2013hydrodynamic}.

First, our results for $v_0$ in the case of no background flow, $u_0 =
0$, are in agreement for the regime where $\mu/T \gg 1$ and
$\mu/T > 0$, but they differ otherwise.
However, this is to be expected: in setting up the problem,
\citet{svintsov2013hydrodynamic} neglect the contribution of holes.
If the contribution of holes is included in their thermodynamic
quantities, then our results are in agreement in both Fermi regimes,
$\abs{\mu/T} \gg 1$.

Nevertheless, the leading-order Dirac-regime speed $v_0$ used by
\citet{svintsov2013hydrodynamic} and derived in
\citet{svintsov2012hydrodynamic} has a minor error.
There, the terms $i k^2 \Sigma_j^2 v_F/\omega \langle p_j^{-1} \rangle$,
with $j = e$ or $h$ for electrons/holes, appear in
\eqsname{} (28) and (29) of Ref.\ \citep{svintsov2012hydrodynamic}.
These terms arise from the $\grad (v_F \langle p_j \rangle)/2$ terms in
the momentum conservation equations, \eqname{} (8) and
(9) of Ref.\ \citep{svintsov2012hydrodynamic}.
This corresponds to our pressure terms $\grad P_j$ (though we combine
$P_e$ and $P_h$ as $P = P_e + P_h$).
The issue arises when \citet{svintsov2012hydrodynamic} restricts to
leading order terms when calculating $v_0$.
As we showed in \cref{eq:K0_def}, $\grad{P}/P \sim \epsilon^2$ in the Dirac
regime (\ie $K_0 = 0$), while the inclusion of these $i k^2 \Sigma_j^2
v_F/\omega \langle p_j^{-1} \rangle$ terms in
\citet{svintsov2012hydrodynamic} implicitly assumes $\grad{P}/P \sim
\epsilon$.
On removing these terms from the leading-order equations, the
results \citet{svintsov2012hydrodynamic} are consistent with ours.

Furthermore, the Fermi-Dirac distribution function chosen by
\citet{svintsov2013hydrodynamic} differs from the one chosen by
\citet{lucas2018hydrodynamics} (and hence, used in this paper):
\citet{svintsov2013hydrodynamic} chose $f(\vec{p})$ as
\begin{equation}
  f(\vec{p}) = \frac{1}{1+\exp((\energy(\vec{p}) - \vec{u} \vdot
    \vec{p} - \mu)/k_B T)} \,,
\end{equation}
while \citet{lucas2018hydrodynamics} chose the manifestly covariant
\begin{equation}
  f(\vec{p}) = \frac{1}{1+\exp((p^{\nu} u_{\nu} - \mu)/k_B T)} \,,
\end{equation}
with $p^{\nu} = (\abs{\vec{p}}, \vec{p})$ and $u^{\nu} =
(1,\vec{u})/\sqrt{1-\abs{\vec{u}}^2/v_F^2}$.
This choice of distribution function is preferable as it preserves the
form of the dispersion relation $\energy = v_F \abs{\vec{p}}$ under
Lorentz boosts (with $\gamma = 1/\sqrt{1-(u/v_F)^2}$).

After accounting for these differences, our results are nearly in
agreement.
A few typographical errors~%
\footnote{
  The sign of the $\beta^2$ term multiplying $u \partial_x u$ in
  \eqname{} (16) of Ref.\ \citep{svintsov2013hydrodynamic} should be
  flipped.
  Additionally, the expression for $F(\nu)$ in \eqname{} (26) should
  read
  \begin{equation}
    \begin{aligned}
      F(\nu) &= \tilde{s}_0^2 -\frac{\beta^2}{2}
        -\frac{\beta_0^2}{1+\nu} \\
      &\qquad - \frac{\beta_0^2 \beta^2}{(1+\nu)^2}
        \frac{5-6\xi}{1-\beta^2} + \frac{\nu \beta_0^2 (3-4
        \xi)}{(1+\nu)^2} \,.
    \end{aligned}
  \end{equation}
  In the KdV equation, \eqname{} (27), the coefficient of the $\nu
  \partial_{\zeta} \nu$ term should be
  \begin{equation}
    (1-\xi) \pqty{2\tilde{s}_0^2-\frac{4}{3}\xi+4\beta_0^2} \,.
  \end{equation}
  Also, the solution to the KdV equation, \eqname{} (28), should be
  \begin{equation}
    \delta n(z) = \delta n_{\text{max}} \cosh^{-2}\bqty{\frac{z}{2}
      \sqrt{\frac{\bm{2}}{d_1 d_2} \frac{s_0^2}{2
      s_0^2-v_F^2}\frac{\delta n_{\text{max}}}{n_0}}} \,,
  \end{equation}
  with \eqname{} (29) changed to
  \begin{equation}
    \delta n_{\text{max}} = \bm{3} \frac{n_0}{2}
      \frac{u_0^2-s_0^2}{s_0^2} \,,
  \end{equation}
  with corrections highlighted in bold.

  For the $u_0 \neq 0$ case, \eqname (34) should be adjusted by
  flipping the sign of the $\gamma$ term multiplying the $u_0 \partial_x
  \delta u$ term.
  Furthermore, the dispersion relation, \eqname (36), should read
  \begin{equation}
    \scriptstyle
    s_{\pm} = \frac{u_0 (2-2 \xi_0+\gamma) \pm \sqrt{s_0^2(1+\gamma) +
      u_0^2 \bqty{(2-2\xi_0+\gamma)^2
      -(1+\gamma)\pqty{3-\frac{10}{3}\xi_0+\gamma}} }}{1+\gamma} \,.
  \end{equation}
}
remain in the KdV equation and corresponding soliton solution and
dispersion relation of \citet{svintsov2013hydrodynamic}.
After repairing these errors, we have consistent solutions and
dispersion relations.

It is worth noting \citet{svintsov2013hydrodynamic} also
use an isothermal assumption, though it is not directly stated;
this assumption is utilized when stating the formula~%
\footnote{
  Note that \citet{svintsov2013hydrodynamic} include factors of
  $\gamma$ in the definitions of $\energy$ and $n$; here, they have been
  factored out to match our definitions.
}
\begin{equation}
  \frac{\dd{\energy}}{\energy} = 2 \xi \frac{\dd{n}}{n} + (3-4\xi)
  \frac{\dd{T}}{T} \,,
\end{equation}
with $\xi \coloneqq n^2/\energy \langle\energy^{-1}\rangle$, and
$\langle \energy^{-1} \rangle \neq \energy^{-1}$ is the average inverse
energy.
While $\energy$ depends on both $n$ and $T$ the corresponding formula
for $\dd{\energy}/\energy$ in \citet{svintsov2013hydrodynamic} only has
the $\dd{n}/n$ term.
In the Fermi regime, $\abs{\mu/T} \gg 1$ and $\xi = 3/4$, so this is a
valid simplification.
However, in the Dirac regime, $\xi \ll 1$, and the $\dd{T}/T$ term
cannot be neglected unless the system is isothermal, $\dd{T} = 0$.

\subsection{Role of gating}
Our setup involves the use of conducting gates to screen the
electrostatic interactions and make the problem local, and hence
more mathematically tractable.
However, \citet{akbari2013universal} instead considered solitons in
ungated graphene; that analysis was restricted to the inviscid, $T=0$
Fermi regime with no background flow~%
\footnote{
  Note that \citet{akbari2013universal} uses a different terminology.
  There, the term ``Dirac fluid'' refers to massless fermions (as in
  graphene) while ``Fermi liquid'' refers to massive fermions.
  Both of these are dealt with in the completely degenerate $T=0$ limit.
  By contrast, we follow the terminology of
  \citet{lucas2018hydrodynamics} to analyze both a ``Fermi liquid''
  ($k_B T \ll \mu$) and ``Dirac fluid'' ($\mu \ll k_B T$) regime for
  massless fermions.
  Therefore, the ``Dirac'' results in \citet{akbari2013universal}
  correspond to our $T=0$ Fermi regime, while the ``Fermi'' results
  correspond to massive fermions not discussed here.
  Interestingly, bilayer graphene can induce such an effective mass for
  the quasiparticle excitations~\citep{mccann2006landau}.
}.
While \citet{akbari2013universal} also derived solitonic solutions,
a number of the properties differed markedly from those derived here.

First, \citet{akbari2013universal} found that there exists a critical
propagation velocity $v_c$ that separates periodic, wavelike solutions
($v<v_c$) and solitonic solutions ($v>v_c$).
This was found to be $v_c = 3/\sqrt{38}$ for $d=2$ and $v_c = 2/3$ for
$d=3$.
However, there appears to be a small error in the derivation:
\eqname{} (7) for $\phi$ involves a term $n^{-2/3}$ which should be
$n^{-3/2}$.
Repeating the derivation with this change shows that the
critical propagation velocity is actually $v_c = 1/\sqrt{d}$.
Our ($u_0=0$, Fermi regime) solutions have velocity
\begin{equation}
  v = \frac{1}{\sqrt{d}} \sqrt{1 + \frac{A n_0^2 d}{P_0 (d+1)}} +
  \epsilon v_1 \ge \frac{1}{\sqrt{d}} = v_c \,,
\end{equation}
where we have used the fact that $\sgn v_1 = \sgn v_0$.
Thus, we see that our soliton's speeds are bounded \emph{below} by
the critical speed, while \citet{akbari2013universal} found that solitons
speeds should be bounded \emph{above} by the critical speed.

Another difference involves the relation between the soliton height and
speed.
Using our expression for $v_1$, we found that the total speed with
$u_0=0$ is
\begin{equation}
  v = v_0 \pqty{ 1 + \epsilon \frac{c_1
  \abs{\mathcal{B}}}{3\abs{v_0 \mathcal{A}}}}
\end{equation}
while the soliton height is $\epsilon c_1$, with a free parameter
$c_1>0$~%
\footnote{
  Here we used the fact that $\sgn(\mathcal{A}'\mathcal{C}') =
  \sgn(v_0)$ for $u_0=0$
}.
Thus, increasing the height corresponds to increasing the speed, and
\viceversa.
However, \citet{akbari2013universal} found that increasing the height
causes the speed to \emph{decrease}.
Nevertheless, we both find the same, inverse relation between the height
and width (as required by total charge conservation).

Furthermore, \citet{akbari2013universal} finds only dark ($n_1/n_0 < 0$)
solitons.
However, our solutions only give \emph{bright} ($n_1/n_0 > 0$) solitons.
Referring to \cref{eq:soliton}, we have $\sgn(n_1) =
\sgn\pqty{\mathcal{B}'\mathcal{C}'}$.
Here, we will consider the Dirac ($m>0$) and Fermi ($m<0$) cases
separately.
For the Dirac regime, with $K_0 = 0$, it is readily apparent that
$\mathcal{B}' \mathcal{C}'$ (\cf \cref{sec:kdv_burgers}) is positive,
yielding bright solitons.

Showing that the same holds true in the Fermi regime, with $K_0 =
(d+1)/d$, is more involved.
Using the expressions for $\mathcal{B}'$ and $\mathcal{C}'$ from
\cref{sec:kdv_burgers}, we see
\begin{equation}
  \sgn\pqty{\frac{n_1}{n_0}} = \sgn\pqty{3d(u_0+v_0)^2-(1 + u_0 v_0)^2}
  \,.
\end{equation}
We see that this is clearly positive when $u_0 = 0$; using the
expression for $v_0$, we find it only crosses zero~%
\footnote{
  Note that this expression has a removable singularity at $u_0 = 0$;
  however, the double-sided limit exists and is $0$.
}
when $u_0$ is given by
\begin{equation}
  \begin{gathered}
    u_0 = \pm 1 \qq{or} \frac{\sqrt{2\lambda(3d-1)+4} \pm
      \lambda\sqrt{3d}}{2-\lambda} \\
    \qq{or} -\frac{\sqrt{2\lambda(3d-1)+4} \pm \lambda\sqrt{3d}
      }{2-\lambda} \,,
  \end{gathered}
\end{equation}
with $\lambda \coloneqq A n_0^2 / P_0(d+1)$ as before.
Finally, it can be checked that each of these solutions are larger (in
magnitude) than unity; that is, $\mathcal{B}'\mathcal{C}'$ does not
cross zero in the range $u_0 \in (-1,1)$.
Thus, for $\abs{u_0} < 1$, we find that $n_1/n_0 > 0$, and only bright
solitons are permitted.
Note that the adiabatic $\mathcal{B}'$ and $\mathcal{C}'$ coefficients
in \cref{sec:kdv_burgers_adi} are identical to their isothermal Fermi
counterparts: therefore, the same reasoning shows the adiabatic system
only has bright solutions, too.

Thus, it appears that a number of our findings are directly opposed to
those of \citet{akbari2013universal}.
While one might be tempted to compare the results of
\citet{akbari2013universal} with our solutions by taking the gating
distance $d_i \to \infty$, various quantities (\eg $v_0$, $W$, \etc)
would no longer be order-1, violating our expansion assumptions.
Instead, it appears that the presence or absence of gates can create
qualitatively different results.
However, this should not be surprising: the electric field with gates is
given by derivatives of the density $E \propto \partial_x n +
(d_1 d_2/3) \partial^3_x n + \ldots$.
On the other hand, the electric field without gates is given by the
\emph{anti-derivative} of $n$: $E(x) \propto \int \dd{y}
n(x)/\abs{x-y}^2$.
More specifically, the $x$-$k$ Fourier transform of the electric
potential with gates is $\hat{\phi} \propto (1-k^2 d_1 d_2/3 + \ldots)
\hat{n}$; highly-dispersive, large $k$-modes \emph{increase} the
electric field's magnitude.
The potential without gates is $\hat{\phi} \propto -\hat{n}/k^2$, so
large $k$-modes \emph{decrease} the electric field's magnitude.
Given that this is the only difference between the setup of the two
problems, it appears that this is the origin of the differences in
the results~%
\footnote{
  A number of other minor differences exist between our work and
  that of \citet{akbari2013universal}: there, velocities were normalized
  by $c$, giving $v_c = c/\sqrt{d}$.
  However, we found it more useful to normalize by $v_F$---yielding $v_c
  = v_F/\sqrt{d}$.
  This difference arose because \citet{akbari2013universal} chose to
  define $u^{\mu} = (c,\vec{u})/\sqrt{1-(u/c)^2}$
  following~\citet{zhu2010relativistic}, while we defined $u^{\mu} =
  (v_F,\vec{u})/\sqrt{1-(u/v_F)^2}$.
  Again, the choice of $v_F$, as opposed to $c$, is preferred since it
  preserves the form of the dispersion relation.
  Replacing the original choice of $u^{\mu}$ (involving $c$) with our
  choice (involving $v_F$) in \citeauthor{akbari2013universal}'s
  derivation yields $v_c = v_F/\sqrt{d}$, \ie our minimum propagation
  speed.

  Finally, our expressions for the pressure differ slightly: it appears
  \citet{akbari2013universal} considered only $g=2$ spin degeneracy in
  \eqname{} (4), rather than graphene's $g=4$ spin/valley degeneracy.
  This only affects the normalization constant ($A_{2D}$ or
  $A_{3D}$ in, for example, \eqname{} (11)), and the subsequent
  conclusions are unaffected.
}.

\subsection{\label{sec:entropy} Energy and Entropy}
It is interesting to determine the rate of energy loss by the soliton to
dissipation.
We can accomplish this by integrating the KdV-Burgers equation
\cref{eq:KdVB}.
Using \cref{eq:u_from_n} to replace $n_1$ with $u_1$, we get (with new
coefficients denoted by primes)
\begin{equation}
  \mathcal{A}' \partial_{t_1}u_1 + \mathcal{F}' \partial_{x} u_1 +
    \mathcal{C}' \partial^3_x u_1 + \mathcal{B}' u_1 \partial_x u_1 =
    \mathcal{G}' \partial^2_x u_1 \,.
\end{equation}
Multiplying this equation by $u_1$ gives
\begin{equation}
  \begin{aligned}
    &\frac{1}{2} \mathcal{A}' \partial_{t_1} u_1^2 + \frac{1}{2}
      \mathcal{F}' \partial_x u_1^2 \\
    &\qquad + \mathcal{C}' \partial_x \pqty{u_1 \partial^2_x u_1} -
      \frac{1}{2} \mathcal{C}' \partial_x \pqty{\partial_x u_1}^2 +
      \frac{1}{3} \mathcal{B}' \partial_x u_1^3 \\
    &\qquad = \mathcal{G}' \partial_x \pqty{u_1 \partial_x u_1} -
      \mathcal{G}' \pqty{\partial_x u_1}^2 \,.
  \end{aligned}
\end{equation}
If we integrate once over all of $x$-space and impose boundary
conditions $u_1 = \partial_x u_1 = 0$ at $x = \pm \infty$, we find
\begin{equation}
  \frac{1}{2} \pqty{ \partial_{t_1} + \frac{\mathcal{F}'}{\mathcal{A}'}
    \partial_x} \int \dd{x} u_1^2 = - \frac{\mathcal{G}'}{\mathcal{A}'}
    \int \dd{x} \pqty{\partial_x u_1}^2 \,.
\end{equation}
The left-hand side represents the time rate-of-change of the kinetic
energy in a moving reference frame; this is more easily seen if the
background current $U_1$ is removed so $\mathcal{F}'=0$.

Using the expressions for $\mathcal{A}'$ and $\mathcal{G}'$ (\cf
\cref{sec:kdv_burgers}), the right-hand side is negative semi-definite
for the case with no background flow $u_0 = 0$.
Thus, we see that---as expected---the viscosity causes the kinetic
energy to decrease.

When $u_0 \neq 0$, it is more difficult to see that $\mathcal{G}'/
\mathcal{A}' \ge 0$, as it must be for viscosity to remove energy.
Here, we will again treat the Dirac and Fermi regimes separately.
Starting with the Dirac case and using the expressions for
$\mathcal{A}'$ and $\mathcal{G}'$ from
\cref{sec:kdv_burgers}, we find
\begin{equation}
  \begin{aligned}
    &\sgn\pqty{\frac{\mathcal{G}'}{\mathcal{A}'}} =
      \sgn \Biggl(\sigma_Q \gamma^2 \pqty{\frac{P_0}{n_0}}^2
      \frac{(u_0+v_0)^4}{1+u_0 v_0} (d+1)^2 \\
    &\qquad + (u_0+v_0)^2 \bqty{\zeta+2 \eta \pqty{1-\frac{1}{d}}} \Biggr)
      \,,
  \end{aligned}
\end{equation}
The only questionable term is $\sigma_Q/(1+u_0 v_0)$.
This term is positive for
\begin{equation}
  \abs{u_0} < \frac{1}{\sqrt{1+[A n_0^2/P_0(d+1)]}} \,.
\end{equation}
However, it blows up when $\abs{u_0} \to 1/\sqrt{1+\lambda}$, with
$\lambda \coloneqq A n_0^2/P_0(d+1)$.
This causes $u_1$ and $P_1$ to become unbounded and invalidates our
perturbation expansion.
Thus, $\abs{u_0} < 1/\sqrt{1+\lambda}$ is a constraint on the
allowed parameters that make our derivation consistent.
Under this constraint, $\mathcal{A}'\mathcal{G}' \ge 0$ in the Dirac
regime, as it must be.

In the Fermi regime, we instead have
\begin{equation}
  \begin{aligned}
    &\sgn\pqty{\frac{\mathcal{G}'}{\mathcal{A}'}} =
      \sgn \vastl( \sigma_Q \gamma^2 \pqty{\frac{P_0}{n_0}}^2 (d+1)
      \frac{(u_0+v_0)^2}{(1+ u_0 v_0)} \\
    & + \frac{d}{d+1} \frac{(u_0+v_0)^2 \bqty{\zeta+2 \eta
      \pqty{1-\frac{1}{d}}}}{(u_0+v_0) \bqty{v_0 \pqty{d-u_0^2}+u_0 (d-1)}}
      \vastr) \,.
  \end{aligned}
\end{equation}
It is easy to show~%
\footnote{
  This can be seen by noting that the expression is positive for $u_0=0$
  and only crosses zero at $\pm 1$, $\pm \sqrt{1+\lambda d}$, or $\pm
  \sqrt{1+\lambda d}/\sqrt{1+\lambda}$, with $\lambda \coloneqq A
  n_0^2/P_0(d+1)$.
  These are each greater than (or equal to) unity for $d\ge 1$;
  therefore, the entire expression is non-negative for $\abs{u_0} \le
  1$.
}
that $(u_0 + v_0) [v_0 (d - u_0^2) + u_0 (d - 1)] > 0$
for $d > 1$ and $\abs{u_0} < 1$; recall that we already required
$\abs{u_0} < 1$, otherwise $\gamma = 1/\sqrt{1-u_0^2}$ would blow up.
Therefore, the $\eta$ and $\zeta$ terms are positive.

As in the Dirac regime, we also have a $\sigma_Q/(1+u_0 v_0)$ term.
Though $v_0$ is different in the Fermi regime, the same reasoning also
shows that this quantity is similarly positive for $\abs{u_0} <
1/\sqrt{1+\lambda}$.
Thus, as long as $\abs{u_0} < 1/\sqrt{1+\lambda}$, we see that
our theory is well-defined, $\mathcal{A}'\mathcal{G}' \ge 0$, and
viscosity causes energy to decrease, as required by the second law of
thermodynamics.
Finally, note that the adiabatic $\mathcal{G}'$ in
\cref{sec:kdv_burgers_adi} differs slightly from this isothermal Fermi
$\mathcal{G}'$; nevertheless, it shares the same questionable terms.
Thus, the same exact reasoning shows $\mathcal{G}'/\mathcal{A}' \ge 0$
for the adiabatic regime~%
\footnote{
  The $\mu_0 n_0$ term is non-negative because $\sgn \mu_0 = \sgn n_0$;
  \cf \cref{eq:dirac_n0_adi}.
}.

To further investigate the soliton's decay, it is helpful to analyze
entropy generation.
\Citet{lucas2018hydrodynamics} provide the following formula~%
\footnote{
  Note that we have added an additional factor to the $\sigma_Q$ term in
  order to account for the electrostatic interactions.
}
for the divergence of the entropy current $s^{\mu}$
\begin{equation}
\begin{aligned}
  \partial_{\mu} s^{\mu} &= \frac{1}{T} \partial_{\mu} u_{\nu}
    \bigl[ \eta  \proj^{\mu \rho} \proj^{\nu \alpha} \bigl(
    \partial_{\rho} u_{\alpha} + \partial_{\alpha} u_{\rho} -
    \frac{2}{d} g_{\rho \alpha} \partial_{\beta} u^{\beta}\bigr) \\
	& \qquad + \zeta \proj^{\mu \nu} \partial_{\alpha} u^{\alpha} \bigr] +
	  \frac{\sigma_Q}{T} \pqty{T\partial_{\mu} \frac{\mu}{T} +F_{\mu
		\rho} u^{\rho} } \\
	& \qquad \times \proj^{\mu \nu} \left( T
	  \partial_{\nu}\frac{\mu}{T}+F_{\nu \rho} u^{\rho} \right) \,.
\end{aligned}
\label{eq:entropy_div}
\end{equation}

For simplicity, consider the case with no background flow, $u_0=U_1=0$.
Upon implementing our usual nondimensionalization in the Dirac regime
(\cf \cref{sec:normalization}) we see the highest-order terms are
\begin{equation}
  \begin{aligned}
    \partial_{\mu} s^{\mu} &=
      \frac{\eta}{T_0} \partial^{i} u_1^{j}
      \bigl[ \partial_{i} (u_1)_{j} + \partial_{j} (u_1)_{i} -
      \frac{2}{d} g_{i j} \partial_{k} u_1^{k}\bigr] \\
    &\qquad + \frac{\zeta}{T_0} \pqty{\partial_{k} u_1^{k}}^2
  \end{aligned}
\end{equation}
Then, restricting to 1-dimensional motion and using our thermodynamic
relations and first-order solutions, we find
\begin{equation}
  \begin{aligned}
    \partial_{\mu} s^{\mu} &= (\partial_x n_1)^2 \bqty{\zeta + 2\eta
      \pqty{1-\frac{1}{d}}} \frac{v_0^2}{T_0 n_0^2} + \order{\epsilon}
      \,.
  \end{aligned}
\end{equation}

\begin{figure*}
  \centering
  \includegraphics{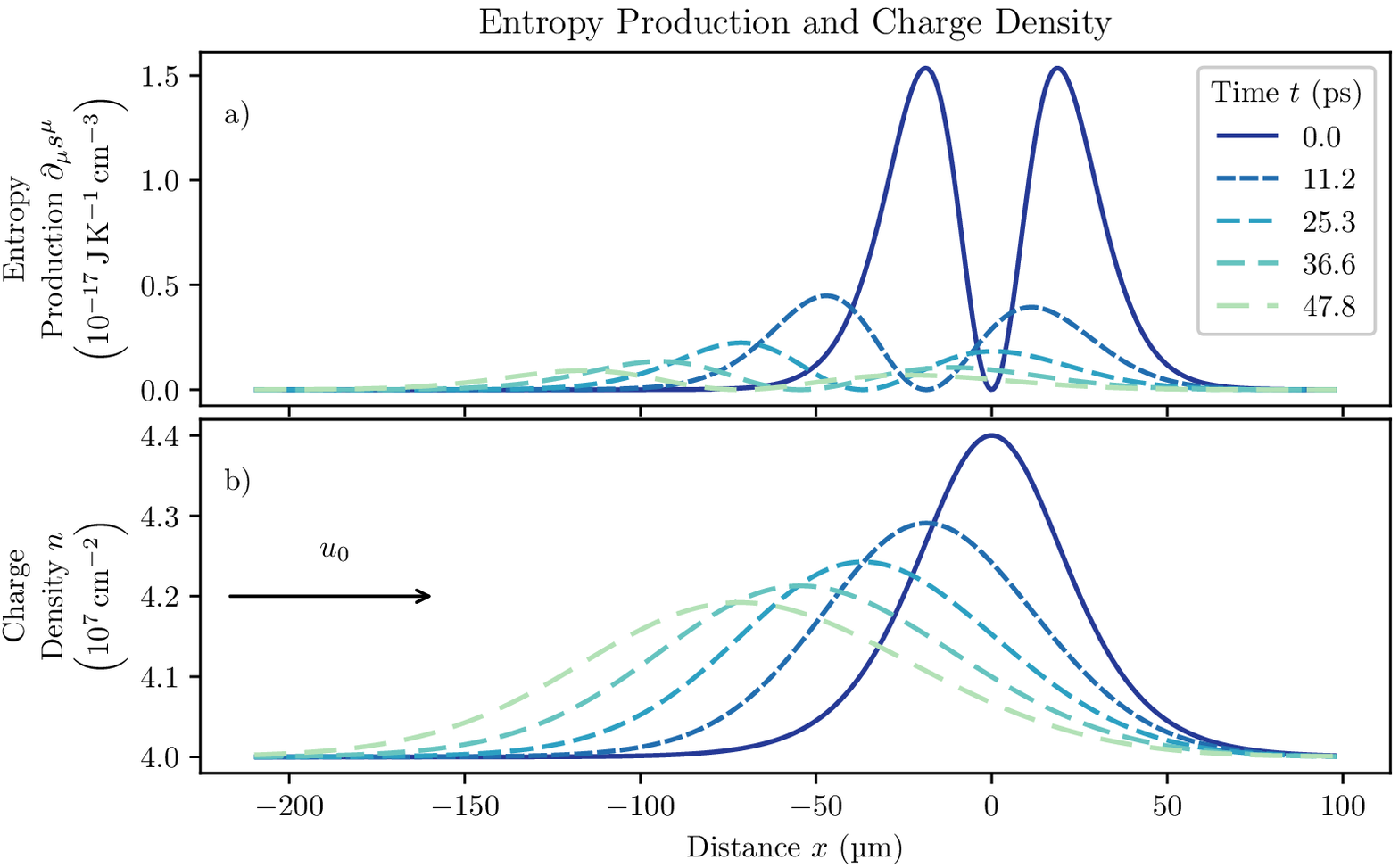}
  \caption{(Color online) The entropy production $\partial_{\mu}
    s^{\mu}$ (a) and soliton charge density $n_1$ (b) at select times.
    Values used were $\mathcal{A}=\AVal$, $\mathcal{B}=\BVal$,
    $\mathcal{C}=\CVal$, $\mathcal{F}=\FVal$, and $\mathcal{G}=\GVal$
    with the height normalized to \solitonHeightDim{}.
    }
  \label{fig:entropy}
\end{figure*}

\begin{figure*}
  \centering
  \includegraphics{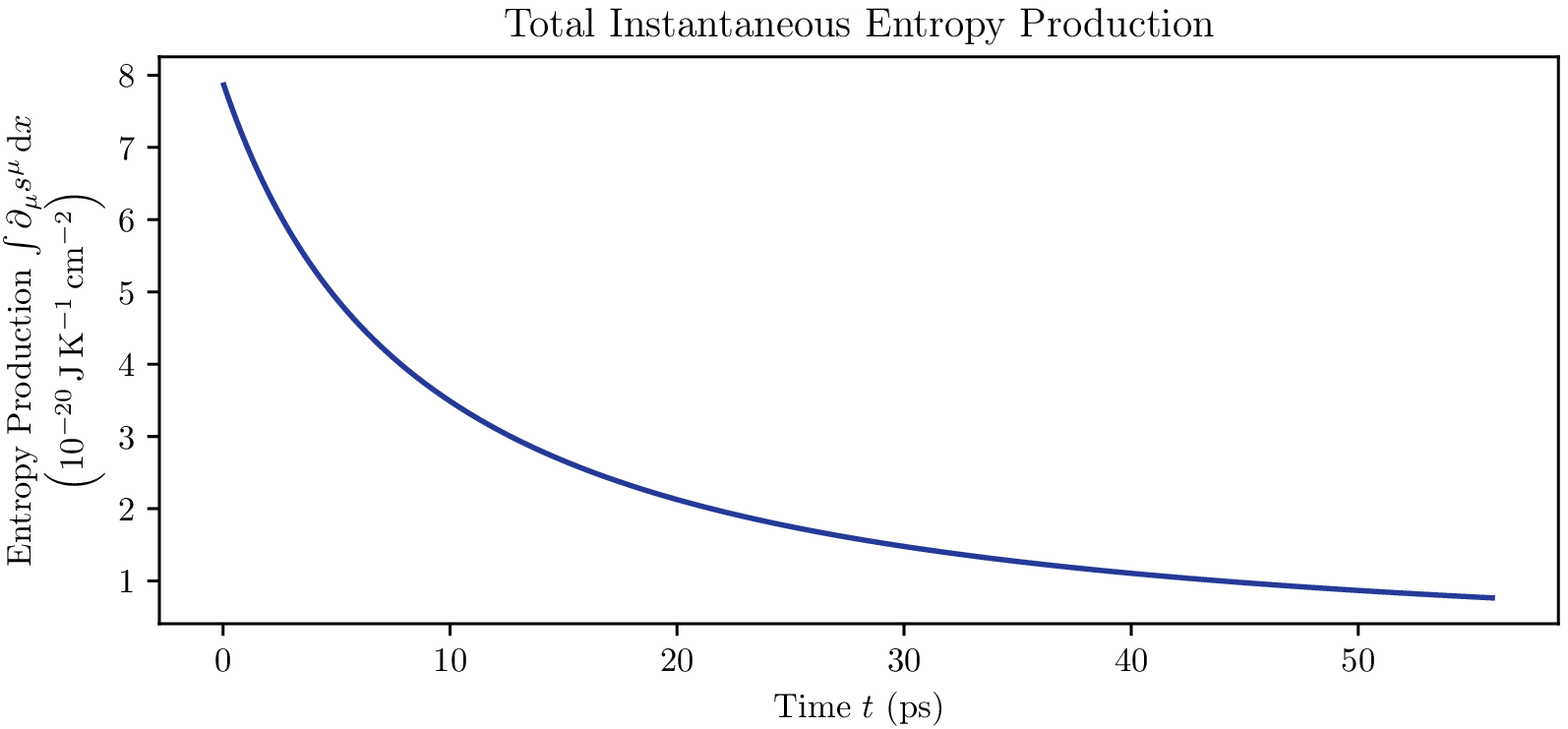}
  \caption{(Color online) The instantaneous entropy production
    $\partial_{\mu} s^{\mu}$ as a function of time.
    Values used were $\mathcal{A}=\AVal$, $\mathcal{B}=\BVal$,
    $\mathcal{C}=\CVal$, $\mathcal{F}=\FVal$, and $\mathcal{G}=\GVal$
    with the height normalized to \solitonHeightDim{}.
    }
  \label{fig:entropy_prod}
\end{figure*}

We see that entropy is generated at locations where the derivative of
$n_1$ is largest: for solitons, this occurs at the leading and trailing
faces (\cref{fig:entropy}).
Further, as the soliton spreads out, the entropy production slows over
time (\cref{fig:entropy_prod}).
Finally, for the Dirac regime, $\sigma_Q$-induced entropy production is
suppressed to sub-leading order; $\eta$ and $\zeta$ are the main
producers of entropy.

\section{\label{sec:experimental} Experimental Proposal}
Here, we will briefly detail the applicability of this theory to
experiment.

\subsection{Values of Parameters}
\begin{table}
  \centering
  \begin{tabular}{Cc @{\hskip 12pt} Cc Cc Cc}
    \hline \hline
    & $\epsilon$-dependence & \shortstack[c]{Sample \\ Nondim. \\ Value} &
      \shortstack[c]{Sample \\ Dim.\ Value} \\
    \hline
    $n_0$ & $\epsilon \hat{n}_0$ \nUnits & \nVal & \nDim \\
    $d_i$ & $\epsilon^{-5/4} \hat{d}_i$ \dUnits & \dVal & \dDim \\
    $A$ & $\epsilon^{-5/4} \hat{A}$ \AAUnits & \AAVal & \AADim \\
    $T_0$ & $\epsilon^{1/4} \hat{T}_0$ \TUnits & \TVal & \TDim \\
    $P_0$ & $\epsilon^{3/4} \hat{T}_0^3$ \PUnits & $\PDivTVal
      \hat{T}_0^3$ & \PDim \\
    $\mu_0$ & $\epsilon^{3/4} \frac{\hat{n}_0}{\hat{T}_0}$
      \muUnits & $\muDivnVal \frac{\hat{n}_0}{\hat{T}_0}$ & \muDim \\
    $\sigma_Q$ & $\epsilon^{1/2} \hat{\sigma}_Q$ \sigmaUnits & \sigmaVal
      & \sigmaDim \\
    $\eta$ & $\hat{\eta}$ \etaUnits & \etaVal & \etaDim \\
    \hline \hline
  \end{tabular}
  \caption{Values of the various parameters in terms of the small
    parameter $\epsilon$.
    Sample values are given for $\epsilon = \epsilonVal$ and dimension $d=2$.
  }
  \label{table:values}
\end{table}

It has been more convenient to deal with nondimensional variables
throughout the derivation.
However, we now convert back to dimensionful quantities to better
understand their physical magnitude.
It is worth emphasizing that this conversion is dependent on the
nondimensionalization we chose.
The values calculated in this section are specific to the Dirac regime
nondimensionalization laid out in \cref{sec:normalization}; a similar
analysis could be performed for the Fermi regime nondimensionalization
specified in \cref{sec:param_choice}.

The dimensional and nondimensional values of the various parameters in
the problem are listed in \cref{table:values}.
For the remainder of this section, we will specialize to dimension
$d=2$.
Note that we are using the values $v_F =
c/300$~\citep{lucas2018hydrodynamics} and $l_{\text{ref}} = \lRefDim$.
For computing the sample values, we have chosen $\epsilon = \epsilonVal$.
We see that all of the nondimensional parameters are approximately equal
to unity, as required.
However, there are a few points to note.

In previous experiments, the distance between the graphene and the gates
$d_i$ ($i = 1,2$) was usually on the order of
\SI{300}{\nano\meter}~\citep{dmitriev2001plasma}.
We require a larger gate distance of $d_i = \dDim$ corresponding to
$\hat{d}_i = \dVal$.
The static dielectric constant $\kappa$ must be chosen relative to $d_1$
and $d_2$.
For the remaining normalizations to be consistent, we require $\kappa
\approx \kappaDim$.
That is, the graphene should be suspended from its contacts with vacuum
filling the gap between the graphene sheet and the conducting gates.

It is important to reiterate the way we nondimensionalized the
intrinsic conductivity.
At a temperature of \TDim{}, $\sigma_Q/e^2$ has a fixed value
of $\sigmaNondim \hbar^{-1}$.
We needed to relate the relative sizes of nondimensional parameters
$\epsilon$ and $\sigma_Q \hbar/e^2$ to solve the problem.
Our derivation assumed $\epsilon \sim \epsilonVal$, so that
$\epsilon^{1/2} \sim \sigma_Q \hbar/e^2$.
This fixes the value of $\hat{\sigma}_Q$ as $\hat{\sigma}_Q =
\sigmaNondim \epsilon^{-1/2}$.

Notice that if $\epsilon$ is increased, then the numerical value of
$\hat{\sigma}_Q$ decreases; hence, the intrinsic conductivity becomes a
higher-order correction and drops out of our first-order solutions.
Conversely, if $\epsilon$ is decreased, $\hat{\sigma}_Q$ could grow
large and require a different nondimensionalization for $\sigma_Q$.
For $\epsilon$ small enough, it would be more appropriate to take
$\sigma_Q = \epsilon^{0} \hat{\sigma}_Q e^2/\hbar$.
This alternative would require different nondimensionalizations for all
variables (\cf \cref{sec:nondim}); nevertheless, similar solutions would
result (though the viscosity would no longer appear in the first-order
corrections).
Similar considerations also apply for $\eta$, though it is considerably
simpler given that $\eta l_{\text{ref}}^d / \hbar \approx 1$.

It is also useful to determine the values of the parameters appearing as
coefficients in the KdV and KdV-Burgers equations (\ie $\mathcal{A}$,
$\mathcal{B}$, $\mathcal{C}$, and $\mathcal{G}$).
For instance, consider the case with $v_0 = 0$, $u_0>0$, and $U_1=0$; we
will also set $\zeta=0$ and choose $c_1 = \cVal$.
Using the above values and the bare thermodynamic coefficients
$\mathcal{C}_0$ and $\mathcal{C}_1$ (\cf \cref{sec:thermo_coeffs}), we
find $\mathcal{A}=\AVal$, $\mathcal{B}=\BVal$, $\mathcal{C}=\CVal$,
$\mathcal{F}=\FVal$, and $\mathcal{G}=\GVal$ (\cf \cref{fig:soliton}).
Importantly, we see that $\mathcal{B}$, $\mathcal{C}$, and $\mathcal{G}$
are all roughly the same order, implying nonlinearity, dispersion, and
dissipation are equally important.

\subsection{\label{sec:source_signal} Source and Signal}
As we discussed in \cref{sec:normalization}, the characteristic length
of the disturbance $\xi$ is related to $l_{\text{ref}}$ as $\xi =
l_{\text{ref}}/\epsilon^{(d+5)/4}$.
For $d=2$ and $\epsilon=\epsilonVal$ with graphene's $l_{\text{ref}} =
\lRefDim$, we find a pulse width of approximately
$\widthDim$.
For the $u_0=0$ case, the propagation speed is approximately $v
= \vVal v_F \sim \vVal c/300$, giving a bandwidth of roughly $v/\xi
= \bandwidthDim$.

If we consider the stationary soliton case $v_0 = 0$, we need to source
a background current $u_0 \neq 0$ to counteract its propagation.
In \cref{sec:stationary}, we found that $u_0 = \uVal v_F = \uDim$; with
a charge density of $n_0 = \nDim$, we need a current density of $K_0 =
\abs{e n_0 u_0} = \currentDensityDim$.

As shown previously, the system has a (dimensional) characteristic decay
time of
\begin{equation}
  t_d = \frac{45 l_{\text{ref}} \mathcal{A} \abs{\mathcal{C}}}{4
    \epsilon^{11/4} v_F \mathcal{G} \abs{\mathcal{B}}} \,.
\end{equation}
Inserting the previously chosen values for these coefficients, we find
$t_d \approx \tdDim$.

To estimate the magnitude of the signal, we first calculate the background
chemical potential $\mu_0 = \epsilon^{3/4} \hbar v_F l_{\text{ref}}^{-1}
\hat{\mu}_0 = \muDim$.
From this, we find the background voltage $V_0 = \mu_0/e =
\voltageDim$.
Then, the signal voltage $V_1 = \mu_1/e$ would be a factor of $\epsilon
\sim \epsilonVal$ smaller, or $\voltagePertDim$.

\subsection{\label{sec:joule} Joule Heating}
For the non-propagating case ($v_0 = 0$), a large uniform background
current $u_0$ flows through the graphene; this will cause Joule heating
of the entire sample due to graphene's resistance.
It is worthwhile to verify that this heating occurs sufficiently slowly
so as not to interfere with the soliton's propagation and decay.

The power produced, per unit area, by Joule heating is
\begin{equation}
  P_J = K_0^2 \rho \,,
\end{equation}
with resistivity $\rho$ and surface current density $K_0$.
As a worst-case scenario, assuming the graphene does not lose any heat
to the environment, this power goes solely towards heating the graphene.

The specific heat of graphene~\citep{popov2002low} at \TDim{} is
approximately \csMassDim{}.
Given an atomic mass of \SI{12.01}{\gram\per\mole} for carbon and an
atomic density of \SI{6.3}{\mole\per\centi\meter\squared} for carbon
atoms in graphene~\citep{bong2015graphene}, we find a specific heat of
$c_s = \csAreaDim$.

Therefore, the soliton's temperature will change at a rate of $P_J/c_s
= \JouleTempRateDim$.
Given that the suggested experiment would be measuring the soliton's
temperature anomaly $T_1 = \epsilon T_0$, it would only be sensitive to
Joule heating after a temperature change of similar magnitude had been
generated.
Hence, it would take approximately $T_1 c_s / P_J = \tJouleDim$
for the system to heat appreciably.
Given that this time is long compared to the characteristic timescales
of the problem ($t_{\text{char}}$ and $t_d$), we are justified in
neglected Joule heating.

Notice that the characteristic Joule-heating time is also long compared
to the electron-phonon scattering time; this implies the electrons and
graphene lattice would thermalize relatively quickly compared to the
Joule heating time.
This is why we utilized the specific heat of the entire graphene system
(electrons and lattice) as opposed to the specific heat of only the
electrons.

\subsection{Experimental Setup}

\begin{figure}
  \centering
  \includegraphics{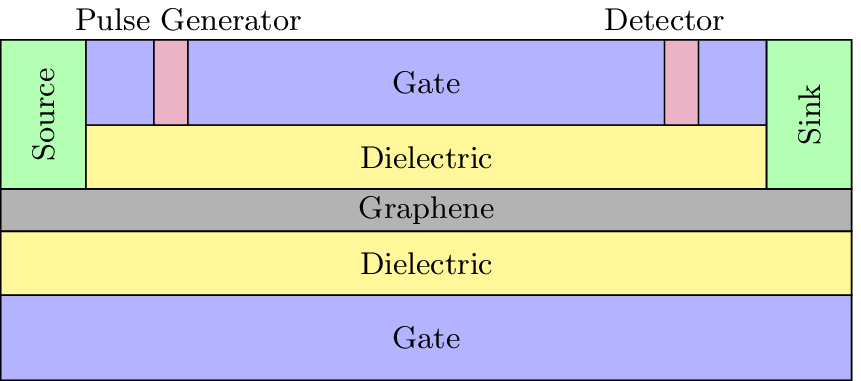}
  \caption{(Color online) Side view of the proposed experimental setup;
    the graphene is sandwiched between two layers of dielectric, and
    further sandwiched between two conducting gates.
    A source and sink on either edge of the graphene generate the
    background current $u_0$.
    The pulse generator produces the soliton and the detector detects
    it.
  }
  \label{fig:setup}
\end{figure}

The solitonic solutions we have derived offer a means to experimentally
measure the viscosity $\eta$ of graphene.
In particular, the viscous coefficients $\sigma_Q$, $\eta$, and $\zeta$
all enter into the coefficient we have denoted $\mathcal{G}$.
Therefore, if the value of $\mathcal{G}$ can be measured, then the
viscosity can be determined.

Referring to the expression for $\mathcal{G}$, we see that $\eta$ only
appears in the combination $\zeta +2 \eta(1-1/d)$; hence, it is this
quantity that can be determined from experiment.
In practice, we expect $\zeta \ll \eta$, and thus this procedure
offers an estimate for $\eta$~\citep{lucas2018hydrodynamics}.
Furthermore, determining $\eta$ from $\mathcal{G}$ requires knowing the
values of all the other parameters $P_0$, $n_0$, \etc.
Most of these are experimentally determined and hence known; the only
other necessary quantity is the intrinsic conductivity $\sigma_Q$.
Previous measurements of this quantity
exist~\citep{novoselov2005two,crossno2016observation};
therefore, it can be treated as a known quantity.

An initial disturbance needs to be generated in the graphene; for
instance, this can be accomplished via a short voltage spike produced by
a thin contact placed laterally atop the sample (\cf \cref{fig:setup}).
It is well known that the KdV equation causes a localized
profile to split into a series of left- and right-moving
solitons~\citep{ablowitz1974inverse} sorted by height.
After the disturbance is allowed to propagate a sufficient distance, the
individual solitons should have separated enough to be separately
distinguished.
The actual population of solitons generated by the pulse will be
dependent on the contact's shape and voltage profile: the distribution
of soliton heights and widths can be determined by the inverse
scattering transform~\citep{gardner1967method}.

Given that the solitons represent a localized change in the charge
density, it should be possible to detect them with a voltmeter; a
voltage time-series could then reconstruct the soliton profile.
The dissipative terms cause two measurable effects: a change in the
propagation speed and a decay of the soliton's height.
This requires measuring either the soliton's speed or amplitude as a
function of time.
Depending on the particular experimental setup, one effect might be more
accessible than the other.
Next, we describe two possible experimental setups.

\subsubsection{No Propagation}
Without a background current $u_0 = 0$, the soliton propagates at a
speed $v \approx v_F \approx c/300$.
Such a fast propagation speed could make measurement difficult.
One way to mitigate this is to impose a counter-current $u_0$ in the
opposite direction of propagation; as detailed in \cref{sec:stationary},
it is possible to choose a background current $u_0 + \epsilon U_1$ such
that the soliton is stationary in the laboratory frame $v_0 + \epsilon v_1
= 0$.
Doing this should make obtaining the height measurements much easier.
In fact, the speed measurements are still feasible in this setup since
the dissipation causes $v_1$, and hence the control current $U_1$, to
decay over time.

One possible barrier to implementation of this method is the boundary
condition of graphene.
So far, we have neglected boundary effects by assuming one-dimensional
propagation; depending on graphene's boundary conditions, this might not
be justified.
Graphene most likely satisfies one of two possible boundary
conditions~\citep{kiselev2018boundary}: either a no-slip boundary
($\vec{u} = 0$) or no-stress (no normal velocity gradient, \ie $[\hat{n}
\vdot \grad] \vec{u} = 0$ with $\hat{n}$ the boundary unit normal).
If the actual boundary is no-slip, our 1-dimensional propagation
assumption is violated; in this case, the sample must be sufficiently
wide to ignore edge effects, or a different experimental setup (\cf the
next section) is needed.
Conversely, a no-stress boundary permits our one-dimensional soliton
solution.
There is some experimental evidence that no-stress boundaries
are the correct boundary type~\citep{lucas2018hydrodynamics}, and theory
predicts that weakly disordered edges at low temperature ($T \lesssim
\SI{40}{\kelvin}$) have a slip-length on the order of
\SI{50}{\micro\meter}.
Therefore, it is plausible that, for graphene samples of width at most
$\sim \SI{100}{\micro\meter}$, a no-stress boundary condition is
appropriate, allowing for large $u_0$ counter-current.

\subsubsection{No Background Current}
If graphene instead possesses a no-slip boundary condition, a
different experimental method will be needed.
For this setup, we will not use a background flow, $u_0 = 0$.
Then, the boundary conditions are mostly irrelevant, since the
fluid velocity is now of order $\order{u_1} = \epsilon v_F$ and can
therefore be made small.
For this setup, height measurements are more suitable; after one decay
period $\tau_0$, the height decreases by a factor of $\frac{1}{2}$ while
the propagation velocity changes by a factor of $\delta v/v_0 =
\frac{1}{2} \epsilon \ll 1$.

Following the method proposed by \citet{coelho2017kelvin}, we
recommend periodically producing a voltage pulse and measuring a set
distance away.
By averaging over many realizations, it should be possible to obtain a
wave profile.
This could be repeated at a few locations, thereby measuring the decay
rate as a function of downstream position.

This method is likely more difficult experimentally given that it
requires taking measurements at multiple locations sequentially.
However, it has the benefit of being theoretically sound regardless of
graphene's boundary conditions.

\section{Conclusion}
Graphene offers a fantastic environment for studying strong-coupling
phenomena.
Hydrodynamic analysis presents a useful set of tools for analyzing the
long-wavelength physics in such a clean, strongly-coupled system.
The Fermi liquid regime has much in common with ordinary metals and has
been the focus of many experiments in graphene; meanwhile, the Dirac
fluid regime hosts a number of intriguing phenomena.
When graphene is placed in a hydrodynamic regime, the electrons obey
relativistic Navier-Stokes equations and can form solitonic solutions.
An ordinary perturbation expansion was used to derive the special case
of a stationary soliton on a background counter flow.
Additionally, a full multiple scales asymptotic analysis was utilized to
treat the general case with arbitrary background flow.
These methods furnished analytic approximations to the shape and speed
of the predicted solitons.
This analysis did not deal with the boundary conditions of the fluid
flow; this offers an interesting avenue for future research.

By including dissipation in our system, we were able to model the decay
of the solitons.
The analysis showed that dissipation causes both a decay of the
soliton's height as well as its speed.
This decay rate offers a means to experimentally measure dissipation in
the hydrodynamic regime of graphene.
The results of this paper help elucidate the connection between solitons
in the Fermi and Dirac regimes of graphene and put forward a new
method for measuring hydrodynamically relevant parameters such as the
intrinsic conductivity and shear viscosity.

\begin{acknowledgments}
  Special thanks to Falk Feddersen for his invaluable support and input.
  The computations in this paper were performed by using
  \textsc{maple}\texttrademark{}~\citep{maple2018}.
  This work was supported in part by funds provided by the U.S.\
  Department of Energy (D.O.E.)\ under cooperative research agreement
  DE-SC0009919.
\end{acknowledgments}

\appendix

\section{\label{sec:thermo_coeffs} Thermodynamic Coefficients}
Following \citet{lucas2018hydrodynamics}, we can derive
the pressure for weak coupling, starting from the grand canonical
ensemble for a free Fermi gas in $d$ dimensions
\begin{align}
  P(\mu, T) &= -\frac{\Phi_G}{V} = \frac{k_B T}{V} \sum_{A,\vec{p}}
    \ln(\mathcal{Z}_{A,\vec{p}}) \nonumber \\
  &= k_B T \sum_A \int \frac{\dd[d]{\vec{p}}}{(2 \pi \hbar)^d}
    \ln(1+e^{(q_A \mu - \energy_A(\vec{p}))/k_B T}) \nonumber \\
  &\begin{aligned}
  &= -\frac{4(k_B T)^{d+1} \Omega_{d-1} (d-1)!}{(2 \pi \hbar v_F)^d}
    \bigl(\Li_{d+1}(-e^{\mu/k_B T}) \\
  &\qquad + \Li_{d+1}(-e^{-\mu/k_B T}) \bigr) \,.
  \end{aligned}
  \label{eq:eqn_state}
\end{align}
Here, we have $\Phi_G$ the grand potential, $\mathcal{Z} =
\exp(-\Phi_G/k_B T)$ the grand partition function, and $V$ the volume.
We made use of the fact that, for a free Fermi gas, the grand partition
function is separable over modes ($A$ and $\vec{p}$): $\mathcal{Z} =
\prod_{A,\vec{p}} \mathcal{Z}_{A,\vec{p}}$.
Additionally, we have the excitation energy $\energy_A(\vec{p}) = v_F
\abs{\vec{p}}$, $\Omega_{d-1} = 2 \pi^{d/2}/\Gamma(d/2)$ the surface
area of a unit $(d-1)$-sphere, $\Gamma$ is the gamma function, and
$\Li_{d}$ the polylogarithm of order $d+1$.
Note that the sum over species runs over spin/valley degeneracy (giving
a factor of 4) as well as electrons/holes with $q_A = \pm 1$.
More specifically, $\sum_A \ln(\mathcal{Z}_A) = 4
\ln(\mathcal{Z}_1(\mu,T)) + 4 \ln(\mathcal{Z}_1(-\mu,T))$.

Likewise, the carrier density is given by
\begin{align}
	n(\mu, T) &= \pdv{P}{\mu} = \frac{4(k_B T)^d \Omega_{d-1} (d-1)!}{(2 \pi
	  \hbar v_F)^d} \nonumber \\
  &\qquad \times \bigl(-\Li_{d}(-e^{\mu/k_B T}) + \Li_{d}(-e^{-\mu/k_B
    T}) \bigr) \,.
\end{align}
We can develop series (asymptotic) expansions in Dirac (Fermi) regimes.

In the Dirac regime ($\mu \ll k_B T$), the polylogarithm can be
approximated as~\citep{wood1992computation}
\begin{equation}
  \Li_s(-e^z) = -\sum_{k=0}^{\infty} \eta(s-k) \frac{z^k}{k!} \,,
\end{equation}
for $\abs{z} < \pi$, with $\eta$ the Dirichlet eta function.
Thus, the pressure is given by
\begin{equation}
  \begin{aligned}
    P(\mu,T) &=  8 \frac{(k_B T)^{d+1} \Omega_{d-1}(d-1)!}{(2 \pi \hbar
      v_F)^d} \\
    &\qquad  \times \sum_{k=0}^{\infty} \frac{\eta(d+1-2k)}{(2k)!}
      \pqty{\frac{\mu}{k_B T}}^{2k} \\
    &= 8 \frac{(k_B T)^{d+1} \Omega_{d-1}(d-1)!}{(2 \pi \hbar v_F)^d}
      \Biggl[\eta(d+1) \\
    &\qquad + \frac{\eta(d-1)}{2}\pqty{\frac{\mu}{k_B
      T}}^2 + \order{\frac{\mu}{k_B T}}^4 \Biggr] \,,
  \end{aligned}
\end{equation}
and the carrier density is
\begin{equation}
  \begin{aligned}
    n(\mu,T) &=  \frac{8 \mu (k_B T)^{d-1} \Omega_{d-1}(d-1)!}{(2 \pi \hbar
      v_F)^d} \\
    &\qquad \times \sum_{k=0}^{\infty} \frac{\eta(d-1-2k)}{(2k+1)!}
      \pqty{\frac{\mu}{k_B T}}^{2k} \\
    &= \frac{8 \mu (k_B T)^{d-1} \Omega_{d-1}(d-1)!}{(2 \pi \hbar v_F)^d}
      \Bigl[\eta(d-1) \\
    &\qquad + \frac{\eta(d-3)}{6}\pqty{\frac{\mu}{k_B
      T}}^2 + \order{\frac{\mu}{k_B T}}^4 \Bigr] \,.
  \end{aligned}
\end{equation}

For instance, for $d=2$, we find
\begin{equation}
  P = \frac{(k_B T)^3}{(\hbar v_F)^2} \Biggl[ \frac{4 \eta(3)}{\pi} +
    \frac{2 \ln(2)}{\pi} \left(\frac{\mu}{k_B T}\right)^2 +
    \order{\frac{\mu}{k_B T}}^4 \Biggr] \,,
\end{equation}
and
\begin{equation}
  n = \frac{\mu (k_B T)}{(\hbar v_F)^2} \left[
    \frac{4 \ln(2)}{\pi} + \frac{1}{6\pi}
    \pqty{\frac{\mu}{k_B T}}^2 + \order{\frac{\mu}{k_B T}}^6 \right] \,.
\end{equation}

Instead, in the Fermi regime ($\mu \gg k_B T$), an asymptotic expansion
of the polylogarithm is given by~\citep{wood1992computation}
\begin{equation}
  \Li_s(-e^z) = -2\sum_{k=0}^{\lfloor {s/2} \rfloor}
    \frac{\eta(2k)}{(s-2k)!} (z)^{s-2k} + \order{e^{-z}} \,,
\end{equation}
for $\Re{z} \gg 1$, while $\Li_s(-\exp(-z))$ is sub-dominant and
therefore can be neglected.
Thus, we find
\begin{equation}
  \begin{aligned}
    P(\mu,T) &=  \frac{8 \abs{\mu}^{d+1} \Omega_{d-1}}{(2 \pi \hbar
      v_F)^d} \\
    &\qquad \times \sum_{k=0}^{\lfloor {(d+1)/2} \rfloor}
      \frac{\eta(2k)(d-1)!}{(d+1-2k)!} \pqty{\frac{k_B T}{\mu}}^{2k} \\
    &= \frac{8 \abs{\mu}^{d+1} \Omega_{d-1}}{(2 \pi \hbar v_F)^d}
      \Biggl[\frac{1}{2(d+1)d} \\
    &\qquad + \frac{\pi^2}{12}\pqty{\frac{k_B T}{\mu}}^2 +
      \order{\frac{k_B T}{\mu}}^4 \Biggr] \,,
  \end{aligned}
\end{equation}
and the carrier density is
\begin{equation}
  \begin{aligned}
    n(\mu,T) &=  \frac{8 \abs{\mu}^{d} \sgn(\mu) \Omega_{d-1}}{(2 \pi
      \hbar v_F)^d} \\
    &\qquad \times \sum_{k=0}^{\lfloor {d/2} \rfloor}
      \frac{ \eta(2k)(d-1)!}{(d-2k)!} \pqty{\frac{k_B
      T}{\mu}}^{2k} \\
    &= \frac{8 \abs{\mu}^{d} \sgn(\mu) \Omega_{d-1}}{(2 \pi \hbar v_F)^d}
      \Biggl[\frac{1}{2d} \\
    &\qquad + \frac{\pi^2(d-1)}{12}\pqty{\frac{k_B T}{\mu}}^2 +
      \order{\frac{k_B T}{\mu}}^4 \Biggr] \,,
  \end{aligned}
\end{equation}
Again, for $d=2$, we have
\begin{equation}
  P = \frac{\abs{\mu}^{3}}{(\hbar v_F)^2} \left[\frac{1}{3\pi} +
    \frac{\pi}{3} \left(\frac{k_B T}{\mu}\right)^2 + \order{\frac{k_B
    T}{\mu}}^4 \right] \,,
\end{equation}
and
\begin{equation}
  n = \frac{\mu^{2}\sgn(\mu)}{(\hbar v_F)^2} \left[\frac{1}{\pi} +
    \frac{\pi}{3} \left(\frac{k_B T}{\mu}\right)^2 + \order{\frac{k_B
    T}{\mu}}^4 \right] \,.
\end{equation}

Thus, we find the following coefficients
\begin{gather}
  \mathcal{C}_0^F = \frac{8 \abs{\mu}^{d+1} \Omega_{d-1}}{(2 \pi \hbar
    v_F)^d} \frac{1}{2 (d+1) d} \\
  \mathcal{C}_1^F = \frac{8 \abs{\mu}^{d+1} \Omega_{d-1}}{(2 \pi \hbar
    v_F)^d} \frac{\pi^2}{12}
\end{gather}
and
\begin{gather}
  \mathcal{C}_0^D = 8 \frac{(k_B T)^{d+1} \Omega_{d-1}(d-1)!}{(2 \pi
    \hbar v_F)^d} \eta(d+1) \\
  \mathcal{C}_1^D = 8 \frac{(k_B T)^{d+1} \Omega_{d-1}(d-1)!}{(2 \pi
    \hbar v_F)^d} \frac{\eta(d-1)}{2} \,.
\end{gather}

When screening is not negligible, these coefficients get renormalized.
For instance, the Dirac coefficients for $d=2$ and $T \to 0$
become~\citep{lucas2016transport}
\begin{gather}
  \mathcal{C}_0^D = 8 \frac{(k_B T)^{3} \Omega_{1}}{(2 \pi
    \hbar v_F)^2} \eta(3) \pqty{\frac{\alpha(T)}{\alpha_0}}^2 \\
  \mathcal{C}_1^D = 8 \frac{(k_B T)^{3} \Omega_{1}}{(2 \pi
    \hbar v_F)^2} \frac{\eta(1)}{2} \pqty{\frac{\alpha(T)}{\alpha_0}}^2
    \,,
\end{gather}
with $\alpha(T)$ given in \cref{eq:alpha_renorm}.

\section{\label{sec:nondim} General Nondimensionalization}
A critical aspect of these derivations was the correct choice of
nondimensionalization scheme.
Depending on the physical regime of interest (Fermi \vs Dirac) as well
as the relative size of terms (\eg how large $\epsilon$ is compared to
$\sigma_Q \hbar/e^2$), different nondimensionalization choices may be
appropriate.
To elucidate the relationship between these various schemes a single,
general nondimensionalization can be performed.
In this section, we will use a unit system in which $\hbar = v_F = k_B =
l_{\text{ref}} = e = 1$.
Note: we are only nondimensionalizing ($\hbar = 1$, \etc), but not
normalizing; \ie we are not requiring that all quantities are unity
(unlike the quantities denoted earlier by carets).

For convenience, the main results are collected here:
\begin{equation}
   \begin{gathered}
     \order{\mu} = \epsilon^{\frac{1}{2} q - \frac{1}{2} p + \frac{1}{2}
       m + \frac{1}{2} \abs{m}}
       \sqrt{\frac{\order{\eta}}{\order{\sigma_Q}}} \,, \\
     \order{T} = \epsilon^{\frac{1}{2} q - \frac{1}{2} p + \frac{1}{2}
       \abs{m}} \sqrt{\frac{\order{\eta}}{\order{\sigma_Q}}} \,, \\
     \order{P} = \epsilon^{\frac{d+1}{2} q - \frac{d+1}{2} p +
       \frac{d+1}{4} m + \frac{d+1}{4} \abs{m}}
       \sqrt{\frac{\order{\eta}}{\order{\sigma_Q}}}^{d+1} \,, \\
     \order{n} = \epsilon^{\frac{d}{2} q - \frac{d}{2} p + \frac{d+1}{4}
       m + \frac{d+1}{4} \abs{m}}
       \sqrt{\frac{\order{\eta}}{\order{\sigma_Q}}}^d \,, \\
     \order{\partial_x} = \epsilon^{1 + \frac{d+1}{2} q - \frac{d-1}{2}
       p + \frac{d+1}{4} m + \frac{d+1}{4} \abs{m}}
       \sqrt{\frac{\order{\eta}^{d-1}}{\order{\sigma_Q}^{d+1}}} \,, \\
     \order{d_i} = \epsilon^{-\frac{1}{2} - \frac{d+1}{2} q + \frac{d-1}{2}
       p - \frac{d+1}{4} m - \frac{d+1}{4} \abs{m}}
       \sqrt{\frac{\order{\sigma_Q}^{d+1}}{\order{\eta}^{d-1}}} \,, \\
     \order{u} = 1 \,, \\
     \order{A} = \epsilon^{ \frac{-d+1}{2} q + \frac{d-1}{2} p -
       \frac{d+1}{4} m - \frac{d+1}{4} \abs{m}}
       \sqrt{\frac{\order{\sigma_Q}}{\order{\eta}}}^{d-1} \,.
   \end{gathered}
   \label[equations]{eq:orders}
\end{equation}
Here, we have defined four parameters~%
\footnote{
  Note that one combination of parameters is not allowed in this
  derivation: $m < -1$ and $q=0$.
  Owing to the thermodynamic relations, $m < -1$ implies that $T_1$ will
  depend on density and pressure of the form $n_{1+\abs{m}}$ and
  $P_{1+\abs{m}}$.
  We are able to manipulate the results for $m=-1$ (\cf
  \cref{sec:MMS_first_order_adi}) to handle these $n_2$ and $P_2$ terms.
  However, for $m < -1$, these terms cannot be eliminated.
  If $q > 0$, then $\mu_1$ and $T_1$ do not appear in our first-order
  corrections, so this is acceptable; if $q = 0$, we would have these
  $n_{1+\abs{m}}$ and $P_{1+\abs{m}}$ terms which cannot be eliminated.
}:
$d$ the spatial dimension, $m \in \mathbb{Z} \setminus \{0\}$, $p \in
\mathbb{N} \ge 0$, and $q \in \mathbb{N} \ge 0$.
The parameter $m$ is defined as
\begin{equation}
  \epsilon^m \coloneqq \order{\frac{\mu}{k_B T}}^2 \,,
\end{equation}
and represents the ``Dirac'' or ``Fermi'' quality of the system: $m>0$
corresponds to increasingly strong ``Dirac''-character while $m<0$ is
more ``Fermi''-like.
The parameter $p$ measures the importance of the shear terms $\eta$: if
$p=0$, the shear terms enter our first-order correction equations while,
for $p>0$, it enters at the $(p+1)$-order correction equations and thus
are not considered in our analysis.
Likewise, the parameter $q$ measures the importance of the conductive
terms $\sigma_Q$: if $q=0$, the conductive terms enter our first-order
correction equations, but they are higher order for $q>0$.

The KdV-Burgers coefficients specified in
\cref{sec:kdv_burgers,sec:kdv_burgers_adi} and throughout the paper
assume $p=q=0$.
When using other choices of $p$ and $q$, it is important to replace
$\eta \to \eta \Kronecker_{p,0}$, $\zeta \to \zeta \Kronecker_{p,0}
\order{\zeta}/\order{\eta}$, and $\sigma_Q \to \sigma \Kronecker_{q,0}$.
This ensures that only the relevant dissipative coefficients appear.

Note that we have specified $\order{u}=1$ to allow for large background
flows $\order{u_0} = 1$.
Nevertheless, these results still apply if $u_0 = 0$ (no background
flow), in which case $u \sim \epsilon u_1$ and $\order{u} = \epsilon$.
Additionally, these nondimensionalizations assume that $u<1$ is small
enough that $\gamma = 1/\sqrt{1-u^2}$ is order $\order{\gamma} = 1$.
Finally, note that we have assumed $\order{\eta} \ge \order{\zeta}$.

\subsection{\label{sec:param_choice} Parameter Choice}
For concreteness, the main paper utilizes a Dirac regime
nondimensionalization of $m=1$ and $p=q=0$ with
$\order{\eta} = 1$ and $\order{\sigma_Q} = \epsilon^{1/2}$.

We also highlight additional terms in the multiple scales expansion
arising from the Fermi regime.
These come about from a nondimensionalization with $m=-1$ and $p=q=0$
with $\order{\eta} = \order{\sigma_Q} = 1$.

The alternate derivation for small $\epsilon$ mentioned in
\cref{sec:experimental} would correspond to $m=p=1$ and $q=0$ with
$\order{\eta} = \order{\sigma_Q} = 1$.

It is worth highlighting that different choices of $\order{\sigma_Q}$
and $\order{\eta}$ do not affect the calculated results or observables
(\cf \cref{sec:diss_order}).
Likewise, the parameters $m$, $p$, and $q$ have minimal, straightforward
effects on the results: $p$ determines whether $\eta$ and $\zeta$ terms
appear in $\mathcal{G}'$; $q$ determines if $\sigma_Q$ appears in
$\mathcal{G}'$; and $m$ determines the form of $P_1$, and thus
$\mathcal{F}'$~%
\footnote{
  Furthermore, $m<-1$ precludes the choice of $q=0$; see
  \footnotename{} \cite{Note19}
}.
Otherwise, the results are independent of the choice of $m$, $p$, and
$q$.
To wit, these choices do not even affect the $\epsilon$-order of
observable quantities; see \cref{sec:analysis}.

Using the definition of $A$, it is easy to check that $\kappa \ge 1$
satisfies $\order{\kappa} = \epsilon^{-1/2 -q}
\order{\sigma_Q} \order{\alpha} \ge 1$; this provides a constraint on
the allowed parameters.
For $\epsilon=\epsilonVal$, $q=0$, $\order{\sigma_Q}=\epsilon^{1/2}$, and
$\order{\alpha}=1$ used throughout the main text, we find
$\order{\kappa} = 1$, consistent with our choice of $\kappa = \kappaDim$.

\subsection{Entropy Divergence}
In \cref{sec:entropy}, we found that the entropy divergence only
depended on the $\eta$ and $\zeta$ terms, to this order.
Using our expressions for the generalized nondimensionalization, we can
investigate what occurs for different parameter regimes.

Recall that \cref{eq:entropy_div} showed that
\begin{equation}
  \begin{aligned}
    & \order{\partial_{\nu} s^{\nu}} = \frac{1}{\order{T}}
      \order{\partial_x u}^2 \bqty{\order{\eta} +
      \order{\zeta}} \\
    & \qquad + \frac{\order{\sigma_Q}}{\order{T}} \bqty{
      \order{\partial_x \mu} + \order{F^{x \rho} u_{\rho}} }^2 \,.
  \end{aligned}
\end{equation}
Restricting our attention, as usual, to $u < 1$ such that $\gamma =
1/\sqrt{1-u^2} \approx 1$, we see that $\order{F_{i \rho} u^{\rho}} =
\order{E_x} = \order{\partial_x \phi} = \order{\partial_x A n}$.
Thus, using the results from \cref{sec:nondim}, we have
\begin{equation}
  \begin{aligned}
    \order{\partial_{\nu} s^{\nu}} &= \frac{\order{\eta}
      \order{\partial_x}^2 \epsilon^{2-p}}{\order{T}} \Bigl[\epsilon^p +
      \frac{\order{\zeta}}{\order{\eta}} \epsilon^p \\
    & \qquad + \epsilon^{q} + \epsilon^{q + \frac{1}{2} m + \frac{1}{2}
      \abs{m}} + \epsilon^{q + m + \abs{m}}\Bigr] \,.
  \end{aligned}
  \label{eq:entropy_div_order}
\end{equation}
Here, the terms in the square brackets represent the $\eta$, $\zeta$,
$\sigma_Q E_x^2 $, $\sigma_Q E_x \partial_x \mu$, and
$\sigma_Q (\partial_x \mu)^2$ terms respectively.
Hence, we recognize that increasing $p$ causes the $\eta$ and $\zeta$
terms to be less relevant, while increasing $q$ does the same to the
$\sigma_Q$ terms.
Furthermore, the leading factor of $\mu/T$ for the $\sigma_Q$ terms in
\cref{eq:entropy_div} causes these terms to be higher order when $m>0$
(\ie when $\mu/T$ is small), as expected.
Finally, note that \cref{eq:orders} were defined under the assumption
$\order{\eta} \ge \order{\zeta}$, so $\order{\zeta}/\order{\eta}$ in
\cref{eq:entropy_div_order} can be, at most, unity.

\subsection{\label{sec:diss_order} Order of Dissipative Coefficients}
Notice that we have left $\order{\sigma_Q}$ and
$\order{\eta}$ undetermined.
There is some subtlety in choosing these parameters.
This most obvious manner to proceed involves using existing theoretical
predictions~\citep{lucas2018hydrodynamics} for their magnitude~%
\footnote{
  Note that the expression for $\sigma_Q$ in the Fermi regime lacks
  numerical factors; see \citet{muller2008quantum} for the exact
  expression for the (screened) Fermi case.
}; for instance, in $d=2$,
\begin{equation}
  \eta \approx
  \begin{cases}
    \frac{0.45 T^2}{\alpha^2} & \text{Dirac} \,. \\
    \frac{3 \mu^2 \abs{n}}{64 \pi \alpha^2 \ln(\alpha^{-1})
      T^2} & \text{Fermi} \,,
  \end{cases}
\end{equation}
and
\begin{equation}
  \sigma_Q =
  \begin{cases}
    \frac{0.12}{\alpha^2} & \text{Dirac} \,, \\
    \hfil 1 & \text{Fermi} \,,
  \end{cases}
\end{equation}
with $\alpha \approx 4/\ln(\SI{e4}{\kelvin}/T)$.
Ignoring logarithmic corrections, these will then generate compatibility conditions on the parameters $m$, $p$, and $q$.
Nevertheless, such a choice is only valid in the infinitesimal
$\epsilon$ limit: we must assume $\epsilon$ is small enough that all the
numerical prefactors---like $3/64\pi \approx 0.015$ for $\eta$ in the
Dirac regime---are considered order-1 (\ie $\order{\epsilon^0}$).
If $\epsilon$ is large enough that, for instance, $3/64 \pi \approx
\epsilon$, then this assumption breaks down.

Alternatively, one could instead calculate the numerical values for
$\sigma_Q$ and $\eta$ from the existing theories.
For instance, in \cref{sec:normalization}, we calculated $\sigma_Q
= \sigmaNondim$ for our choices of parameters.
This value can then be compared to the expected value of $\epsilon$ to
determine the correct scaling.
Continuing our example, assuming $\epsilon \approx \epsilonVal$, we found
$\sigma_Q \approx \epsilon^{1/2}$.
While this method is somewhat more \adhoc than the previously described
one, it has the benefit that it is now valid in a neighborhood of the
desired $\epsilon$ rather than for solely infinitesimal $\epsilon$.
This is the method used in the main text since we are considering
$\epsilon$ small but finite.

\subsection{\label{sec:dom_bal} Derivation: Dominant Balance}
Now, we will derive the results given at the beginning of
\cref{sec:nondim}.
These results follow from the application of dominant balance.

First, we define a small nondimensional parameter $\epsilon \ll 1$ as
our expansion parameter: that is, all terms will be expanded in integer
powers of $\epsilon$ as $y = y_0 + \epsilon y_1 + \ldots$.
Further, we will assume that all leading-order quantities are uniform in
space and constant in time (\ie $y(x,t) = y_0 + \epsilon y_1(x,t) +
\ldots$).
This implies that derivatives will always generate one extra factor of
$\epsilon$: $\order{\partial_{\mu} y(x,t)} = \order{\partial_{\mu}
\epsilon y_1(x,t)} = \epsilon \order{\partial_{\mu}} \order{y}$.

Next, we introduce the parameter $m \in \mathbb{Z} \setminus \{0\}$ as
\begin{equation}
  \epsilon^m \coloneqq \order{\frac{\mu}{T}}^2 \,.
\end{equation}
We require that $m$ be an integer since it enters in an asymptotic
expansion of the equation of state $P(\mu,T)$; since our main equations
are expanded in integer powers of $\epsilon$, we must also have this
asymptotic expansion in integer powers of $\epsilon$.
Also, notice we used the square of $\mu/T$; it is easily seen that the
asymptotic expansion of $P(\mu, T)$ only involves even powers of $\mu/T$
since it is an even function of $\mu/T$~%
\footnote{
  Equivalently, \citet{lucas2016transport} prove $P(\mu,T)$ only
  involves even powers by recognizing that the equation of state is
  charge conjugation invariant.
}.
Thus, we see that the Dirac regime follows when $m > 0$ and the Fermi
case corresponds to $m<0$; the $m=0$ case is excluded because then the
thermodynamic equation of state (\cf \cref{eq:eqn_state}) cannot be
expanded in a series/asymptotic expansion.

With this definition, we are able to collapse the two different
nondimensionalizations of the pressure.
From the thermodynamic equation of state \cref{eq:eqn_state}, we see
that $\order{P} = \order{T}^{d+1}$ for Dirac and $\order{P} =
\order{\mu}^{d+1}$ for Fermi.
Therefore, we have $\order{P} = \epsilon^{(d+1)m/4+(-d-1)\absm/4}
\order{T}^{d+1}$ in general.
Likewise, the charge density can be nondimensionalized as $\order{n} =
\epsilon^{\absm/2}\order{P}/\order{T} =
\epsilon^{(d+1)m/4+(-d+1)\absm/4}\order{T}^d$.

Now, we begin using dominant balance to impose restrictions based on our
desire that certain terms appear at certain orders.
Here, we must use some foresight about which terms the equations will
contain.
To ensure that we have wavelike solutions, we want the terms appearing
in the leading order equations to match those in \cref{eq:MMS_1}.
Since we want the dispersive electromagnetic terms $d_1 d_2
\partial^3_x n$ to appear at as first-order corrections, this means the
nondispersive electromagnetic term $\partial_x n$ must appear at leading
order.
Thus, the two electromagnetic terms must differ by one factor of
$\epsilon$: this imposes $\order{d_i} =
\epsilon^{1/2}/\order{\partial_x}$; this is our first assumption.
Requiring the nondispersive electromagnetic term to enter at leading
order enforces $\order{\partial_x P} = \order{A n \partial_x n}$
yielding our second assumption: $\order{A} =
\epsilon^{(-d-1)m/4+(d-3)\absm/4}\order{T}^{-d+1}$.

Next, we wish the leading order equations to be satisfied even if $u_0 =
0$.
Setting $u_0 = 0$ and performing a dominant balance on the leading charge
conservation equation \cref{eq:MMS_charge_1} gives $\order{\partial_t} =
\order{u} \order{\partial_x}$, our third requirement.
Another dominant balance on the leading momentum conservation equation
\cref{eq:MMS_mom_1} yields $\order{u} = 1$, our fourth and final
requirement.

Moving onto the shear- and bulk-viscosity terms, we introduce a second
parameter $p \in \mathbb{N} \ge 0$.
This parameter is defined such that $p=0$ ensures that the shear/bulk
viscosities appear in our first-order correction equations, $p=1$ would
push these terms to second-order corrections, and so on.
Since we are only concerned with first-order corrections, this means
shear/bulk viscosity is relevant for $p=0$ and irrelevant for $p>0$.
This is implemented by imposing $\order{\epsilon \partial_x P} =
\epsilon^{-p} \order{\eta \partial_x^2 n}$, yielding
$\order{\partial_x} = \epsilon^{p+1} \order{P}/\order{\eta}$.

Finally, we introduce one more parameter $q \in \mathbb{N} \ge 0$
controlling the order at which the intrinsic conductivity $\sigma_Q$
appears.
Similar to the parameter $p$, the parameter $q=0$ yields $\sigma_Q$
terms at first-order while $q>0$ corresponds to higher-order terms
(which will be neglected in this analysis).
It is easy to check that of the two $\sigma_Q$ terms, the
electromagnetic term $\order{F^{\nu \rho} u_{\rho}} = \order{A
\partial_x n}$ is always larger than the thermoelectric term $\order{T
\partial_x(\mu/T)} \le \order{A \partial_x n}$.
Thus, we introduce the parameter $q$ as $\order{\epsilon \partial_t n} =
\epsilon^{-q} \order{\sigma_Q \partial^2_x A n}$.
This implies that $\order{T} = \epsilon^{q/2 -p/2+\absm/2}
\sqrt{\order{\eta}/\order{\sigma_Q}}$.
Using these various relations reproduces the results given at the
beginning of \cref{sec:nondim}.

\section{\label{sec:adiabatic} Adiabatic System}
Here, we can utilize the same nondimensionalization laid out in
\cref{sec:nondim} for the isothermal system.
This follows because the derivation in \cref{sec:dom_bal} required that
the leading order equations still be satisfied when $u_0 = 0$.
However, it is easy to show that, when $u_0=0$, the leading order energy
conservation equation \cref{eq:MMS_energy_1_adi} is equivalent to the
leading order charge conservation equation \cref{eq:MMS_charge_1_adi}
combined with the isothermal relation between $P$ and $n$.
Thus, the leading order, $u_0=0$ adiabatic system is equivalent to the
leading order, $u_0=0$ isothermal system, and the previous
nondimensionalization carries over.

Here, we will redo the multiple scales derivation using the adiabatic
assumption.
Therefore, we will now include the energy conservation equation
\cref{eq:energy_cons} and allow $T$ to vary dynamically.
As we did in \cref{sec:multiple_scales}, we expand all of the dynamic
variables (including $T$) in a perturbation expansion.

\subsection{Perturbative Thermodynamics}
We will be using the thermodynamic relationships of \cref{sec:thermo} to
write $\mu$ and $T$ in terms of $n$ and $P$.
Expanding the thermodynamic variables and collecting powers of
$\epsilon$ yields the following relations for the Dirac regime:

{
\renewcommand{\theequation}{Dirac: \Alph{section}\arabic{equation}}
\begin{gather}
  P_0 = T_0^{d+1} \mathcal{C}_0 \,, \\
  n_0 = 2 T_0^{d-1} \mu_0 \mathcal{C}_1 \,, \label{eq:dirac_n0_adi} \\
  P_1 = P_0 \bqty{\frac{T_1}{T_0} (d+1) +
    \frac{\mathcal{C}_1}{\mathcal{C}_0} \pqty{\frac{\mu_0}{T_0}}^2
    \Kronecker_{m,1}} \,, \displaybreak[0] \\
  n_1 = n_0 \bqty{\frac{\mu_1}{\mu_0} + \frac{T_1}{T_0} (d-1) + 2
    \frac{\mathcal{C}_2}{\mathcal{C}_1} \pqty{\frac{\mu_0}{T_0}}^2
    \Kronecker_{m,1}} \,, \displaybreak[0] \\\
  \begin{aligned}
    &P_2 = P_0 \Biggl[\frac{T_2}{T_0} (d+1) + \frac{T_1^2}{T_0^2}
      \frac{(d+1) d}{2} \\
    &\qquad + \frac{\mathcal{C}_1}{\mathcal{C}_0} \pqty{2
      \frac{\mu_1}{\mu_0} + (d-1) \frac{T_1}{T_0}}
      \pqty{\frac{\mu_0}{T_0}}^2 \Kronecker_{m,1} \\
    &\qquad +
      \frac{\mathcal{C}_2}{\mathcal{C}_0} \pqty{\frac{\mu_0}{T_0}}^4
      \Kronecker_{m,1} + \frac{\mathcal{C}_1}{\mathcal{C}_0}
      \pqty{\frac{\mu_0}{T_0}}^2 \Kronecker_{m,2} \Biggr] \,,
  \end{aligned} \displaybreak[0] \\
  \begin{aligned}
    &n_2 = n_0 \Biggl[\frac{\mu_2}{\mu_0} + \frac{T_2}{T_0} (d-1) +
    \frac{T_1^2}{T_0^2} \frac{(d-1) (d-2)}{2} \\
   &\qquad + \frac{\mu_1}{\mu_0} \frac{T_1}{T_0} (d-1) \\
   &\qquad + 2 \frac{\mathcal{C}_2}{\mathcal{C}_1} \pqty{3
    \frac{\mu_1}{\mu_0} + (d-3) \frac{T_1}{T_0}}
    \pqty{\frac{\mu_0}{T_0}}^2 \Kronecker_{m,1} \\
   &\qquad + 3
    \frac{\mathcal{C}_3}{\mathcal{C}_1} \pqty{\frac{\mu_0}{T_0}}^4
    \Kronecker_{m,1} + 2 \frac{\mathcal{C}_2}{\mathcal{C}_1}
    \pqty{\frac{\mu_0}{T_0}}^2 \Kronecker_{m,2} \Biggr]
  \end{aligned}
\end{gather}
}

{
\renewcommand{\theequation}{Fermi: \Alph{section}\arabic{equation}}
Similarly, for the Fermi regime, we find
\begin{gather}
  P_0 = \abs{\mu_0}^{d+1} \mathcal{C}_0 \,, \\
  n_0 = \abs{\mu_0}^d \sgn(\mu_0) \mathcal{C}_0 (d+1) \,, \\
  P_1 = P_0 \bqty{\frac{\mu_1}{\mu_0} (d+1) +
    \frac{\mathcal{C}_1}{\mathcal{C}_0} \pqty{\frac{T_0}{\mu_0}}^2
    \Kronecker_{m,-1}} \,, \displaybreak[0] \\
  n_1 = n_0 \bqty{\frac{\mu_1}{\mu_0} d +
    \frac{\mathcal{C}_1}{\mathcal{C}_0} \frac{d-1}{d+1}
    \pqty{\frac{T_0}{\mu_0}}^2 \Kronecker_{m,-1}} \,, \displaybreak[0] \\
  \begin{aligned}
    &P_2 = P_0 \Biggl[\frac{\mu_2}{\mu_0} (d+1) + \frac{\mu_1^2}{\mu_0^2}
      \frac{(d+1) d}{2} \\
    &\qquad + \frac{\mathcal{C}_1}{\mathcal{C}_0} \pqty{2
      \frac{T_1}{T_0} + (d-1) \frac{\mu_1}{\mu_0}}
      \pqty{\frac{T_0}{\mu_0}}^2 \Kronecker_{m,-1} \\
    &\qquad + \frac{\mathcal{C}_2}{\mathcal{C}_0} \pqty{\frac{T_0}{\mu_0}}^4
      \Kronecker_{m,-1} + \frac{\mathcal{C}_1}{\mathcal{C}_0}
      \pqty{\frac{T_0}{\mu_0}}^2 \Kronecker_{m,-2} \Biggr] \,,
  \end{aligned} \displaybreak[0] \\
  \begin{aligned}
    &n_2 = n_0 \Biggl[\frac{\mu_2}{\mu_0} d + \frac{\mu_1^2}{\mu_0^2}
      \frac{d (d-1)}{2} \\
    &\qquad + \frac{\mathcal{C}_1}{\mathcal{C}_0}
      \frac{d-1}{d+1} \pqty{2 \frac{T_1}{T_0} + (d-2)
      \frac{\mu_1}{\mu_0}} \pqty{\frac{T_0}{\mu_0}}^2 \Kronecker_{m,-1} \\
    &\qquad +
      \frac{\mathcal{C}_2}{\mathcal{C}_0} \frac{d-3}{d+1}
      \pqty{\frac{T_0}{\mu_0}}^4 \Kronecker_{m,-1} +
      \frac{\mathcal{C}_1}{\mathcal{C}_0} \frac{d-1}{d+1}
      \pqty{\frac{T_0}{\mu_0}}^2 \Kronecker_{m,-2} \Biggr]
  \end{aligned}
\end{gather}

In the Dirac regime, we can invert these relations to write $\mu$ and
$T$ in terms of $P$ and $n$, treating these as the independent
variables at each order.
However, in the Fermi regime, this perturbation expansion introduces a
peculiarity.
The $P_0$ and $n_0$ equations do not contain $T_0$; therefore, rather
than giving the value of $T_0$, these equations provide a constraint on
$P_0$ and $n_0$:
\begin{equation}
  P_0 = \frac{\abs{n_0}^{(d+1)/d}}{\abs{\mathcal{C}_0}^{1/d}
    (d+1)^{(d+1)/d}} \sgn{C_0} \,.
\end{equation}
}
Similarly, the $P_1(x,t)$ and $n_1(x,t)$ equations only depend on a single
dynamical variable $\mu_1(x,t)$ (but not $T_1(x,t)$);
therefore, these also give a restriction on $P_1$ and $n_1$ to ensure
that $T_0(x,t) = T_0$ is independent of $x$ and $t$:
\begin{equation}
  \frac{P_1}{P_0} = \frac{n_1}{n_0}\frac{d+1}{d} +
    \frac{\mathcal{C}_1}{\mathcal{C}_0} \frac{1}{d}
    \pqty{\frac{T_0}{\mu_0}}^2 \Kronecker_{m,-1} \,.
\end{equation}
This requirement will be utilized later.

\subsection{Conservation Equations}
If we again restrict to 1D motion and collect terms by powers of
$\epsilon$ we get the following equations:
\begin{widetext}
Leading Order:
\begin{subequations}
\begin{align}
  \pdv{n_1}{t_0} + \gamma^2 n_0 u_0 \pdv{u_1}{t_0} + u_0
    \pdv{n_1}{x} + n_0 \gamma^2 \pdv{u_1}{x} &= 0 \,,
    \label{eq:MMS_charge_1_adi} \\
  \gamma^2 \pdv{\energy_1}{t_0} + \gamma^2 u_0^2 \pdv{P_1}{t_0} + 2 u_0
    (\energy_0 + P_0) \gamma^4 \pdv{u_1}{t_0} + (1 + u_0^2) (\energy_0 +
    P_0) \gamma^4 \pdv{u_1}{x} + u_0 \gamma^2 \pdv{x}(\energy_1 +
    P_1) & \nonumber \\
  + A n_0 u_0 \gamma^2 \pdv{n_1}{x} + A n_0^2 u_0^2 \gamma^4
    \pdv{u_1}{x} &= 0 \,, \label{eq:MMS_energy_1_adi} \\
  \gamma^3 (\energy_0 + P_0) \pdv{u_1}{t_0} + \gamma u_0
    \pdv{P_1}{t_0} + u_0 \gamma^3 (\energy_0 + P_0) \pdv{u_1}{x} +
    \gamma \pdv{P_1}{x} + A n_0 \gamma \pdv{n_1}{x} + A n_0^2 u_0
    \gamma^3 \pdv{u_1}{x} &= 0 \,, \label{eq:MMS_mom_1_adi}
\end{align}
\end{subequations}
\label{eq:MMS_1_adi}
First-Order Correction:
\begin{subequations}
\begin{align}
  \pdv{n_2}{t_0} + \gamma^2 n_0 u_0 \pdv{u_2}{t_0} + u_0
    \pdv{n_2}{x} + n_0 \gamma^2 \pdv{u_2}{x} &= \text{RHS} \,,
    \label{eq:MMS_charge_2_adi} \\
  \gamma^2 \pdv{\energy_2}{t_0} + \gamma^2 u_0^2 \pdv{P_2}{t_0} + 2 u_0
    (\energy_0 + P_0) \gamma^4 \pdv{u_2}{t_0} + (1 + u_0^2) (\energy_0 +
    P_0) \gamma^4 \pdv{u_2}{x} + u_0 \gamma^2 \pdv{x}(\energy_2 +
    P_2) + A n_0 u_0 \gamma \pdv{n_2}{x} &= \text{RHS} \,,
    \label{eq:MMS_energy_2_adi} \\
  \gamma^3 (\energy_0 + P_0) \pdv{u_2}{t_0} + \gamma u_0
    \pdv{P_2}{t_0} + u_0 \gamma^3 (\energy_0 + P_0) \pdv{u_2}{x} +
    \gamma \pdv{P_2}{x} + A n_0 \pdv{n_2}{x} &= \text{RHS} \,.
    \label{eq:MMS_mom_2_adi}
\end{align}
\end{subequations}
\end{widetext}
Again, we have used the electrostatic coupling $A$ according to
\cref{eq:electric_potential}.
See \cref{sec:full_equations} for the terms on the right-hand side.

\subsection{Leading Order Equations}
Using $\energy = P d$ and combining equations like \pagebreak[0]
\begin{align*}
  & \pqty{\pdv{t_0}+u_0\pdv{x}} \Biggl\{ A d n_0 \gamma^2 \pdv{x}
    \mbox{$\bigl[$\cref{eq:MMS_charge_1_adi}$\bigr]$} \\
  &\qquad + \pqty{u_0 \pdv{t_0} + \pdv{x} }
    \mbox{$\bigl[$\cref{eq:MMS_energy_1_adi}$\bigr]$} \\
  &\qquad - \gamma \pqty{(d+u_0^2) \pdv{t_0} + u_0 (d+1) \pdv{x}}
  \mbox{$\bigl[$\cref{eq:MMS_mom_1_adi}$\bigr]$} \Biggr\}
\end{align*}
gives
\begin{equation}
  \begin{aligned}
    0 &= \gamma^2 \pqty{\pdv{t_0} + u_0 \pdv{x}} \Biggl\{\gamma^2 (d+1) P_0
    (u_0^2 -d) \pdv[2]{u_1}{t_0} \\
    &\quad -2 \gamma^2 (d+1) P_0 u_0 (d-1) \pdv{u_1}{t_0}{x} \\
    &\quad + \bqty{ A d n_0^2 + \gamma^2 (d+1) P_0 (1-d u_0^2) }
    \pdv[2]{u_1}{x} \Biggr\} \,.
  \end{aligned}
\end{equation}
This wave equation has solutions $f(x+v_0 t_0) + g(x-v_0 t_0)$ with
$v_0$ given by
\begin{equation}
  v_0^{(\pm)} = -\frac{u_0 (d-1)}{d-u_0^2} \pm
    \frac{\sqrt{d}}{(d-u_0^2) \gamma^2}
    \sqrt{1 + \frac{A n_0^2 \gamma^2 (d-u_0^2)}{P_0 (d+1)}} \,.
\end{equation}
We will take the $(+)$ sign so that $v_0 = v_0^{(+)}$; the other can be
recovered by taking $u_0 \to -u_0$ and $v_0 \to -v_0$.
Further, we restrict to unidirectional solutions $u_1(x,t_0,t_1) = f(x
\pm v_0 t_0, t_1)$ for a definite choice of $\pm$; here, we choose $(+)$
as well---the other propagation direction can be recovered by taking
$v_0 \to -v_0$.

For stationary perturbations ($v_0 = 0$), we can solve for $u_0$:
\begin{equation}
  u_0 = \pm \sqrt{\frac{(1/d)+[A n_0^2/P_0
    (d+1)]}{1+[A n_0^2/P_0 (d+1)]}} \,.
\end{equation}
For reference, the velocity of propagation in the absence of a
background flow ($u_0 = 0$) is
\begin{equation}
  v_0 = \pm \frac{1}{\sqrt{d}} \sqrt{1 + \frac{A d n_0^2}{(d+1) P_0}} \,.
\end{equation}

In general, $n_1$, $u_1$, and $P_1$ have traveling wave solutions;
neglecting solutions of the form $f(x-u_0 t_0,t_1)$ that are simply
advected by the background current, we find solutions given by
\begin{subequations}
\begin{gather}
  n_1(x,t_0,t_1) = n_1(x + v_0 t_0, t_1) + F_1(t_1) \,, \\
  \begin{aligned}
    u_1(x,t_0,t_1) &= -\frac{(u_0 + v_0)}{n_0 \gamma^2 (1+u_0 v_0)} n_1(x
      + v_0 t_0, t_1) \\
    &\qquad + F_2(t_1) \label{eq:u_from_n_adi} \,,
  \end{aligned} \\
  \frac{P_1(x,t_0,t_1)}{P_0} = \frac{d+1}{d} \frac{n_1(x+ v_0
    t_0,t_1)}{n_0} + F_3(t_1) \,.
\end{gather}
\end{subequations}
Here, we have arbitrary functions $F_1(t_1)$, $F_2(t_2)$, and
$F_3(t_2)$; by imposing boundary conditions $n_1 = 0$ at $x = \pm
\infty$, we set $F_1 = 0$.
We will allow $U_1(t_2) \coloneqq F_2(t_2)$ to remain arbitrary; this
uniform background current can be superimposed on the soliton solution
as in \cref{sec:stationary} if desired~%
\footnote{
  Note that it is possible to generate a stationary soliton by
  appropriate choice of $F_1$ or $F_3$ instead, though the resulting
  coefficients will be different.
}.
In the Dirac regime, we can impose $P_1 = 0$ at $x=\pm \infty$ to set
$F_3=0$; however, for the Fermi regime, requiring that $T_0(x,t) = T_0$
independent of $(x,t)$ restricts the relationship between $P_1$ and
$n_1$.
Hence, we will write $F_3$ as
\begin{equation}
  F_3(t_1) = \Kronecker_{m,-1} \frac{1}{d}
    \frac{\mathcal{C}_1}{\mathcal{C}_0} \pqty{\frac{T_0}{\mu_0}}^2 \,.
\end{equation}

\subsection{\label{sec:MMS_first_order_adi} First-Order Corrections}
Now considering the first-order corrections, preventing
secular growth of the higher-order terms (\ie $n_2$, $u_2$, \etc)
requires imposing a compatibility condition on the lower-order terms
(\ie $n_1$, $u_1$, \etc).
We can manipulate the system as
\begin{align*}
  & \pqty{\pdv{t_0}+u_0\pdv{x}}\Biggl\{A d n_0 \gamma^2 \pdv{x}
    \mbox{$\bigl[$\cref{eq:MMS_charge_2_adi}$\bigr]$} \\
  &\qquad + \pqty{u_0 \pdv{t_0} + \pdv{x} }
    \mbox{$\bigl[$\cref{eq:MMS_energy_2_adi}$\bigr]$} \\
  &\qquad - \gamma \bqty{(d+u_0^2) \pdv{t_0} + u_0 (d+1) \pdv{x}}
    \mbox{$\bigl[$\cref{eq:MMS_mom_2_adi}$\bigr]$} \Biggr\}
    \displaybreak[0] \\
  & +\Kronecker_{m,-1} \gamma \frac{1}{2} \frac{\mathcal{C}_0}{\mathcal{C}_1}
    \frac{\sigma_Q (d+1)}{n_0} \frac{\mu_0^3}{T_0^2}
    \pqty{\pdv{x}+u_0\pdv{t_0}}^2 \Biggl( \\
  &\qquad -\pqty{\pdv{x}+u_0\pdv{t_0}}
    \mbox{$\bigl[$\cref{eq:MMS_energy_2_adi}$\bigr]$} \\
  &\qquad + \gamma\bqty{(d+u_0^2)\pdv{t_0} + u_0(d+1)\pdv{x}}
    \mbox{$\bigl[$\cref{eq:MMS_mom_2_adi}$\bigr]$} \\
  &\qquad + \frac{Adn_0^2}{P_0(d+1)} \pdv{x} \bigl\{\gamma u_0
    \mbox{$\bigl[$\cref{eq:MMS_mom_2_adi}$\bigr]$} -
    \mbox{$\bigl[$\cref{eq:MMS_energy_2_adi}$\bigr]$}\bigr\}
    \Biggr)
\end{align*}
to obtain
\begin{equation}
  \begin{aligned}
    &\gamma^4 P_0 (d+1) (d-u_0^2) \Biggl[\pqty{\pdv{t_0} + u_0 \pdv{x}}
      \\
    &\qquad - \Kronecker_{m,1} \gamma \frac{1}{2}
      \frac{\mathcal{C}_0}{\mathcal{C}_1} \frac{\sigma_Q (d+1)}{n_0}
      \frac{\mu_0^3}{T_0^2}\pqty{\pdv{x}+u_0\pdv{t_0}}^2 \Biggr] \\
    &\qquad \times \pqty{v_0^{(+)} \pdv{x} - \pdv{t_0}} \pqty{v_0^{(-)}
      \pdv{x} - \pdv{t_0}} u_2 \\
    & = \text{LOT} \,,
  \end{aligned}
  \label{eq:HOT_inhom_adi}
\end{equation}
where LOT represents lower-order terms (\ie $n_1$, $u_1$, \etc).

It is instructive here to change variables to
$\chi^{(\pm)}_0=x+v_0^{(\pm)} t_0$.
Then, the equation becomes
\begin{equation}
  \begin{aligned}
    &\gamma^4 P_0 (d+1) \pqty{d-u_0^2} \pqty{v_0^{(+)}-v_0^{(-)}}^2 \\
    &\qquad \times \Biggl\{
     \bqty{\sum_{\pm} \pqty{u_0 + v_0^{(\pm)}}
      \pdv{\chi_0^{(\pm)}}} \\
    &\qquad - \Kronecker_{m,-1} \gamma \frac{3\sigma_Q}{\pi^2 n_0 d}
      \frac{\mu_0^3}{T_0^2}\bqty{\sum_{\pm} \pqty{1+u_0 v_0^{(\pm)}}
      \pdv{\chi_0^{(\pm)}}}^2 \Biggr\} \\
    &\qquad \times \pdv{\chi^{(-)}_0} \pdv{\chi^{(+)}_0} u_2 \\
    & = \text{LOT}
  \end{aligned}
\end{equation}
This is where we encounter an apparent problem.
Upon inserting our solutions for the lower-order terms, we find the
right-hand side depends on products and derivatives of
$f\pqty{\chi^{(+)}_0}$.
This implies that the LOT is solely a function of $\chi^{(+)}_0$.

However, we see that functions of the form $f(\chi^{(+)})$ are also
solutions to the homogeneous equation in \cref{eq:HOT_inhom_adi} due to
the presence of the $\partial_{\chi_0^{(-)}}$ operator.

So, products and derivatives of $f(\chi^{(+)}_0)$ appear as inhomogeneous
forcing terms that give rise to secular terms.
For instance, terms proportional to $f^{(4)}\pqty{\chi^{(+)}_0}$ give
rise to solutions of the form $\chi^{(-)}_0 f^{(3)}\pqty{\chi^{(+)}_0}$.
This grows unbounded in $\chi^{(-)}_0$---and hence, in time $t$.
This will eventually cause $\abs{u_2} > \abs{u_1}$, invalidating the
perturbation expansion.
Thus, unless LOT vanishes identically, it will give rise to
$\chi^{(\pm)}_0$-secular terms in $u_2$---\ie solutions growing
unbounded in $t_0$ or $x$.

Hence, we require the right-hand side to vanish and we are left with the
desired compatibility equation:
\begin{align}
  0 &= (u_0 + v_0) \pdv[2]{{\chi_0^{(+)}}} (\text{KdVB}[n_1]) \nonumber \\
  &\qquad- \Kronecker_{m,-1} \gamma \frac{1}{2}
    \frac{\mathcal{C}_0}{\mathcal{C}_1} \frac{\sigma_Q (d+1)(1+u_0
    v_0)^2}{n_0} \frac{\mu_0^3}{T_0^2} \nonumber \\
  &\qquad \times \pdv[3]{{\chi_0^{(+)}}}
    \eval{(\text{KdVB}[n_1])}_{\sigma_Q=0} \,.
  \label{eq:MMS_compat_adi}
\end{align}
Here, (KdVB$[n_1]$) represents the Korteweg-de Vries-Burgers equation,
discussed earlier, acting on $n_1$:
\begin{equation}
  \begin{aligned}
    &\mathcal{A}' \pdv{n_1}{t_1} + \mathcal{F}' \pdv{n_1}{\chi^{(+)}_0}
      + \mathcal{B}' n_1 \pdv{n_1}{\chi^{(+)}_0} \\
    &\qquad + \mathcal{C}' \pdv[3]{n_1}{{\chi^{(+)}_0}} -
      \mathcal{G}' \pdv[2]{n_1}{{\chi^{(+)}_0}} n_1  = 0 \,;
  \end{aligned}
  \label{eq:kdv_burgers_adi}
\end{equation}
see \cref{sec:kdv_burgers_adi} for the functional form of the coefficients.
Likewise, ($\eval{\text{KdVB}[n_1]}_{\sigma_Q=0}$) represents the
Korteweg-de Vries-Burgers equation without $\sigma_Q$ terms.

It is interesting to note the similarities and differences between the
adiabatic KdV-Burgers coefficients (\cref{sec:kdv_burgers_adi}) and the
isothermal coefficients (\cref{sec:kdv_burgers}).
For most of the coefficients ($\mathcal{A}'$, $\mathcal{B}'$, and
$\mathcal{C}'$), the adiabatic coefficients are identical to the
isothermal Fermi ($m=-1$) coefficients.
The $(1+u_0 v_0) \mathcal{C}_1 /\mathcal{C}_0$ term in $\mathcal{F}'$
differs slightly between the adiabatic Fermi case (coefficient
$(d+1)/d^2$) and isothermal Fermi case (coefficient $(d-1)/d^2$); the
adiabatic Dirac case is completely absent ($\Kronecker_{m,-1}$) compared to
the isothermal Dirac case.
Interestingly, the adiabatic $\eta$ and $\zeta$ terms in $\mathcal{G}'$
matches the isothermal Fermi terms, while the adiabatic $\sigma_Q$ term
matches the isothermal \emph{Dirac} one.

\subsection{Solving the Compatibility Equation}
In the Fermi regime ($m=-1$), the compatibility equation
\cref{eq:MMS_compat_adi} no longer has the simple, decaying soliton
solution derived in \cref{sec:dissipation}.
This can certainly be solved numerically.
Additionally, we can generate an approximate solution if we assume that
$\order{\sigma_Q} \ll 1$,
(but $\gg \epsilon$ to prevent them from falling to the next order in
our perturbation expansion) and use the same trick as we did in
\cref{sec:dissipation}.
Namely, we factor out a small parameter $\delta \sim
\order{\sigma_Q}$ from $\sigma_Q = \delta \tilde{\sigma}_Q$.
Then, $\order{\tilde{\sigma}_Q} = 1$, and we can expand in factors of
$\delta$.

Then, another short multiple scales expansion for $n_1$ can be done in
$\delta = \order{\mathcal{G}/\mathcal{A}}$.
To be consistent with our original perturbation series, we require that
$\epsilon \ll \delta \ll 1$.
As usual, we expand $n_1$ as $n_1 = n_1^{(0)} + \delta n_1^{(1)}$ and
$\partial_{t_1} = \partial_{\tau_0} + \delta \partial_{\tau_1}$.
Then, to leading order, we have
\begin{equation}
  (u_0 + v_0) \partial_{\chi_0^{(+)}}^2
    \eval{(\text{KdVB}[n_1^{(0)}])}_{\tilde{\sigma}_Q=0} = 0 \,.
\end{equation}
This is satisfied by the KdVB equation,
\begin{equation}
  \begin{aligned}
    \mathcal{L}_0 n_1^{(0)} &\coloneqq \mathcal{A}' \partial_{\tau_0}
      n_1^{(0)} + \mathcal{F}' \partial_{\chi_0^{(+)}}
      n_1^{(0)} + \frac{\mathcal{B}'}{2}
      \partial_{\chi_0^{(+)}} \pqty{ n_1^{(0)} }^2 \\
    &\quad + \mathcal{C}'
      \partial_{\chi_0^{(+)}}^3 n_1^{(0)}  -
      \eval{\mathcal{G}'}_{\tilde{\sigma}_Q=0} \partial_{\chi_0^{(+)}}^2
      n_1^{(0)} \\
    &= 0 \,.
  \end{aligned}
\end{equation}
Now, we further assume that $\eta$ and $\zeta$ are small; specifically,
we assume $\order{\eta} \ll 1, \epsilon \ll \order{\delta} \ll
\order{\zeta}$.
Then, the solution was found in \cref{sec:dissipation} upon replacing
$\mathcal{G}'$ with $\eval{\mathcal{G}'}_{\tilde{\sigma}_Q=0}$:
\begin{align}
  &n_1^{(0)}\pqty{\chi_0^{(+)},\tau_0} = c_1\pqty{\tau_0}
    \sgn(\mathcal{B'C'}) \sech^2
    \Biggl( \sqrt{\frac{c_1 \abs{\mathcal{B}'}}{12 \abs{\mathcal{C}'}}}
      \nonumber \\
  &\qquad \times \biggl[\chi_0^{(+)} - \pqty{\frac{c_1
    \abs{\mathcal{B}'}}{3 \abs{\mathcal{A}'}} \sgn(\mathcal{A'C'})  +
    \frac{\mathcal{F}'}{\mathcal{A}'}} \tau_0 \biggr] \Biggr) \,,
    \label{eq:soliton_adi}
\end{align}
where
\begin{equation}
  c_1(\tau_0) = \frac{c_1(0)}{1 + \tau_0/\tau^{(0)}_d}
  \label{eq:MMS_c1_adi}
\end{equation}
with
\begin{equation}
  \tau^{(0)}_d = \frac{45 \mathcal{A}' \abs{\mathcal{C}'}}{4 c_1(0)
    \abs{\mathcal{B}'} \eval{\mathcal{G}'}_{\tilde{\sigma}_Q=0}} \,.
  \label{eq:MMS_tau_d_adi}
\end{equation}
As mentioned above, we have assumed $\order{\delta} \ll
\order{\eval{\mathcal{G}'/\mathcal{A}'}_{\tilde{\sigma}_Q=0}} \ll
1$, so $1/\tau^{(0)}_d \ll 1$.

At the next order in $\delta$, we must allow the constant $c_1(0)$ to
become time-dependent on a slow time-scale $c_1(0) = c_1(0,\tau_1)$.
Now, our equation is
\begin{equation}
  \begin{aligned}
    &(u_0+v_0)\partial_{\chi_0^{(+)}}^2 \Bigl( \mathcal{A}'
      \partial_{\tau_0}
      n_1^{(1)} + \mathcal{F}' \partial_{\chi_0^{(+)}}
      n_1^{(1)} \\
    &\qquad +\mathcal{B}' \partial_{\chi_0^{(+)}} \pqty{n_1^{(0)}
      n_1^{(1)} }+ \mathcal{C}' \partial_{\chi_0^{(+)}}^3 n_1^{(1)} \Bigr)
      \\
    &\quad= (u_0+v_0)\partial_{\chi_0^{(+)}}^2 \pqty{-\partial_{t_1}
      \mathcal{A}' n_1^{(0)}
      + \partial_{\chi_0^{(+)}}^4 \mathcal{G}' n_1^{(0)}} \\
    &\qquad - \Kronecker_{m,-1} \gamma \frac{1}{2}
      \frac{\mathcal{C}_0}{\mathcal{C}_1} \frac{\sigma_Q (d+1)(1+u_0
      v_0)^2}{n_0} \frac{\mu_0^3}{T_0^2} \\
    &\qquad \times \partial{\chi_0^{(+)}}^3
      \eval{(\text{KdVB}[n_1^{(0)}])}_{\tilde{\sigma}_Q=0} \\
    &\quad= (u_0+v_0)\partial_{\chi_0^{(+)}}^2 \pqty{-\partial_{t_1}
      \mathcal{A}' n_1^{(0)} +\partial_{\chi_0^{(+)}}^2 \mathcal{G}'
      n_1^{(0)}} \,.
  \end{aligned}
\end{equation}
In the last line, we used the fact that $n_1^{(0)}$ satisfies the
$\eval{\text{KdVB}}_{\tilde{\sigma}_Q=0}$ equation to simplify the
right-hand side.
Integrating twice and dropping constants of integration (we want
$n_1^{(0)} =n_1^{(1)}=0$ to be a solution) gives
\begin{equation}
  \begin{aligned}
    \mathcal{L}_1 n_1^{(1)} &\coloneqq \mathcal{A}' \partial_{\tau_0}
      n_1^{(1)} + \mathcal{F}' \partial_{\chi_0^{(+)}}
      n_1^{(1)}+\mathcal{B}' \partial_{\chi_0^{(+)}} \pqty{n_1^{(0)}
      n_1^{(1)} } \\
    &\quad + \mathcal{C}' \partial_{\chi_0^{(+)}}^3 n_1^{(1)} \\
    &= -\partial_{t_1} \mathcal{A}' n_1^{(0)}
      +\partial_{\chi_0^{(+)}}^2 \mathcal{G}' n_1^{(0)} \,.
  \end{aligned}
\end{equation}

As before, we note that $\mathcal{L}_0$ and $-\mathcal{L}_1$ are
adjoints:
\begin{equation}
  \int \dd{\chi_0^{(+)}} \pqty{ n_1^{(1)} \mathcal{L}_0 n_1^{(0)} +
    n_1^{(0)} \mathcal{L}_1 n_1^{(1)} } = 0 \,.
\end{equation}
Thus, we get the compatibility condition
\begin{equation}
  (u_0+v_0) \int n_1^{(0)} \pqty{\mathcal{A}'\partial_{\tau_1}n_1^{(0)} -
    \mathcal{G}'\partial_{x}^2 n_1^{(0)}} \dd{\chi_0^{(+)}} = 0
    \,,
\end{equation}
which yields the equation
\begin{equation}
  \partial_{\tau_1} c_1(0,\tau_1) = -\frac{c_1(0,\tau_1)^2
    \abs{\mathcal{B}'}\tilde{\mathcal{G}}'}{\abs{\mathcal{C}'}\mathcal{A}'}
    \frac{4}{45} \,.
\end{equation}
Then, solving this equation and converting back to time $t_1$ gives
\begin{equation}
  c_1(0,t_1) = \frac{c_1(0,0)}{1 + t_1/t^{(1)}_d}
\end{equation}
with
\begin{equation}
  t^{(1)}_d = \frac{45 \mathcal{A}' \abs{\mathcal{C}'}}{4 c_1(0,0)
     \abs{\mathcal{B}'} \mathcal{G}'} \,,
\end{equation}
with $c_1(0,0)$ the initial value of the parameter $c_1(t_0,t_1)$.
Combined with the result for $c_1(t_0, t_1)$
(\cref{eq:MMS_c1_adi,eq:MMS_tau_d_adi}),
\begin{equation}
  c_1(t_0,t_1) = \frac{c_1(0,t_1)}{1 + t_0/t^{(0)}_d}
\end{equation}
\begin{minipage}{\linewidth}
with
\begin{equation}
  t^{(0)}_d = \frac{45 \mathcal{A}' \abs{\mathcal{C}'}}{4 c_1(0,t_1)
    \abs{\mathcal{B}'} \eval{\mathcal{G}'}_{\sigma_Q=0}} \,,
\end{equation}
we now have a complete solution.
\end{minipage}

\begin{widetext}

\section{\label{sec:full_equations} Full Equations}
\noindent
\begin{minipage}{\linewidth}
All quantities are expressed in normalized, nondimensional form
according to the procedures laid out in \cref{sec:normalization} and
\cref{sec:nondim}.
The energy conservation equations (\cref{eq:energy_eq_1,eq:energy_eq_2})
are only used for the adiabatic setup.

Leading Order:
\begin{subequations}
\begin{align}
  &\pdv{n_1}{t_0} + \gamma^2 n_0 u_0 \pdv{u_1}{t_0} + u_0
    \pdv{n_1}{x} + n_0 \gamma^2 \pdv{u_1}{x} = 0 \,,\\
  &\gamma^2 \pdv{\energy_1}{t_0} + \gamma^2 u_0^2 \pdv{P_1}{t_0} + 2 u_0
    (\energy_0 + P_0) \gamma^4 \pdv{u_1}{t_0} + (1 + u_0^2) (\energy_0 +
    P_0) \gamma^4 \pdv{u_1}{x} + u_0 \gamma^2 \pdv{x}(\energy_1 +
    P_1) \nonumber \\
  &\qquad + A n_0 u_0 \gamma^2 \pdv{n_1}{x} + A n_0^2 u_0^2 \gamma^4
    \pdv{u_1}{x} = 0 \label{eq:energy_eq_1} \,,\\
  &\gamma^3 (\energy_0 + P_0) \pdv{u_1}{t_0} + \gamma u_0
    \pdv{P_1}{t_0} + u_0 \gamma^3 (\energy_0 + P_0) \pdv{u_1}{x} +
    \gamma \pdv{P_1}{x} + A n_0 \gamma \pdv{n_1}{x} + A n_0^2 u_0
    \gamma^3 \pdv{u_1}{x} = 0 \,.
\end{align}
\end{subequations}
First-Order Corrections:
\begin{subequations}
\begin{align}
  &\begin{aligned}
  &\pdv{n_2}{t_0} + \gamma^2 n_0 u_0 \pdv{u_2}{t_0} + u_0
    \pdv{n_2}{x} + n_0 \gamma^2 \pdv{u_2}{x} = \\
  &\qquad -\pdv{n_1}{t_1} - n_0 u_0 \gamma^2 \pdv{u_1}{t_1} - u_0 u_1
    \gamma^2 \pdv{n_1}{t_0} - \gamma^2 \bqty{\gamma^2(1+2u_0^2) n_0 u_1 +
    u_0 n_1} \pdv{u_1}{t_0} \\
  &\qquad - \gamma^2 \bqty{u_0 \gamma^2 (2+u_0^2) n_0 u_1 + n_1 + n_0 u_0
    u_1} \pdv{u_1}{x} - \gamma^2 u_1 \pdv{n_1}{x} \\
  &\qquad + \gamma u_0 A \sigma_Q
    \pdv{n_1}{t_0}{x} + \gamma A \sigma_Q \pdv[2]{n_1}{x} + \gamma^3
    u_0 A \sigma_Q n_0 \pqty{\pdv[2]{u_1}{x} + u_0 \pdv{u_1}{t_0}{x}} \\
  &\qquad + \Theta(-m) \gamma \sigma_Q
    \bqty{\pqty{u_0^2\pdv[2]{\mu_1}{t_0}+2u_0\pdv{\mu_1}{t_0}{x}+\pdv[2]{\mu_1}{x}}
    - \frac{\mu_0}{T_0}
    \pqty{u_0^2\pdv[2]{T_1}{t_0}+2u_0\pdv{T_1}{t_0}{x}+\pdv[2]{T_1}{x}}}
    \,,
  \end{aligned} \\
  &\begin{aligned}
  &\gamma^2 \pdv{\energy_2}{t_0} + \gamma^2 u_0^2 \pdv{P_2}{t_0} + 2 u_0
    (\energy_0 + P_0) \gamma^4 \pdv{u_2}{t_0} + (1 + u_0^2) (\energy_0 +
    P_0) \gamma^4 \pdv{u_2}{x} + u_0 \gamma^2 \pdv{x}(\energy_2 +
    P_2) \\
  &\qquad + A n_0 u_0 \gamma^2 \pdv{n_2}{x} + A n_0^2 u_0^2 \gamma^4
    \pdv{u_2}{x} = \\
  &\qquad -2(\energy_0 + P_0) u_0 \gamma^4 \pdv{u_1}{t_1} - \gamma^2
    \pdv{\energy_1}{t_1} - \gamma^2 u_0^2 \pdv{P_1}{t_1} -2 (\energy_0 +
    P_0) \gamma^6(1+3u_0^2) u_1 \pdv{u_1}{t_0} -2 (\energy_1 +
    P_1) \gamma^4 u_0 \pdv{u_1}{t_0} \\
  &\qquad -2u_0 u_1 \gamma^4 \pdv{t_0}(\energy_1 + P_1) -2 (3+ u_0^2)
    u_0 u_1 \gamma^6 (\energy_0 + P_0) \pdv{u_1}{x} - \gamma^4 (1+u_0^2)
    (\energy_1 + P_1) \pdv{u_1}{x} - (1+u_0^2) u_1 \gamma^4 \pdv{x}
    (\energy_1 + P_1) \\
  &\qquad + \gamma^5 u_0 \bqty{\zeta + 2 \eta \pqty{1-\frac{1}{d}}}
    \pqty{u_0^2 \pdv[2]{u_1}{t_0} + 2 u_0 \pdv{u_1}{t_0}{x} +
    \pdv[2]{u_1}{x}} \\
  &\qquad - A n_0 u_1 \gamma^4 \pdv{n_1}{x} - A u_0 n_1 \gamma^2
    \pdv{n_1}{x} - A n_0 u_0 \gamma^2 \frac{d_1 d_2}{3} \pdv[3]{n_1}{x}
    - 2 A n_0^2 u_0 (1+u_0^2) u_1 \gamma^6 \pdv{u_1}{x} \\
  &\qquad - 2 A n_0 u_0^2 n_1 \gamma^4 \pdv{u_1}{x} - A n_0^2 u_0^2
    \gamma^4 \frac{d_1 d_2}{3} \pdv[3]{u_1}{x} -A n_0 u_0^2 u_1 \gamma^4
    \pdv{n_1}{x} \,,
  \end{aligned} \label{eq:energy_eq_2} \\
  &\begin{aligned}
  &\gamma^3 (\energy_0 + P_0) \pdv{u_2}{t_0} + \gamma u_0
    \pdv{P_2}{t_0} + u_0 \gamma^3 (\energy_0 + P_0) \pdv{u_2}{x} +
    \gamma \pdv{P_2}{x} + A n_0 \gamma \pdv{n_2}{x} + A n_0^2 u_0
    \gamma^3 \pdv{u_2}{x} = \\
  &\qquad -(\energy_0 + P_0) \gamma^3 \pdv{u_1}{t_1} - u_0 \gamma
    \pdv{P_1}{t_1} - \gamma^3 \bqty{2 u_0 u_1 \gamma^2 (\energy_0 + P_0)
    + (\energy_1 + P_1)}\pdv{u_1}{t_0} - u_1 \gamma \pdv{P_1}{t_0} \\
  &\qquad -\gamma^3 \bqty{ u_0 (\energy_1 + P_1) + (1+u_0^2) u_1 \gamma^2
    (\energy_0 + P_0)} \pdv{u_1}{x} - A n_1 \gamma \pdv{n_1}{x} - A n_0
    \gamma \frac{d_1 d_2}{3} \pdv[3]{n_1}{x} \\
  &\qquad - A n_0^2 (1+u_0^2) u_1 \gamma^5 \pdv{u_1}{x} - A n_0^2 u_0
    \gamma^3 \frac{d_1 d_2}{3} \pdv[3]{u_1}{x} - 2 A n_0 u_0 n_1
    \gamma^3 \pdv{u_1}{x} \\
  &\qquad +\gamma^4 \bqty{\zeta + 2 \eta \pqty{1 - \frac{1}{d}}} \pqty{u_0^2
    \pdv[2]{u_1}{t_0} + 2 u_0 \pdv{u_1}{t_0}{x} + \pdv[2]{u_1}{x}} \,.
  \end{aligned}
\end{align}
\end{subequations}
\end{minipage}

\section{\label{sec:kdv_burgers} Isothermal KdV-Burgers}
All quantities are expressed in dimensional form; to get the
dimensionless expressions, simply set $v_F = \hbar = l_{\text{ref}} = k_B = e =
1$ and remove all factors of $\epsilon$.
See \cref{sec:thermo_coeffs} for the values of $\mathcal{C}_0$ and
$\mathcal{C}_1$ and \cref{sec:nondim} for the $\ordersymbol$
expressions.
The KdV-Burgers equation is given by
\begin{equation}
  \frac{\mathcal{A}'}{v_F} \pdv{n_1}{t_1} + \mathcal{F}' \pdv{n_1}{x} +
    \mathcal{B}' \frac{l_{\text{ref}}^d}{\order{n l_{\text{ref}}^d}} n_1 \pdv{n_1}{x} +
    \mathcal{C}' \frac{l_{\text{ref}}^2}{\order{\partial_x l_{\text{ref}}}^2}
    \pdv[3]{n_1}{x} = \mathcal{G}' \frac{l_{\text{ref}}}{\order{\partial_x
    l_{\text{ref}}}} \pdv[2]{n_1}{x} \,,
\end{equation}
with
\begin{gather}
  \mathcal{A}' = 2 \gamma^2 \frac{u_0+v_0}{v_F^2+u_0 v_0} \frac{P_0
    l_{\text{ref}} d}{n_0 \hbar v_F^3} \Bqty{v_0 \bqty{v_F^2 (d+1)-u_0^2 K_0}+u_0
    v_F^2 (d+1-K_0)} \,, \\
  \begin{aligned}
    \mathcal{B}' &= -\gamma^2 \frac{u_0+v_0}{v_F^2+u_0 v_0}
      \frac{P_0 l_{\text{ref}}^{1-d}}{n_0^2 d \hbar v_F^4} \bigl\{d^2 v_F^2
      (u_0+v_0)^2 [4(d+1)-K_0(d+3)] \\
    &\qquad + (v_F^2+u_0 v_0)^2 [(d+1)\Theta(-m) - K_0 d^2] \bigr\}
      \order{n l_{\text{ref}}^d} \,,
  \end{aligned} \\
  \mathcal{C}' = -\frac{n_0}{\epsilon l_{\text{ref}} \hbar} A d \frac{d_1
    d_2}{3} \frac{u_0+v_0}{v_F^2+u_0 v_0}
    \order{\partial_x l_{\text{ref}}}^2 \,, \\
  \begin{aligned}
  \mathcal{F}' &= \gamma^2 \frac{P_0 l_{\text{ref}} d}{\epsilon n_0 \hbar v_F^4}
    \frac{u_0+v_0}{v_F^2+u_0 v_0} \Biggl\{2 U_1 \gamma^2 (d+1-K_0)
    (u_0+v_0) (v_F^2+u_0 v_0) \\
  &\qquad + \frac{\mathcal{C}_1}{\mathcal{C}_0}
    \pqty{\frac{\mu_0}{k_B T_0}}^{2m}
    \bqty{v_F^2 (u_0+v_0)^2 (d+1) \pqty{\frac{1}{d} \Kronecker_{m,-1} +
    \Kronecker_{m,1}} - (v_F^2 + u_0 v_0)^2 \pqty{\frac{d-1}{d^2}
    \Kronecker_{m,-1} +2 \Kronecker_{m,1}}} \Biggr\} \,,
  \end{aligned}\\
  \begin{aligned}
    \mathcal{G}' &= \frac{\gamma^3 (v_F^2+u_0 v_0)}{\epsilon n_0 \hbar
      v_F^{10}} \Biggl\{ \frac{\sigma_Q}{e^2}
      \gamma^2 \pqty{\frac{P_0}{n_0}}^2 (u_0+v_0)
      (v_F^2+u_0 v_0) (d+1) \bqty{v_0 \pqty{v_F^2(d+1)-u_0^2 K_0} + u_0
      v_F^2 (d+1-K_0)} \\
    &\qquad \times \bqty{\frac{d(u_0+v_0)^2}{(v_F^2 + u_0 v_0)^2} +
      \underbrace{\Theta(-m) - K_0 \frac{d}{d+1}}_{=0}}
      + v_F^6 d (u_0+v_0)^2 \bqty{\zeta+2 \eta
      \pqty{1-\frac{1}{d}}} \Biggr\} \order{\partial_x l_{\text{ref}}} \,,
  \end{aligned}
\end{gather}
and
\begin{equation}
  v_0^{(\pm)} = v_F^2 \frac{-u_0 (d+1-K_0) \pm \frac{1}{\gamma}
    \sqrt{[K_0 (d+1) v_F^2]/\gamma^2 + (A n_0^2/P_0)
    \bqty{v_F^2(d+1)-u_0^2 K_0}}}{v_F^2(d+1)-u_0^2 K_0} \,.
\end{equation}

If we impose $v_0 =0$, then the coefficients take the form given in
\cref{eq:stationary_coeffs}.
If instead we impose $u_0 = U_1 = 0$, they take the form
\begin{gather}
  \mathcal{A}' = 2 \frac{P_0 (d+1) v_0^2 l_{\text{ref}} d}{n_0 \hbar v_F^3} \,, \\
  \mathcal{B}' = -v_0 \frac{P_0 l_{\text{ref}}^{1-d}}{n_0^2 d \hbar v_F^4}
    \Bqty{d^2 v_0^2 [4(d+1)-K_0(d+3)] + v_F^2\bqty{(d+1)\Theta(-m) - K_0
    d^2}} \order{n l_{\text{ref}}^d} \,, \\
  \mathcal{C}' = -\frac{n_0}{\epsilon l_{\text{ref}} \hbar v_F^2} A d
    \frac{d_1 d_2}{3} v_0 \order{\partial_x l_{\text{ref}}}^2 \,,
    \displaybreak[0] \\
  \mathcal{F}' = \frac{P_0 l_{\text{ref}} d}{\epsilon n_0 \hbar v_F^4}
    v_0 \frac{\mathcal{C}_1}{\mathcal{C}_0}
    \pqty{\frac{\mu_0}{k_B T_0}}^{2m}
    \bqty{v_0^2 (d+1) \pqty{\frac{1}{d} \Kronecker_{m,-1} +
    \Kronecker_{m,1}} - v_F^2 \pqty{\frac{d-1}{d^2} \Kronecker_{m,-1}
    +2 \Kronecker_{m,1}}} \,, \\
  \mathcal{G}' = \frac{1}{\epsilon n_0 \hbar v_F^4} \Biggl\{
    \frac{\sigma_Q}{e^2} \pqty{\frac{P_0}{n_0}}^2 v_0^2
    (d+1)^2 \bqty{\frac{v_0^2 d}{v_F^2} +
    \underbrace{\Theta(-m) - K_0 \frac{d}{d+1}}_{=0}}
    + v_0^2 v_F^2 d \bqty{\zeta+2 \eta \pqty{1-\frac{1}{d}}}
    \Biggr\} \order{\partial_x l_{\text{ref}}} \,,
\end{gather}
with
\begin{equation}
  v_0 = \pm \frac{v_F}{\sqrt{d+1}} \sqrt{K_0 + \frac{A n_0^2}{P_0}} \,.
\end{equation}

\section{\label{sec:kdv_burgers_adi} Adiabatic KdV-Burgers}
All quantities are expressed in dimensional form; to get the
dimensionless expressions, simply set $v_F = \hbar = l_{\text{ref}} = k_B = e =
1$ and remove all factors of $\epsilon$.
See \cref{sec:thermo_coeffs} for the values of $\mathcal{C}_0$ and
$\mathcal{C}_1$ and \cref{sec:nondim} for the $\ordersymbol$
expressions.
The KdV-Burgers equation is given by
\begin{equation}
  \frac{\mathcal{A}'}{v_F} \pdv{n_1}{t_1} + \mathcal{F}' \pdv{n_1}{x} +
    \mathcal{B}' \frac{l_{\text{ref}}^d}{\order{n} l_{\text{ref}}^d} n_1 \pdv{n_1}{x} +
    \mathcal{C}' \frac{l_{\text{ref}}^2}{\order{\partial_x l_{\text{ref}}}^2}
    \pdv[3]{n_1}{x} = \mathcal{G}' \frac{l_{\text{ref}}}{\order{\partial_x
    l_{\text{ref}}}} \pdv[2]{n_1}{x} \,,
\end{equation}
with
\begin{gather}
  \mathcal{A}' = 2 \gamma^2 \frac{u_0+v_0}{v_F^2+u_0 v_0} \frac{P_0
    (d+1) l_{\text{ref}}}{n_0 \hbar v_F^3} \bqty{v_0 \pqty{v_F^2 d-u_0^2}+u_0
    v_F^2 (d-1)} \,, \\
  \mathcal{B}' = -\gamma^2 \frac{u_0+v_0}{v_F^2+u_0 v_0} \frac{P_0
    (d+1) l_{\text{ref}}^{1-d}}{n_0^2 \hbar v_F^4} \frac{d-1}{d} \bqty{3d v_F^2
    (u_0+v_0)^2 - (v_F^2+u_0 v_0)^2}  \order{n l_{\text{ref}}^d} \,, \\
  \mathcal{C}' = -\frac{n_0}{\epsilon l_{\text{ref}} \hbar} A d \frac{d_1
    d_2}{3} \frac{u_0+v_0}{v_F^2+u_0 v_0}
    \order{\partial_x l_{\text{ref}}}^2 \,, \\
  \begin{aligned}
  \mathcal{F}' &= \gamma^2 \frac{P_0 (d+1) l_{\text{ref}}}{\epsilon n_0 \hbar
    v_F^4} \frac{u_0+v_0}{v_F^2+u_0 v_0} \Bigl\{ 2 U_1 \gamma^2 (d-1)
    (u_0+v_0) (v_F^2+u_0 v_0) \\
  &\qquad + \Kronecker_{m,-1} \frac{1}{d}
    \frac{\mathcal{C}_1}{\mathcal{C}_0}
    \pqty{\frac{k_B T_0}{\mu_0}}^2 \bqty{d v_F^2 (u_0+v_0)^2 - (v_F^2+u_0
    v_0)^2} \Bigr\} \,,
  \end{aligned}\\
  \mathcal{G}' = \frac{d \gamma (v_F^2+u_0 v_0)}{\epsilon n_0 \hbar
      v_F^4} \Biggl\{ \frac{\sigma_Q}{e^2} \frac{A^2
      n_0^2 v_F^2}{v_F^2+u_0 v_0} + \gamma^2 (u_0+v_0)^2
      \bqty{\zeta+2 \eta \pqty{1-\frac{1}{d}}} \Biggr\}
      \order{\partial_x l_{\text{ref}}} \,,
\end{gather}
and
\begin{equation}
  v_0^{(\pm)} = -\frac{u_0 v_F^2 (d-1)}{v_F^2 d-u_0^2} \pm
    \frac{v_F^2 \sqrt{d}}{(v_F^2 d-u_0^2) \gamma^2} \sqrt{v_F^2 +
    \frac{A n_0^2 \gamma^2 (v_F^2 d-u_0^2)}{P_0 (d+1)}} \,.
\end{equation}

If we impose $v_0 =0$, then the coefficients take the form given in
\cref{eq:stationary_coeffs}.
If instead we impose $u_0 = U_1 = 0$, they take the form
\begin{gather}
  \mathcal{A}' = 2 \frac{P_0 (d+1) v_0^2 l_{\text{ref}} d}{n_0 \hbar v_F^3} \,,
    \\
  \mathcal{B}' = -\frac{P_0 (d+1) l_{\text{ref}}^{1-d}}{n_0^2 \hbar v_F^4}
    \frac{d-1}{d} v_0 \pqty{3d v_0^2-v_F^2} \order{n l_{\text{ref}}^d} \,, \\
  \mathcal{C}' = - \frac{n_0}{\epsilon l_{\text{ref}} \hbar v_F^2} A d
    \frac{d_1 d_2}{3} v_0 \order{\partial_x l_{\text{ref}}}^2 \,,
    \displaybreak[0] \\
  \mathcal{F}' = \Kronecker_{m,-1} \frac{A n_0 v_0 l_{\text{ref}}}{\epsilon
    \hbar v_F^2} \frac{\mathcal{C}_1}{\mathcal{C}_0}
    \pqty{\frac{k_B T_0}{\mu_0}}^2 \,, \\
  \mathcal{G}' = \frac{d}{\epsilon n_0 \hbar v_F^2}
    \Bqty{\frac{\sigma_Q}{e^2} A^2 n_0^2
    + v_0^2 \bqty{\zeta+2 \eta \pqty{1-\frac{1}{d}}}}
    \order{\partial_x l_{\text{ref}}} \,,
\end{gather}
with
\begin{equation}
  v_0 = \pm \frac{v_F}{\sqrt{d}} \sqrt{1 + \frac{A d n_0^2}{(d+1) P_0}}
    \,.
\end{equation}

\end{widetext}



\input{GrapheneSolitons.bbl}

\end{document}

%% file: GrapheneSolitons.bbl
%

%% file: GrapheneSolitons.bbl
\begin{thebibliography}{63}%
\makeatletter
\providecommand \@ifxundefined [1]{%
 \@ifx{#1\undefined}
}%
\providecommand \@ifnum [1]{%
 \ifnum #1\expandafter \@firstoftwo
 \else \expandafter \@secondoftwo
 \fi
}%
\providecommand \@ifx [1]{%
 \ifx #1\expandafter \@firstoftwo
 \else \expandafter \@secondoftwo
 \fi
}%
\providecommand \natexlab [1]{#1}%
\providecommand \enquote  [1]{``#1''}%
\providecommand \bibnamefont  [1]{#1}%
\providecommand \bibfnamefont [1]{#1}%
\providecommand \citenamefont [1]{#1}%
\providecommand \href@noop [0]{\@secondoftwo}%
\providecommand \href [0]{\begingroup \@sanitize@url \@href}%
\providecommand \@href[1]{\@@startlink{#1}\@@href}%
\providecommand \@@href[1]{\endgroup#1\@@endlink}%
\providecommand \@sanitize@url [0]{\catcode `\\12\catcode `\$12\catcode
  `\&12\catcode `\#12\catcode `\^12\catcode `\_12\catcode `\%12\relax}%
\providecommand \@@startlink[1]{}%
\providecommand \@@endlink[0]{}%
\providecommand \url  [0]{\begingroup\@sanitize@url \@url }%
\providecommand \@url [1]{\endgroup\@href {#1}{\urlprefix }}%
\providecommand \urlprefix  [0]{URL }%
\providecommand \Eprint [0]{\href }%
\providecommand \doibase [0]{https://doi.org/}%
\providecommand \selectlanguage [0]{\@gobble}%
\providecommand \bibinfo  [0]{\@secondoftwo}%
\providecommand \bibfield  [0]{\@secondoftwo}%
\providecommand \translation [1]{[#1]}%
\providecommand \BibitemOpen [0]{}%
\providecommand \bibitemStop [0]{}%
\providecommand \bibitemNoStop [0]{.\EOS\space}%
\providecommand \EOS [0]{\spacefactor3000\relax}%
\providecommand \BibitemShut  [1]{\csname bibitem#1\endcsname}%
\let\auto@bib@innerbib\@empty
\bibitem [{\citenamefont {Lucas}\ and\ \citenamefont
  {Fong}(2018)}]{lucas2018hydrodynamics}%
  \BibitemOpen
  \bibfield  {author} {\bibinfo {author} {\bibfnamefont {A.}~\bibnamefont
  {Lucas}}\ and\ \bibinfo {author} {\bibfnamefont {K.~C.}\ \bibnamefont
  {Fong}},\ }\bibfield  {title} {\bibinfo {title} {Hydrodynamics of electrons
  in graphene},\ }\href {https://doi.org/10.1088/1361-648X/aaa274} {\bibfield
  {journal} {\bibinfo  {journal} {Journal of Physics: Condensed Matter}\
  }\textbf {\bibinfo {volume} {30}},\ \bibinfo {pages} {053001} (\bibinfo
  {year} {2018})},\ \Eprint {https://arxiv.org/abs/1710.08425}
  {arXiv:1710.08425 [cond-mat]} \BibitemShut {NoStop}%
\bibitem [{\citenamefont {Landau}\ and\ \citenamefont
  {Lifshitz}(1959)}]{landau1959fluid}%
  \BibitemOpen
  \bibfield  {author} {\bibinfo {author} {\bibfnamefont {L.~D.}\ \bibnamefont
  {Landau}}\ and\ \bibinfo {author} {\bibfnamefont {E.~M.}\ \bibnamefont
  {Lifshitz}},\ }\bibfield  {title} {\bibinfo {title} {Fluid mechanics, 1959},\
  }\href@noop {} {\bibfield  {journal} {\bibinfo  {journal} {Course of
  theoretical physics}\ } (\bibinfo {year} {1959})}\BibitemShut {NoStop}%
\bibitem [{\citenamefont {Landau}(1956)}]{landau1956theory}%
  \BibitemOpen
  \bibfield  {author} {\bibinfo {author} {\bibfnamefont {L.~D.}\ \bibnamefont
  {Landau}},\ }\bibfield  {title} {\bibinfo {title} {The theory of a fermi
  liquid},\ }\href {http://www.jetp.ac.ru/cgi-bin/dn/e_003_06_0920.pdf}
  {\bibfield  {journal} {\bibinfo  {journal} {Journal of Experimental and
  Theoretical Physics}\ }\textbf {\bibinfo {volume} {3}},\ \bibinfo {pages}
  {920} (\bibinfo {year} {1956})}\BibitemShut {NoStop}%
\bibitem [{\citenamefont {Novoselov}\ \emph {et~al.}(2005)\citenamefont
  {Novoselov}, \citenamefont {Geim}, \citenamefont {Morozov}, \citenamefont
  {Jiang}, \citenamefont {Katsnelson}, \citenamefont {Grigorieva},
  \citenamefont {Dubonos},\ and\ \citenamefont {Firsov}}]{novoselov2005two}%
  \BibitemOpen
  \bibfield  {author} {\bibinfo {author} {\bibfnamefont {K.~S.}\ \bibnamefont
  {Novoselov}}, \bibinfo {author} {\bibfnamefont {A.~K.}\ \bibnamefont {Geim}},
  \bibinfo {author} {\bibfnamefont {S.~V.}\ \bibnamefont {Morozov}}, \bibinfo
  {author} {\bibfnamefont {D.}~\bibnamefont {Jiang}}, \bibinfo {author}
  {\bibfnamefont {M.~I.}\ \bibnamefont {Katsnelson}}, \bibinfo {author}
  {\bibfnamefont {I.~V.}\ \bibnamefont {Grigorieva}}, \bibinfo {author}
  {\bibfnamefont {S.~V.}\ \bibnamefont {Dubonos}},\ and\ \bibinfo {author}
  {\bibfnamefont {A.~A.}\ \bibnamefont {Firsov}},\ }\bibfield  {title}
  {\bibinfo {title} {Two-dimensional gas of massless dirac fermions in
  graphene},\ }\href {https://doi.org/10.1038/nature04233} {\bibfield
  {journal} {\bibinfo  {journal} {Nature (London)}\ }\textbf {\bibinfo {volume}
  {438}},\ \bibinfo {pages} {197} (\bibinfo {year} {2005})},\ \Eprint
  {https://arxiv.org/abs/cond-mat/0509330} {cond-mat/0509330} \BibitemShut
  {NoStop}%
\bibitem [{\citenamefont {Crossno}\ \emph {et~al.}(2016)\citenamefont
  {Crossno}, \citenamefont {Shi}, \citenamefont {Wang}, \citenamefont {Liu},
  \citenamefont {Harzheim}, \citenamefont {Lucas}, \citenamefont {Sachdev},
  \citenamefont {Kim}, \citenamefont {Taniguchi}, \citenamefont {Watanabe}
  \emph {et~al.}}]{crossno2016observation}%
  \BibitemOpen
  \bibfield  {author} {\bibinfo {author} {\bibfnamefont {J.}~\bibnamefont
  {Crossno}}, \bibinfo {author} {\bibfnamefont {J.~K.}\ \bibnamefont {Shi}},
  \bibinfo {author} {\bibfnamefont {K.}~\bibnamefont {Wang}}, \bibinfo {author}
  {\bibfnamefont {X.}~\bibnamefont {Liu}}, \bibinfo {author} {\bibfnamefont
  {A.}~\bibnamefont {Harzheim}}, \bibinfo {author} {\bibfnamefont
  {A.}~\bibnamefont {Lucas}}, \bibinfo {author} {\bibfnamefont
  {S.}~\bibnamefont {Sachdev}}, \bibinfo {author} {\bibfnamefont
  {P.}~\bibnamefont {Kim}}, \bibinfo {author} {\bibfnamefont {T.}~\bibnamefont
  {Taniguchi}}, \bibinfo {author} {\bibfnamefont {K.}~\bibnamefont {Watanabe}},
  \emph {et~al.},\ }\bibfield  {title} {\bibinfo {title} {Observation of the
  dirac fluid and the breakdown of the wiedemann-franz law in graphene},\
  }\href {https://doi.org/10.1126/science.aad0343} {\bibfield  {journal}
  {\bibinfo  {journal} {Science}\ }\textbf {\bibinfo {volume} {351}},\ \bibinfo
  {pages} {1058} (\bibinfo {year} {2016})},\ \Eprint
  {https://arxiv.org/abs/1509.04713} {arXiv:1509.04713 [cond-mat]} \BibitemShut
  {NoStop}%
\bibitem [{\citenamefont {Torre}\ \emph {et~al.}(2015)\citenamefont {Torre},
  \citenamefont {Tomadin}, \citenamefont {Geim},\ and\ \citenamefont
  {Polini}}]{torre2015nonlocal}%
  \BibitemOpen
  \bibfield  {author} {\bibinfo {author} {\bibfnamefont {I.}~\bibnamefont
  {Torre}}, \bibinfo {author} {\bibfnamefont {A.}~\bibnamefont {Tomadin}},
  \bibinfo {author} {\bibfnamefont {A.~K.}\ \bibnamefont {Geim}},\ and\
  \bibinfo {author} {\bibfnamefont {M.}~\bibnamefont {Polini}},\ }\bibfield
  {title} {\bibinfo {title} {Nonlocal transport and the hydrodynamic shear
  viscosity in graphene},\ }\href {https://doi.org/10.1103/PhysRevB.92.165433}
  {\bibfield  {journal} {\bibinfo  {journal} {Physical Review B}\ }\textbf
  {\bibinfo {volume} {92}},\ \bibinfo {pages} {165433} (\bibinfo {year}
  {2015})},\ \Eprint {https://arxiv.org/abs/1508.00363} {arXiv:1508.00363
  [cond-mat]} \BibitemShut {NoStop}%
\bibitem [{\citenamefont {Tomadin}\ \emph {et~al.}(2014)\citenamefont
  {Tomadin}, \citenamefont {Vignale},\ and\ \citenamefont
  {Polini}}]{tomadin2014corbino}%
  \BibitemOpen
  \bibfield  {author} {\bibinfo {author} {\bibfnamefont {A.}~\bibnamefont
  {Tomadin}}, \bibinfo {author} {\bibfnamefont {G.}~\bibnamefont {Vignale}},\
  and\ \bibinfo {author} {\bibfnamefont {M.}~\bibnamefont {Polini}},\
  }\bibfield  {title} {\bibinfo {title} {Corbino disk viscometer for 2d quantum
  electron liquids},\ }\href {https://doi.org/10.1103/PhysRevLett.113.235901}
  {\bibfield  {journal} {\bibinfo  {journal} {Physical review letters}\
  }\textbf {\bibinfo {volume} {113}},\ \bibinfo {pages} {235901} (\bibinfo
  {year} {2014})},\ \Eprint {https://arxiv.org/abs/1401.0938} {arXiv:1401.0938
  [cond-mat]} \BibitemShut {NoStop}%
\bibitem [{\citenamefont {Levitov}\ and\ \citenamefont
  {Falkovich}(2016)}]{levitov2016electron}%
  \BibitemOpen
  \bibfield  {author} {\bibinfo {author} {\bibfnamefont {L.}~\bibnamefont
  {Levitov}}\ and\ \bibinfo {author} {\bibfnamefont {G.}~\bibnamefont
  {Falkovich}},\ }\bibfield  {title} {\bibinfo {title} {Electron viscosity,
  current vortices and negative nonlocal resistance in graphene},\ }\href
  {https://doi.org/10.1038/nphys3667} {\bibfield  {journal} {\bibinfo
  {journal} {Nature Physics}\ }\textbf {\bibinfo {volume} {12}},\ \bibinfo
  {pages} {672} (\bibinfo {year} {2016})},\ \Eprint
  {https://arxiv.org/abs/1508.00836} {arXiv:1508.00836 [cond-mat]} \BibitemShut
  {NoStop}%
\bibitem [{\citenamefont {Dyakonov}\ and\ \citenamefont
  {Shur}(1993)}]{dyakonov1993shallow}%
  \BibitemOpen
  \bibfield  {author} {\bibinfo {author} {\bibfnamefont {M.}~\bibnamefont
  {Dyakonov}}\ and\ \bibinfo {author} {\bibfnamefont {M.}~\bibnamefont
  {Shur}},\ }\bibfield  {title} {\bibinfo {title} {Shallow water analogy for a
  ballistic field effect transistor: New mechanism of plasma wave generation by
  dc current},\ }\href {https://doi.org/10.1103/PhysRevLett.71.2465} {\bibfield
   {journal} {\bibinfo  {journal} {Physical review letters}\ }\textbf {\bibinfo
  {volume} {71}},\ \bibinfo {pages} {2465} (\bibinfo {year}
  {1993})}\BibitemShut {NoStop}%
\bibitem [{\citenamefont {Bandurin}\ \emph {et~al.}(2016)\citenamefont
  {Bandurin}, \citenamefont {Torre}, \citenamefont {Kumar}, \citenamefont
  {Shalom}, \citenamefont {Tomadin}, \citenamefont {Principi}, \citenamefont
  {Auton}, \citenamefont {Khestanova}, \citenamefont {Novoselov}, \citenamefont
  {Grigorieva} \emph {et~al.}}]{bandurin2016negative}%
  \BibitemOpen
  \bibfield  {author} {\bibinfo {author} {\bibfnamefont {D.~A.}\ \bibnamefont
  {Bandurin}}, \bibinfo {author} {\bibfnamefont {I.}~\bibnamefont {Torre}},
  \bibinfo {author} {\bibfnamefont {R.~K.}\ \bibnamefont {Kumar}}, \bibinfo
  {author} {\bibfnamefont {M.~B.}\ \bibnamefont {Shalom}}, \bibinfo {author}
  {\bibfnamefont {A.}~\bibnamefont {Tomadin}}, \bibinfo {author} {\bibfnamefont
  {A.}~\bibnamefont {Principi}}, \bibinfo {author} {\bibfnamefont {G.~H.}\
  \bibnamefont {Auton}}, \bibinfo {author} {\bibfnamefont {E.}~\bibnamefont
  {Khestanova}}, \bibinfo {author} {\bibfnamefont {K.~S.}\ \bibnamefont
  {Novoselov}}, \bibinfo {author} {\bibfnamefont {I.~V.}\ \bibnamefont
  {Grigorieva}}, \emph {et~al.},\ }\bibfield  {title} {\bibinfo {title}
  {Negative local resistance caused by viscous electron backflow in graphene},\
  }\href {https://doi.org/10.1126/science.aad0201} {\bibfield  {journal}
  {\bibinfo  {journal} {Science}\ }\textbf {\bibinfo {volume} {351}},\ \bibinfo
  {pages} {1055} (\bibinfo {year} {2016})},\ \Eprint
  {https://arxiv.org/abs/1509.04165} {arXiv:1509.04165 [cond-mat]} \BibitemShut
  {NoStop}%
\bibitem [{\citenamefont {Kumar}\ \emph {et~al.}(2017)\citenamefont {Kumar},
  \citenamefont {Bandurin}, \citenamefont {Pellegrino}, \citenamefont {Cao},
  \citenamefont {Principi}, \citenamefont {Guo}, \citenamefont {Auton},
  \citenamefont {Shalom}, \citenamefont {Ponomarenko}, \citenamefont
  {Falkovich} \emph {et~al.}}]{kumar2017superballistic}%
  \BibitemOpen
  \bibfield  {author} {\bibinfo {author} {\bibfnamefont {R.~K.}\ \bibnamefont
  {Kumar}}, \bibinfo {author} {\bibfnamefont {D.~A.}\ \bibnamefont {Bandurin}},
  \bibinfo {author} {\bibfnamefont {F.~M.~D.}\ \bibnamefont {Pellegrino}},
  \bibinfo {author} {\bibfnamefont {Y.}~\bibnamefont {Cao}}, \bibinfo {author}
  {\bibfnamefont {A.}~\bibnamefont {Principi}}, \bibinfo {author}
  {\bibfnamefont {H.}~\bibnamefont {Guo}}, \bibinfo {author} {\bibfnamefont
  {G.~H.}\ \bibnamefont {Auton}}, \bibinfo {author} {\bibfnamefont {M.~B.}\
  \bibnamefont {Shalom}}, \bibinfo {author} {\bibfnamefont {L.~A.}\
  \bibnamefont {Ponomarenko}}, \bibinfo {author} {\bibfnamefont
  {G.}~\bibnamefont {Falkovich}}, \emph {et~al.},\ }\bibfield  {title}
  {\bibinfo {title} {Superballistic flow of viscous electron fluid through
  graphene constrictions},\ }\href {https://doi.org/10.1038/nphys4240}
  {\bibfield  {journal} {\bibinfo  {journal} {Nature Physics}\ }\textbf
  {\bibinfo {volume} {13}},\ \bibinfo {pages} {1182} (\bibinfo {year}
  {2017})},\ \Eprint {https://arxiv.org/abs/1703.06672} {arXiv:1703.06672
  [cond-mat]} \BibitemShut {NoStop}%
\bibitem [{\citenamefont {Lucas}\ and\ \citenamefont
  {Das~Sarma}(2018)}]{lucas2018electronic}%
  \BibitemOpen
  \bibfield  {author} {\bibinfo {author} {\bibfnamefont {A.}~\bibnamefont
  {Lucas}}\ and\ \bibinfo {author} {\bibfnamefont {S.}~\bibnamefont
  {Das~Sarma}},\ }\bibfield  {title} {\bibinfo {title} {Electronic sound modes
  and plasmons in hydrodynamic two-dimensional metals},\ }\href
  {https://doi.org/10.1103/PhysRevB.97.115449} {\bibfield  {journal} {\bibinfo
  {journal} {Physical Review B}\ }\textbf {\bibinfo {volume} {97}},\ \bibinfo
  {pages} {115449} (\bibinfo {year} {2018})},\ \Eprint
  {https://arxiv.org/abs/1801.01495} {arXiv:1801.01495 [cond-mat]} \BibitemShut
  {NoStop}%
\bibitem [{\citenamefont {Sun}\ \emph {et~al.}(2016)\citenamefont {Sun},
  \citenamefont {Basov},\ and\ \citenamefont {Fogler}}]{sun2016adiabatic}%
  \BibitemOpen
  \bibfield  {author} {\bibinfo {author} {\bibfnamefont {Z.}~\bibnamefont
  {Sun}}, \bibinfo {author} {\bibfnamefont {D.~N.}\ \bibnamefont {Basov}},\
  and\ \bibinfo {author} {\bibfnamefont {M.~M.}\ \bibnamefont {Fogler}},\
  }\bibfield  {title} {\bibinfo {title} {Adiabatic amplification of plasmons
  and demons in 2d systems},\ }\href
  {https://doi.org/10.1103/PhysRevLett.117.076805} {\bibfield  {journal}
  {\bibinfo  {journal} {Physical review letters}\ }\textbf {\bibinfo {volume}
  {117}},\ \bibinfo {pages} {076805} (\bibinfo {year} {2016})},\ \Eprint
  {https://arxiv.org/abs/1601.02722} {arXiv:1601.02722 [cond-mat]} \BibitemShut
  {NoStop}%
\bibitem [{\citenamefont {Akbari-Moghanjoughi}(2013)}]{akbari2013universal}%
  \BibitemOpen
  \bibfield  {author} {\bibinfo {author} {\bibfnamefont {M.}~\bibnamefont
  {Akbari-Moghanjoughi}},\ }\bibfield  {title} {\bibinfo {title} {Universal
  aspects of localized excitations in graphene},\ }\href
  {https://doi.org/10.1063/1.4818707} {\bibfield  {journal} {\bibinfo
  {journal} {Journal of Applied Physics}\ }\textbf {\bibinfo {volume} {114}},\
  \bibinfo {pages} {073302} (\bibinfo {year} {2013})}\BibitemShut {NoStop}%
\bibitem [{\citenamefont {Svintsov}\ \emph {et~al.}(2013)\citenamefont
  {Svintsov}, \citenamefont {Vyurkov}, \citenamefont {Ryzhii},\ and\
  \citenamefont {Otsuji}}]{svintsov2013hydrodynamic}%
  \BibitemOpen
  \bibfield  {author} {\bibinfo {author} {\bibfnamefont {D.}~\bibnamefont
  {Svintsov}}, \bibinfo {author} {\bibfnamefont {V.}~\bibnamefont {Vyurkov}},
  \bibinfo {author} {\bibfnamefont {V.}~\bibnamefont {Ryzhii}},\ and\ \bibinfo
  {author} {\bibfnamefont {T.}~\bibnamefont {Otsuji}},\ }\bibfield  {title}
  {\bibinfo {title} {Hydrodynamic electron transport and nonlinear waves in
  graphene},\ }\href {https://doi.org/10.1103/PhysRevB.88.245444} {\bibfield
  {journal} {\bibinfo  {journal} {Physical Review B}\ }\textbf {\bibinfo
  {volume} {88}},\ \bibinfo {pages} {245444} (\bibinfo {year} {2013})},\
  \Eprint {https://arxiv.org/abs/1310.3963} {arXiv:1310.3963 [cond-mat]}
  \BibitemShut {NoStop}%
\bibitem [{Note1()}]{Note1}%
  \BibitemOpen
  \bibinfo {note} {Note that some of our variable definitions differ from those
  of \protect \citet {lucas2018hydrodynamics} to better match usual
  conventions. The relevant changes (with the variables of \protect \citet
  {lucas2018hydrodynamics} subscripted with L) are $J^{\mu } = -e J^{\mu
  }_{\protect \text {L}}$, $F^{\mu ,\nu } = -F^{\mu \nu }_{\protect \text
  {L}}/e$, and $\sigma _Q = e^2 \sigma _{Q, \protect \text {L}}$.}\BibitemShut
  {Stop}%
\bibitem [{Note2()}]{Note2}%
  \BibitemOpen
  \bibinfo {note} {Note that, as mentioned previously, $\delta \sim \partial $;
  the factor of $l_{\protect \text {ee}}$ is implicit in the definitions of the
  dissipative coefficients $\sigma _Q$, $\eta $, and $\zeta $~\protect \citep
  {lucas2018hydrodynamics}.}\BibitemShut {Stop}%
\bibitem [{\citenamefont {Gurzhi}(1963)}]{gurzhi1963minimum}%
  \BibitemOpen
  \bibfield  {author} {\bibinfo {author} {\bibfnamefont {R.~N.}\ \bibnamefont
  {Gurzhi}},\ }\bibfield  {title} {\bibinfo {title} {Minimum of resistance in
  impurity-free conductors},\ }\href
  {http://www.jetp.ac.ru/cgi-bin/dn/e_017_02_0521.pdf} {\bibfield  {journal}
  {\bibinfo  {journal} {Journal of Experimental and Theoretical Physics}\
  }\textbf {\bibinfo {volume} {17}},\ \bibinfo {pages} {521} (\bibinfo {year}
  {1963})}\BibitemShut {NoStop}%
\bibitem [{\citenamefont {Bandurin}\ \emph {et~al.}(2018)\citenamefont
  {Bandurin}, \citenamefont {Shytov}, \citenamefont {Levitov}, \citenamefont
  {Kumar}, \citenamefont {Berdyugin}, \citenamefont {Shalom}, \citenamefont
  {Grigorieva}, \citenamefont {Geim},\ and\ \citenamefont
  {Falkovich}}]{bandurin2018fluidity}%
  \BibitemOpen
  \bibfield  {author} {\bibinfo {author} {\bibfnamefont {D.~A.}\ \bibnamefont
  {Bandurin}}, \bibinfo {author} {\bibfnamefont {A.~V.}\ \bibnamefont
  {Shytov}}, \bibinfo {author} {\bibfnamefont {L.~S.}\ \bibnamefont {Levitov}},
  \bibinfo {author} {\bibfnamefont {R.~K.}\ \bibnamefont {Kumar}}, \bibinfo
  {author} {\bibfnamefont {A.~I.}\ \bibnamefont {Berdyugin}}, \bibinfo {author}
  {\bibfnamefont {M.~B.}\ \bibnamefont {Shalom}}, \bibinfo {author}
  {\bibfnamefont {I.~V.}\ \bibnamefont {Grigorieva}}, \bibinfo {author}
  {\bibfnamefont {A.~K.}\ \bibnamefont {Geim}},\ and\ \bibinfo {author}
  {\bibfnamefont {G.}~\bibnamefont {Falkovich}},\ }\bibfield  {title} {\bibinfo
  {title} {Fluidity onset in graphene},\ }\bibfield  {journal} {\bibinfo
  {journal} {Nature communications}\ }\textbf {\bibinfo {volume} {9}},\ \href
  {https://doi.org/10.1038/s41467-018-07004-4} {10.1038/s41467-018-07004-4}
  (\bibinfo {year} {2018}),\ \Eprint {https://arxiv.org/abs/1806.03231}
  {arXiv:1806.03231 [cond-mat]} \BibitemShut {NoStop}%
\bibitem [{\citenamefont {Stauber}\ \emph {et~al.}(2007)\citenamefont
  {Stauber}, \citenamefont {Peres},\ and\ \citenamefont
  {Guinea}}]{stauber2007electronic}%
  \BibitemOpen
  \bibfield  {author} {\bibinfo {author} {\bibfnamefont {T.}~\bibnamefont
  {Stauber}}, \bibinfo {author} {\bibfnamefont {N.~M.~R.}\ \bibnamefont
  {Peres}},\ and\ \bibinfo {author} {\bibfnamefont {F.}~\bibnamefont
  {Guinea}},\ }\bibfield  {title} {\bibinfo {title} {Electronic transport in
  graphene: A semiclassical approach including midgap states},\ }\href
  {https://doi.org/10.1103/PhysRevB.76.205423} {\bibfield  {journal} {\bibinfo
  {journal} {Physical Review B}\ }\textbf {\bibinfo {volume} {76}},\ \bibinfo
  {pages} {205423} (\bibinfo {year} {2007})},\ \Eprint
  {https://arxiv.org/abs/0707.3004} {arXiv:0707.3004 [cond-mat]} \BibitemShut
  {NoStop}%
\bibitem [{Note3()}]{Note3}%
  \BibitemOpen
  \bibinfo {note} {The Stefan-Boltzmann law would give a power loss rate of
  $P_r = \sigma \epsilon \bqty {(T_0+ T_1)^4-T_0^4} \approx 4 \sigma \epsilon
  T_0^3 T_1$, with $\sigma = \SI {5.67e-8}{\watt \per \meter \squared \per
  \kelvin \tothe {4}}$ and $\epsilon \le 1$ graphene's emissivity. Using
  $\epsilon \approx \SI {1}{\percent }$~\protect \citep {freitag2010thermal},
  $T_0 = \SI {60}{\kelvin }$, and $T_1 = 0.1T_0 = \SI {6.0}{\kelvin }$, we find
  a power loss density of $P_r = \SI {2.9e-7}{\kilo \watt \per \centi \meter
  \squared }$. \par As we will calculate in \protect \cref {sec:joule},
  graphene has a specific heat of $c_s = \SI {4.5e-9}{\joule \per \centi \meter
  \squared \per \kelvin }$. Therefore, the soliton's temperature will change at
  a rate of $P_r/c_s = \SI {65}{\kelvin \per \second }$. Hence, it would take
  approximately $T_1 c_s / P_r = \SI {93}{\milli \second }$ for the system to
  thermalize with the environment via radiation.}\BibitemShut {Stop}%
\bibitem [{\citenamefont {Ong}\ and\ \citenamefont
  {Pop}(2011)}]{ong2011effect}%
  \BibitemOpen
  \bibfield  {author} {\bibinfo {author} {\bibfnamefont {Z.-Y.}\ \bibnamefont
  {Ong}}\ and\ \bibinfo {author} {\bibfnamefont {E.}~\bibnamefont {Pop}},\
  }\bibfield  {title} {\bibinfo {title} {Effect of substrate modes on thermal
  transport in supported graphene},\ }\href
  {https://doi.org/10.1103/PhysRevB.84.075471} {\bibfield  {journal} {\bibinfo
  {journal} {Physical Review B}\ }\textbf {\bibinfo {volume} {84}},\ \bibinfo
  {pages} {075471} (\bibinfo {year} {2011})},\ \Eprint
  {https://arxiv.org/abs/1101.2463} {arXiv:1101.2463 [cond-mat]} \BibitemShut
  {NoStop}%
\bibitem [{\citenamefont {Chen}\ \emph {et~al.}(2017)\citenamefont {Chen},
  \citenamefont {Yan},\ and\ \citenamefont {Kumar}}]{chen2017coupled}%
  \BibitemOpen
  \bibfield  {author} {\bibinfo {author} {\bibfnamefont {L.}~\bibnamefont
  {Chen}}, \bibinfo {author} {\bibfnamefont {Z.}~\bibnamefont {Yan}},\ and\
  \bibinfo {author} {\bibfnamefont {S.}~\bibnamefont {Kumar}},\ }\bibfield
  {title} {\bibinfo {title} {Coupled electron-phonon transport and heat
  transfer pathways in graphene nanostructures},\ }\href
  {https://doi.org/10.1016/j.carbon.2017.07.095} {\bibfield  {journal}
  {\bibinfo  {journal} {Carbon}\ }\textbf {\bibinfo {volume} {123}},\ \bibinfo
  {pages} {525} (\bibinfo {year} {2017})}\BibitemShut {NoStop}%
\bibitem [{\citenamefont {Virtanen}(2014)}]{virtanen2014energy}%
  \BibitemOpen
  \bibfield  {author} {\bibinfo {author} {\bibfnamefont {P.}~\bibnamefont
  {Virtanen}},\ }\bibfield  {title} {\bibinfo {title} {Energy transport via
  multiphonon processes in graphene},\ }\href
  {https://doi.org/10.1103/PhysRevB.89.245409} {\bibfield  {journal} {\bibinfo
  {journal} {Physical Review B}\ }\textbf {\bibinfo {volume} {89}},\ \bibinfo
  {pages} {245409} (\bibinfo {year} {2014})},\ \Eprint
  {https://arxiv.org/abs/1312.3833} {arXiv:1312.3833 [cond-mat]} \BibitemShut
  {NoStop}%
\bibitem [{\citenamefont {Govorov}\ \emph {et~al.}(1999)\citenamefont
  {Govorov}, \citenamefont {Kovalev},\ and\ \citenamefont
  {Chaplik}}]{govorov1999solitons}%
  \BibitemOpen
  \bibfield  {author} {\bibinfo {author} {\bibfnamefont {A.~O.}\ \bibnamefont
  {Govorov}}, \bibinfo {author} {\bibfnamefont {V.~M.}\ \bibnamefont
  {Kovalev}},\ and\ \bibinfo {author} {\bibfnamefont {A.~V.}\ \bibnamefont
  {Chaplik}},\ }\bibfield  {title} {\bibinfo {title} {Solitons in semiconductor
  microstructures with a two-dimensional electron gas},\ }\href
  {https://doi.org/10.1134/1.568201} {\bibfield  {journal} {\bibinfo  {journal}
  {JETP Letters}\ }\textbf {\bibinfo {volume} {70}},\ \bibinfo {pages} {488}
  (\bibinfo {year} {1999})}\BibitemShut {NoStop}%
\bibitem [{Note4()}]{Note4}%
  \BibitemOpen
  \bibinfo {note} {After choosing $\hbar =v_F=k_B=e=1$, all quantities will be
  expressed in various powers of length. If the parameters have been chosen
  correctly, there will exist a characteristic length $\Xi $ shared by all
  quantities. It is most convienent to choose $l_{\protect \text {ref}}=\Xi $,
  though it is not strictly necessary---choosing $l_{\protect \text {ref}}$
  otherwise will multiply all terms in each equation by the same factor of
  $l_{\protect \text {ref}}/\Xi $.}\BibitemShut {Stop}%
\bibitem [{Note5()}]{Note5}%
  \BibitemOpen
  \bibinfo {note} {As discussed in \protect \cref {sec:diss_order}, we could
  introduce three additional microscopic equations and eliminate $\eta $,
  $\zeta $, and $\sigma _Q/e^2$ as independent quantities. However, we will
  refrain from doing so.}\BibitemShut {Stop}%
\bibitem [{\citenamefont {Fritz}\ \emph {et~al.}(2008)\citenamefont {Fritz},
  \citenamefont {Schmalian}, \citenamefont {M{\"u}ller},\ and\ \citenamefont
  {Sachdev}}]{fritz2008quantum}%
  \BibitemOpen
  \bibfield  {author} {\bibinfo {author} {\bibfnamefont {L.}~\bibnamefont
  {Fritz}}, \bibinfo {author} {\bibfnamefont {J.}~\bibnamefont {Schmalian}},
  \bibinfo {author} {\bibfnamefont {M.}~\bibnamefont {M{\"u}ller}},\ and\
  \bibinfo {author} {\bibfnamefont {S.}~\bibnamefont {Sachdev}},\ }\bibfield
  {title} {\bibinfo {title} {Quantum critical transport in clean graphene},\
  }\href {https://doi.org/10.1103/PhysRevB.78.085416} {\bibfield  {journal}
  {\bibinfo  {journal} {Physical Review B}\ }\textbf {\bibinfo {volume} {78}},\
  \bibinfo {pages} {085416} (\bibinfo {year} {2008})},\ \Eprint
  {https://arxiv.org/abs/0802.4289} {arXiv:0802.4289 [cond-mat]} \BibitemShut
  {NoStop}%
\bibitem [{\citenamefont {Mei}\ \emph {et~al.}(2005)\citenamefont {Mei},
  \citenamefont {Stiassnie},\ and\ \citenamefont {Yue}}]{mei2005theory}%
  \BibitemOpen
  \bibfield  {author} {\bibinfo {author} {\bibfnamefont {C.~C.}\ \bibnamefont
  {Mei}}, \bibinfo {author} {\bibfnamefont {M.}~\bibnamefont {Stiassnie}},\
  and\ \bibinfo {author} {\bibfnamefont {D.~K.-P.}\ \bibnamefont {Yue}},\
  }\href@noop {} {\emph {\bibinfo {title} {Theory and applications of ocean
  surface waves: nonlinear aspects}}},\ Vol.~\bibinfo {volume} {23}\ (\bibinfo
  {publisher} {World scientific},\ \bibinfo {year} {2005})\BibitemShut
  {NoStop}%
\bibitem [{\citenamefont {Akbari-Moghanjoughi}(2012)}]{akbari2012higher}%
  \BibitemOpen
  \bibfield  {author} {\bibinfo {author} {\bibfnamefont {M.}~\bibnamefont
  {Akbari-Moghanjoughi}},\ }\bibfield  {title} {\bibinfo {title} {Higher-order
  nonlinear electron-acoustic solitary excitations in partially degenerate
  quantum electron-ion plasmas},\ }\href
  {https://doi.org/10.1007/s12648-012-0071-9} {\bibfield  {journal} {\bibinfo
  {journal} {Indian Journal of Physics}\ }\textbf {\bibinfo {volume} {86}},\
  \bibinfo {pages} {413} (\bibinfo {year} {2012})},\ \Eprint
  {https://arxiv.org/abs/1109.1847} {arXiv:1109.1847 [astro-ph]} \BibitemShut
  {NoStop}%
\bibitem [{Note6()}]{Note6}%
  \BibitemOpen
  \bibinfo {note} {Note that it is possible to generate a stationary soliton by
  appropriate choice of $F_1$ instead, though the resulting coefficients will
  be different.}\BibitemShut {Stop}%
\bibitem [{Note7()}]{Note7}%
  \BibitemOpen
  \bibinfo {note} {A few terms were simplified using Kronecker deltas in
  \protect \cref {sec:kdv_burgers,sec:kdv_burgers_adi}. For instance,
  substituting the dimensional expressions into $\protect \mathcal {G}'$
  generates an $\epsilon ^{-q}$ term multiplying $\sigma _Q$ and an $\epsilon
  ^{-p}$ term multiplying $\eta $ and $\zeta $. However, these can be
  neglected: as mentioned at the end of \protect \cref {sec:nondim}, $\sigma
  _Q$ carries an implicit $\delta _{q,0}$ while $\eta $ and $\zeta $ have
  implicit $\delta _{q,0}$ and $\delta _{q,0} \order {\zeta }/\order {\eta }$,
  respectively. Similarly, the thermodynamic contribution of $\protect \mathcal
  {F}'$ has a factor of $\epsilon ^{-m^2}$; however, given the presence of the
  Kronecker deltas, this is equivalent to $\epsilon ^{-1}$.}\BibitemShut
  {Stop}%
\bibitem [{Note8()}]{Note8}%
  \BibitemOpen
  \bibinfo {note} {Actually, as written, the coefficients in \protect \cref
  {sec:kdv_burgers,sec:kdv_burgers_adi} have all had a common factor of
  $\epsilon ^{p/2-q/2} \protect \sqrt {\order {\sigma _Q}/\order {\eta }}$
  removed for brevity.}\BibitemShut {Stop}%
\bibitem [{Note9()}]{Note9}%
  \BibitemOpen
  \bibinfo {note} {Hence, $c_1$ is the normalized, order-unity analog of
  $n_{\protect \text {max}}$.}\BibitemShut {Stop}%
\bibitem [{\citenamefont {Svintsov}\ \emph {et~al.}(2012)\citenamefont
  {Svintsov}, \citenamefont {Vyurkov}, \citenamefont {Yurchenko}, \citenamefont
  {Otsuji},\ and\ \citenamefont {Ryzhii}}]{svintsov2012hydrodynamic}%
  \BibitemOpen
  \bibfield  {author} {\bibinfo {author} {\bibfnamefont {D.}~\bibnamefont
  {Svintsov}}, \bibinfo {author} {\bibfnamefont {V.}~\bibnamefont {Vyurkov}},
  \bibinfo {author} {\bibfnamefont {S.}~\bibnamefont {Yurchenko}}, \bibinfo
  {author} {\bibfnamefont {T.}~\bibnamefont {Otsuji}},\ and\ \bibinfo {author}
  {\bibfnamefont {V.}~\bibnamefont {Ryzhii}},\ }\bibfield  {title} {\bibinfo
  {title} {Hydrodynamic model for electron-hole plasma in graphene},\ }\href
  {https://doi.org/10.1063/1.4705382} {\bibfield  {journal} {\bibinfo
  {journal} {Journal of Applied Physics}\ }\textbf {\bibinfo {volume} {111}},\
  \bibinfo {pages} {083715} (\bibinfo {year} {2012})},\ \Eprint
  {https://arxiv.org/abs/1201.0592} {arXiv:1201.0592 [cond-mat]} \BibitemShut
  {NoStop}%
\bibitem [{Note10()}]{Note10}%
  \BibitemOpen
  \bibinfo {note} {The sign of the $\beta ^2$ term multiplying $u \partial _x
  u$ in Eq.{} (16) of Ref.\ \protect \citep {svintsov2013hydrodynamic} should
  be flipped. Additionally, the expression for $F(\nu )$ in Eq.{} (26) should
  read \begin {equation} \begin {aligned} F(\nu ) &= \protect \mathaccentV
  {tilde}07E{s}_0^2 -\protect \frac {\beta ^2}{2} -\protect \frac {\beta
  _0^2}{1+\nu } \\ &\hskip 2em\relax - \protect \frac {\beta _0^2 \beta
  ^2}{(1+\nu )^2} \protect \frac {5-6\xi }{1-\beta ^2} + \protect \frac {\nu
  \beta _0^2 (3-4 \xi )}{(1+\nu )^2} \protect \tmspace +\thinmuskip {.1667em}.
  \end {aligned} \end {equation} In the KdV equation, Eq.{} (27), the
  coefficient of the $\nu \partial _{\zeta } \nu $ term should be \begin
  {equation} (1-\xi ) \pqty {2\protect \mathaccentV {tilde}07E{s}_0^2-\protect
  \frac {4}{3}\xi +4\beta _0^2} \protect \tmspace +\thinmuskip {.1667em}. \end
  {equation} Also, the solution to the KdV equation, Eq.{} (28), should be
  \begin {equation} \delta n(z) = \delta n_{\protect \text {max}} \cosh
  ^{-2}\bqty {\protect \frac {z}{2} \protect \sqrt {\protect \frac {\protect
  \bm {2}}{d_1 d_2} \protect \frac {s_0^2}{2 s_0^2-v_F^2}\protect \frac {\delta
  n_{\protect \text {max}}}{n_0}}} \protect \tmspace +\thinmuskip {.1667em},
  \end {equation} with Eq.{} (29) changed to \begin {equation} \delta
  n_{\protect \text {max}} = \protect \bm {3} \protect \frac {n_0}{2} \protect
  \frac {u_0^2-s_0^2}{s_0^2} \protect \tmspace +\thinmuskip {.1667em}, \end
  {equation} with corrections highlighted in bold. \par For the $u_0 \not =0$
  case, Eq.(34) should be adjusted by flipping the sign of the $\gamma $ term
  multiplying the $u_0 \partial _x \delta u$ term. Furthermore, the dispersion
  relation, Eq.(36), should read \begin {equation} \scriptstyle s_{\pm } =
  \protect \frac {u_0 (2-2 \xi _0+\gamma ) \pm \protect \sqrt {s_0^2(1+\gamma )
  + u_0^2 \bqty {(2-2\xi _0+\gamma )^2 -(1+\gamma )\pqty {3-\protect \frac
  {10}{3}\xi _0+\gamma }} }}{1+\gamma } \protect \tmspace +\thinmuskip
  {.1667em}. \end {equation}}\BibitemShut {NoStop}%
\bibitem [{Note11()}]{Note11}%
  \BibitemOpen
  \bibinfo {note} {Note that \protect \citet {svintsov2013hydrodynamic} include
  factors of $\gamma $ in the definitions of $\varepsilon $ and $n$; here, they
  have been factored out to match our definitions.}\BibitemShut {Stop}%
\bibitem [{Note12()}]{Note12}%
  \BibitemOpen
  \bibinfo {note} {Note that \protect \citet {akbari2013universal} uses a
  different terminology. There, the term ``Dirac fluid'' refers to massless
  fermions (as in graphene) while ``Fermi liquid'' refers to massive fermions.
  Both of these are dealt with in the completely degenerate $T=0$ limit. By
  contrast, we follow the terminology of \protect \citet
  {lucas2018hydrodynamics} to analyze both a ``Fermi liquid'' ($k_B T \ll \mu
  $) and ``Dirac fluid'' ($\mu \ll k_B T$) regime for massless fermions.
  Therefore, the ``Dirac'' results in \protect \citet {akbari2013universal}
  correspond to our $T=0$ Fermi regime, while the ``Fermi'' results correspond
  to massive fermions not discussed here. Interestingly, bilayer graphene can
  induce such an effective mass for the quasiparticle excitations~\protect
  \citep {mccann2006landau}.}\BibitemShut {Stop}%
\bibitem [{Note13()}]{Note13}%
  \BibitemOpen
  \bibinfo {note} {Here we used the fact that $\protect \sgn (\protect \mathcal
  {A}'\protect \mathcal {C}') = \protect \sgn (v_0)$ for $u_0=0$}\BibitemShut
  {NoStop}%
\bibitem [{Note14()}]{Note14}%
  \BibitemOpen
  \bibinfo {note} {Note that this expression has a removable singularity at
  $u_0 = 0$; however, the double-sided limit exists and is $0$.}\BibitemShut
  {Stop}%
\bibitem [{Note15()}]{Note15}%
  \BibitemOpen
  \bibinfo {note} {A number of other minor differences exist between our work
  and that of \protect \citet {akbari2013universal}: there, velocities were
  normalized by $c$, giving $v_c = c/\protect \sqrt {d}$. However, we found it
  more useful to normalize by $v_F$---yielding $v_c = v_F/\protect \sqrt {d}$.
  This difference arose because \protect \citet {akbari2013universal} chose to
  define $u^{\mu } = (c,\protect \mathbf {\protect \bm {u}})/\protect \sqrt
  {1-(u/c)^2}$ following~\protect \citet {zhu2010relativistic}, while we
  defined $u^{\mu } = (v_F,\protect \mathbf {\protect \bm {u}})/\protect \sqrt
  {1-(u/v_F)^2}$. Again, the choice of $v_F$, as opposed to $c$, is preferred
  since it preserves the form of the dispersion relation. Replacing the
  original choice of $u^{\mu }$ (involving $c$) with our choice (involving
  $v_F$) in \protect \citeauthor {akbari2013universal}'s derivation yields $v_c
  = v_F/\protect \sqrt {d}$, \protect \ie our minimum propagation speed. \par
  Finally, our expressions for the pressure differ slightly: it appears
  \protect \citet {akbari2013universal} considered only $g=2$ spin degeneracy
  in Eq.{} (4), rather than graphene's $g=4$ spin/valley degeneracy. This only
  affects the normalization constant ($A_{2D}$ or $A_{3D}$ in, for example,
  Eq.{} (11)), and the subsequent conclusions are unaffected.}\BibitemShut
  {Stop}%
\bibitem [{Note16()}]{Note16}%
  \BibitemOpen
  \bibinfo {note} {This can be seen by noting that the expression is positive
  for $u_0=0$ and only crosses zero at $\pm 1$, $\pm \protect \sqrt {1+\lambda
  d}$, or $\pm \protect \sqrt {1+\lambda d}/\protect \sqrt {1+\lambda }$, with
  $\lambda \mathrel {\mathop :}\mathrel {\mkern -1.2mu}=A n_0^2/P_0(d+1)$.
  These are each greater than (or equal to) unity for $d\ge 1$; therefore, the
  entire expression is non-negative for $\abs {u_0} \le 1$.}\BibitemShut
  {Stop}%
\bibitem [{Note17()}]{Note17}%
  \BibitemOpen
  \bibinfo {note} {The $\mu _0 n_0$ term is non-negative because $\protect \sgn
  \mu _0 = \protect \sgn n_0$; \protect \cf \protect \cref
  {eq:dirac_n0_adi}.}\BibitemShut {Stop}%
\bibitem [{Note18()}]{Note18}%
  \BibitemOpen
  \bibinfo {note} {Note that we have added an additional factor to the $\sigma
  _Q$ term in order to account for the electrostatic interactions.}\BibitemShut
  {Stop}%
\bibitem [{\citenamefont {Dmitriev}\ \emph {et~al.}(2001)\citenamefont
  {Dmitriev}, \citenamefont {Kachorovskii},\ and\ \citenamefont
  {Shur}}]{dmitriev2001plasma}%
  \BibitemOpen
  \bibfield  {author} {\bibinfo {author} {\bibfnamefont {A.~P.}\ \bibnamefont
  {Dmitriev}}, \bibinfo {author} {\bibfnamefont {V.~Y.}\ \bibnamefont
  {Kachorovskii}},\ and\ \bibinfo {author} {\bibfnamefont {M.~S.}\ \bibnamefont
  {Shur}},\ }\bibfield  {title} {\bibinfo {title} {Plasma wave instability in
  gated collisionless two-dimensional electron gas},\ }\href
  {https://doi.org/10.1063/1.1391395} {\bibfield  {journal} {\bibinfo
  {journal} {Applied Physics Letters}\ }\textbf {\bibinfo {volume} {79}},\
  \bibinfo {pages} {922} (\bibinfo {year} {2001})}\BibitemShut {NoStop}%
\bibitem [{\citenamefont {Popov}(2002)}]{popov2002low}%
  \BibitemOpen
  \bibfield  {author} {\bibinfo {author} {\bibfnamefont {V.~N.}\ \bibnamefont
  {Popov}},\ }\bibfield  {title} {\bibinfo {title} {Low-temperature specific
  heat of nanotube systems},\ }\href
  {https://doi.org/10.1103/PhysRevB.66.153408} {\bibfield  {journal} {\bibinfo
  {journal} {Physical Review B}\ }\textbf {\bibinfo {volume} {66}},\ \bibinfo
  {pages} {153408} (\bibinfo {year} {2002})}\BibitemShut {NoStop}%
\bibitem [{\citenamefont {Bong}\ \emph {et~al.}(2015)\citenamefont {Bong},
  \citenamefont {Jo}, \citenamefont {Kang}, \citenamefont {Lee}, \citenamefont
  {Kim}, \citenamefont {Lee},\ and\ \citenamefont {Cho}}]{bong2015graphene}%
  \BibitemOpen
  \bibfield  {author} {\bibinfo {author} {\bibfnamefont {H.}~\bibnamefont
  {Bong}}, \bibinfo {author} {\bibfnamefont {S.~B.}\ \bibnamefont {Jo}},
  \bibinfo {author} {\bibfnamefont {B.}~\bibnamefont {Kang}}, \bibinfo {author}
  {\bibfnamefont {S.~K.}\ \bibnamefont {Lee}}, \bibinfo {author} {\bibfnamefont
  {H.~H.}\ \bibnamefont {Kim}}, \bibinfo {author} {\bibfnamefont {S.~G.}\
  \bibnamefont {Lee}},\ and\ \bibinfo {author} {\bibfnamefont {K.}~\bibnamefont
  {Cho}},\ }\bibfield  {title} {\bibinfo {title} {Graphene growth under knudsen
  molecular flow on a confined catalytic metal coil},\ }\href
  {https://doi.org/10.1039/C4NR04153D} {\bibfield  {journal} {\bibinfo
  {journal} {Nanoscale}\ }\textbf {\bibinfo {volume} {7}},\ \bibinfo {pages}
  {1314} (\bibinfo {year} {2015})}\BibitemShut {NoStop}%
\bibitem [{\citenamefont {Ablowitz}\ \emph {et~al.}(1974)\citenamefont
  {Ablowitz}, \citenamefont {Kaup}, \citenamefont {Newell},\ and\ \citenamefont
  {Segur}}]{ablowitz1974inverse}%
  \BibitemOpen
  \bibfield  {author} {\bibinfo {author} {\bibfnamefont {M.~J.}\ \bibnamefont
  {Ablowitz}}, \bibinfo {author} {\bibfnamefont {D.~J.}\ \bibnamefont {Kaup}},
  \bibinfo {author} {\bibfnamefont {A.~C.}\ \bibnamefont {Newell}},\ and\
  \bibinfo {author} {\bibfnamefont {H.}~\bibnamefont {Segur}},\ }\bibfield
  {title} {\bibinfo {title} {The inverse scattering transform-fourier analysis
  for nonlinear problems},\ }\href {https://doi.org/10.1002/sapm1974534249}
  {\bibfield  {journal} {\bibinfo  {journal} {Studies in Applied Mathematics}\
  }\textbf {\bibinfo {volume} {53}},\ \bibinfo {pages} {249} (\bibinfo {year}
  {1974})}\BibitemShut {NoStop}%
\bibitem [{\citenamefont {Gardner}\ \emph {et~al.}(1967)\citenamefont
  {Gardner}, \citenamefont {Greene}, \citenamefont {Kruskal},\ and\
  \citenamefont {Miura}}]{gardner1967method}%
  \BibitemOpen
  \bibfield  {author} {\bibinfo {author} {\bibfnamefont {C.~S.}\ \bibnamefont
  {Gardner}}, \bibinfo {author} {\bibfnamefont {J.~M.}\ \bibnamefont {Greene}},
  \bibinfo {author} {\bibfnamefont {M.~D.}\ \bibnamefont {Kruskal}},\ and\
  \bibinfo {author} {\bibfnamefont {R.~M.}\ \bibnamefont {Miura}},\ }\bibfield
  {title} {\bibinfo {title} {Method for solving the korteweg-devries
  equation},\ }\href {https://doi.org/10.1103/PhysRevLett.19.1095} {\bibfield
  {journal} {\bibinfo  {journal} {Physical review letters}\ }\textbf {\bibinfo
  {volume} {19}},\ \bibinfo {pages} {1095} (\bibinfo {year}
  {1967})}\BibitemShut {NoStop}%
\bibitem [{\citenamefont {Kiselev}\ and\ \citenamefont
  {Schmalian}(2019)}]{kiselev2018boundary}%
  \BibitemOpen
  \bibfield  {author} {\bibinfo {author} {\bibfnamefont {E.~I.}\ \bibnamefont
  {Kiselev}}\ and\ \bibinfo {author} {\bibfnamefont {J.}~\bibnamefont
  {Schmalian}},\ }\bibfield  {title} {\bibinfo {title} {Boundary conditions of
  viscous electron flow},\ }\href {https://doi.org/10.1103/PhysRevB.99.035430}
  {\bibfield  {journal} {\bibinfo  {journal} {Phys. Rev. B}\ }\textbf {\bibinfo
  {volume} {99}},\ \bibinfo {pages} {035430} (\bibinfo {year} {2019})},\
  \Eprint {https://arxiv.org/abs/1806.03933} {arXiv:1806.03933 [cond-mat]}
  \BibitemShut {NoStop}%
\bibitem [{\citenamefont {Coelho}\ \emph {et~al.}(2017)\citenamefont {Coelho},
  \citenamefont {Mendoza}, \citenamefont {Doria},\ and\ \citenamefont
  {Herrmann}}]{coelho2017kelvin}%
  \BibitemOpen
  \bibfield  {author} {\bibinfo {author} {\bibfnamefont {R.~C.~V.}\
  \bibnamefont {Coelho}}, \bibinfo {author} {\bibfnamefont {M.}~\bibnamefont
  {Mendoza}}, \bibinfo {author} {\bibfnamefont {M.~M.}\ \bibnamefont {Doria}},\
  and\ \bibinfo {author} {\bibfnamefont {H.~J.}\ \bibnamefont {Herrmann}},\
  }\bibfield  {title} {\bibinfo {title} {Kelvin-helmholtz instability of the
  dirac fluid of charge carriers on graphene},\ }\href
  {https://doi.org/10.1103/PhysRevB.96.184307} {\bibfield  {journal} {\bibinfo
  {journal} {Physical Review B}\ }\textbf {\bibinfo {volume} {96}},\ \bibinfo
  {pages} {184307} (\bibinfo {year} {2017})},\ \Eprint
  {https://arxiv.org/abs/1706.00801} {arXiv:1706.00801 [cond-mat]} \BibitemShut
  {NoStop}%
\bibitem [{\citenamefont {a~division~of Waterloo
  Maple~Inc.}(2018)}]{maple2018}%
  \BibitemOpen
  \bibfield  {author} {\bibinfo {author} {\bibfnamefont {M.}~\bibnamefont
  {a~division~of Waterloo Maple~Inc.}},\ }\href@noop {} {\bibinfo {title}
  {Maple 2018}} (\bibinfo {year} {2018}),\ \bibinfo {note} {waterloo,
  Ontario}\BibitemShut {NoStop}%
\bibitem [{\citenamefont {Wood}(1992)}]{wood1992computation}%
  \BibitemOpen
  \bibfield  {author} {\bibinfo {author} {\bibfnamefont {D.}~\bibnamefont
  {Wood}},\ }\href {https://www.cs.kent.ac.uk/pubs/1992/110} {\emph {\bibinfo
  {title} {The Computation of Polylogarithms}}},\ \bibinfo {type} {Tech. Rep.}\
  \bibinfo {number} {15-92*}\ (\bibinfo  {institution} {University of Kent,
  Computing Laboratory},\ \bibinfo {address} {University of Kent, Canterbury,
  UK},\ \bibinfo {year} {1992})\BibitemShut {NoStop}%
\bibitem [{\citenamefont {Lucas}\ \emph {et~al.}(2016)\citenamefont {Lucas},
  \citenamefont {Crossno}, \citenamefont {Fong}, \citenamefont {Kim},\ and\
  \citenamefont {Sachdev}}]{lucas2016transport}%
  \BibitemOpen
  \bibfield  {author} {\bibinfo {author} {\bibfnamefont {A.}~\bibnamefont
  {Lucas}}, \bibinfo {author} {\bibfnamefont {J.}~\bibnamefont {Crossno}},
  \bibinfo {author} {\bibfnamefont {K.~C.}\ \bibnamefont {Fong}}, \bibinfo
  {author} {\bibfnamefont {P.}~\bibnamefont {Kim}},\ and\ \bibinfo {author}
  {\bibfnamefont {S.}~\bibnamefont {Sachdev}},\ }\bibfield  {title} {\bibinfo
  {title} {Transport in inhomogeneous quantum critical fluids and in the dirac
  fluid in graphene},\ }\href {https://doi.org/10.1103/PhysRevB.93.075426}
  {\bibfield  {journal} {\bibinfo  {journal} {Physical Review B}\ }\textbf
  {\bibinfo {volume} {93}},\ \bibinfo {pages} {075426} (\bibinfo {year}
  {2016})},\ \Eprint {https://arxiv.org/abs/1510.01738} {arXiv:1510.01738
  [cond-mat]} \BibitemShut {NoStop}%
\bibitem [{Note19()}]{Note19}%
  \BibitemOpen
  \bibinfo {note} {Note that one combination of parameters is not allowed in
  this derivation: $m < -1$ and $q=0$. Owing to the thermodynamic relations, $m
  < -1$ implies that $T_1$ will depend on density and pressure of the form
  $n_{1+\abs {m}}$ and $P_{1+\abs {m}}$. We are able to manipulate the results
  for $m=-1$ (\protect \cf \protect \cref {sec:MMS_first_order_adi}) to handle
  these $n_2$ and $P_2$ terms. However, for $m < -1$, these terms cannot be
  eliminated. If $q > 0$, then $\mu _1$ and $T_1$ do not appear in our
  first-order corrections, so this is acceptable; if $q = 0$, we would have
  these $n_{1+\abs {m}}$ and $P_{1+\abs {m}}$ terms which cannot be
  eliminated.}\BibitemShut {Stop}%
\bibitem [{Note20()}]{Note20}%
  \BibitemOpen
  \bibinfo {note} {Furthermore, $m<-1$ precludes the choice of $q=0$; see
  footnote{} \cite {Note19}}\BibitemShut {NoStop}%
\bibitem [{Note21()}]{Note21}%
  \BibitemOpen
  \bibinfo {note} {Note that the expression for $\sigma _Q$ in the Fermi regime
  lacks numerical factors; see \protect \citet {muller2008quantum} for the
  exact expression for the (screened) Fermi case.}\BibitemShut {Stop}%
\bibitem [{Note22()}]{Note22}%
  \BibitemOpen
  \bibinfo {note} {Equivalently, \protect \citet {lucas2016transport} prove
  $P(\mu ,T)$ only involves even powers by recognizing that the equation of
  state is charge conjugation invariant.}\BibitemShut {Stop}%
\bibitem [{Note23()}]{Note23}%
  \BibitemOpen
  \bibinfo {note} {Note that it is possible to generate a stationary soliton by
  appropriate choice of $F_1$ or $F_3$ instead, though the resulting
  coefficients will be different.}\BibitemShut {Stop}%
\bibitem [{\citenamefont {Freitag}\ \emph {et~al.}(2010)\citenamefont
  {Freitag}, \citenamefont {Chiu}, \citenamefont {Steiner}, \citenamefont
  {Perebeinos},\ and\ \citenamefont {Avouris}}]{freitag2010thermal}%
  \BibitemOpen
  \bibfield  {author} {\bibinfo {author} {\bibfnamefont {M.}~\bibnamefont
  {Freitag}}, \bibinfo {author} {\bibfnamefont {H.-Y.}\ \bibnamefont {Chiu}},
  \bibinfo {author} {\bibfnamefont {M.}~\bibnamefont {Steiner}}, \bibinfo
  {author} {\bibfnamefont {V.}~\bibnamefont {Perebeinos}},\ and\ \bibinfo
  {author} {\bibfnamefont {P.}~\bibnamefont {Avouris}},\ }\bibfield  {title}
  {\bibinfo {title} {Thermal infrared emission from biased graphene},\ }\href
  {https://doi.org/10.1038/nnano.2010.90} {\bibfield  {journal} {\bibinfo
  {journal} {Nature nanotechnology}\ }\textbf {\bibinfo {volume} {5}},\
  \bibinfo {pages} {497} (\bibinfo {year} {2010})},\ \Eprint
  {https://arxiv.org/abs/1004.0369} {arXiv:1004.0369 [cond-mat]} \BibitemShut
  {NoStop}%
\bibitem [{\citenamefont {McCann}\ and\ \citenamefont
  {Fal’ko}(2006)}]{mccann2006landau}%
  \BibitemOpen
  \bibfield  {author} {\bibinfo {author} {\bibfnamefont {E.}~\bibnamefont
  {McCann}}\ and\ \bibinfo {author} {\bibfnamefont {V.~I.}\ \bibnamefont
  {Fal’ko}},\ }\bibfield  {title} {\bibinfo {title} {Landau-level degeneracy
  and quantum hall effect in a graphite bilayer},\ }\href
  {https://doi.org/10.1103/PhysRevLett.96.086805} {\bibfield  {journal}
  {\bibinfo  {journal} {Physical Review Letters}\ }\textbf {\bibinfo {volume}
  {96}},\ \bibinfo {pages} {086805} (\bibinfo {year} {2006})},\ \Eprint
  {https://arxiv.org/abs/cond-mat/0510237} {cond-mat/0510237} \BibitemShut
  {NoStop}%
\bibitem [{\citenamefont {Zhu}\ and\ \citenamefont
  {Ji}(2010)}]{zhu2010relativistic}%
  \BibitemOpen
  \bibfield  {author} {\bibinfo {author} {\bibfnamefont {J.}~\bibnamefont
  {Zhu}}\ and\ \bibinfo {author} {\bibfnamefont {P.}~\bibnamefont {Ji}},\
  }\bibfield  {title} {\bibinfo {title} {Relativistic quantum corrections to
  laser wakefield acceleration},\ }\href
  {https://doi.org/10.1103/PhysRevE.81.036406} {\bibfield  {journal} {\bibinfo
  {journal} {Physical Review E}\ }\textbf {\bibinfo {volume} {81}},\ \bibinfo
  {pages} {036406} (\bibinfo {year} {2010})}\BibitemShut {NoStop}%
\bibitem [{\citenamefont {M{\"u}ller}\ \emph {et~al.}(2008)\citenamefont
  {M{\"u}ller}, \citenamefont {Fritz},\ and\ \citenamefont
  {Sachdev}}]{muller2008quantum}%
  \BibitemOpen
  \bibfield  {author} {\bibinfo {author} {\bibfnamefont {M.}~\bibnamefont
  {M{\"u}ller}}, \bibinfo {author} {\bibfnamefont {L.}~\bibnamefont {Fritz}},\
  and\ \bibinfo {author} {\bibfnamefont {S.}~\bibnamefont {Sachdev}},\
  }\bibfield  {title} {\bibinfo {title} {Quantum-critical relativistic
  magnetotransport in graphene},\ }\href
  {https://doi.org/10.1103/PhysRevB.78.115406} {\bibfield  {journal} {\bibinfo
  {journal} {Physical Review B}\ }\textbf {\bibinfo {volume} {78}},\ \bibinfo
  {pages} {115406} (\bibinfo {year} {2008})},\ \Eprint
  {https://arxiv.org/abs/0805.1413} {arXiv:0805.1413 [cond-mat]} \BibitemShut
  {NoStop}%
\end{thebibliography}
